\documentclass[reprint,aps,superscriptaddress,nofootinbib,preprintnumbers,longbibliography]{revtex4-1}
\usepackage{graphicx, amsmath, amssymb, amstext}
\usepackage{relsize, float, multirow}
\usepackage{array, epstopdf, hyperref, pbox,makecell,booktabs}
\usepackage{mathtools,pbox}
\usepackage{dcolumn,color}
\usepackage{soul,lipsum}
\usepackage{hyperref,colortbl,subcaption}
\usepackage[utf8]{inputenc}

\newcommand{\id}{{\rm d}}

\newcommand{\hmpci}{  h \ \text{Mpc}^{-1}}
\newcommand{\km}{k_\text{max}}
\newcommand{\hunit}{\rm ~km\,s^{-1}Mpc^{-1}}
\newcommand{\tast}{\theta_{\ast}}
\newcommand{\rast}{r_{\ast}}
\newcommand{\zast}{z_{\ast}}
\newcommand{\Om}{\Omega_{\rm m}}

\newcommand{\Or}{\Omega_{\rm r}}
\newcommand{\OK}{\Omega_{\rm K}}

\newcommand{\kp}{k_{\rm peak}}
\newcommand{\DA}{D_{\rm A}}
\newcommand{\ellp}{\ell_{\rm peak}}
\newcommand{\kpspm}{\rm km/s/Mpc}
\newcommand{\dmu}{\dot{\mu}}

\def\mnras{MNRAS}
\def\apj{ApJ}

\def\prd{Phys. Rev. D }

\def\apss{Astrophysics and Space Science}
\def\jcap{JCAP}

\begin{document}

\title{Dark neutrino interactions phase out the Hubble tension}


\author{Subhajit Ghosh} 
\email{subhajit@theory.tifr.res.in}
\author{Rishi Khatri}
\email{khatri@theory.tifr.res.in}
\author{and Tuhin S.~Roy}
\email{tuhin@theory.tifr.res.in}
\affiliation{Department of Theoretical Physics, Tata Institute of Fundamental Research, Mumbai 400005, India}

\preprint{TIFR/TH/19-31}

\begin{abstract}
	New interactions of neutrinos can stop them from free streaming even after
the weak interaction freeze-out. This results in a phase shift in the cosmic
microwave background (CMB) acoustic peaks which can alleviate the Hubble
tension. In addition, the perturbations in neutrinos do not decay away on horizon entry and contribute to metric perturbation enhancing the matter power spectrum. We demonstrate 
that this acoustic phase shift 
 can be achieved using new interactions of standard left-handed neutrinos
with dark matter
without changing the number of effective relativistic degrees of
	freedom.
Using Planck CMB and the WiggleZ galaxy survey $ (k\le 0.12 \hmpci ) $ data, we demonstrate that in this model the Hubble tension reduces to approximately $ 2.1 \sigma$.
Our model predicts potentially observable modifications of the CMB B-modes and the matter power spectrum that can be observed in future data sets.

\end{abstract}


\maketitle

\section{Introduction} The values of the Hubble constant ($H_0$) inferred from  cosmic microwave
background (CMB) anisotropies ($ 67.5 \pm 0.6
\hunit$~\cite{Ade:2015xua,Aghanim:2018eyx}) and 
baryon acoustic oscillations
(BAO) measurements ($66.98 \pm 1.18 \hunit$~\cite{Beutler:2011hx,Font-Ribera:2013wce,Delubac:2014aqe,Ross:2014qpa,Addison:2017fdm,alam2017})
are significantly smaller than the measurements from observations  of the
nearby Universe using the distance ladder ($74.03 \pm  1.42
\hunit$~\cite{riess2016,riess2018,Riess:2019cxk}).  The
gravitational lensing time delay measurements in multiply imaged quasar
systems  which are independent of the cosmic distance ladder also gives a
higher value ($72.5^{+2.1}_{-2.3}\hunit$~\cite{h0licow2017,h0licow2019}).
Recently, an independent calibration of distance ladder without using the cepheids but using the tip of the red giant branch gives a value of Hubble $ H_0 = 69.8 \pm 0.8 (\text{stat.}) \pm 1.7(\text{sys.})  $, which is in between the Planck and cepheid based distance ladder values~\cite{freedman2019}; see also Refs.~\cite{yuan2019,Freedman:2020dne}.  
 This tension, calculated
using Gaussian error bars, between the Planck CMB and local Hubble
measurement  stands {at approximately $ 4\,\sigma$
  \cite{Riess:2019cxk,Aghanim:2018eyx}.}
  Increasingly, this
tension is being seen as a hint of physics beyond the  $\Lambda$CDM
cosmology~\cite{barreira,umilta2015,Lesgourgues:2015wza,Alam:2016wpf,DiValentino:2016hlg, Qing-Guo:2016ykt,Ko:2016uft,Karwal:2016vyq,Kumar:2016zpg,renk2017,DiValentino:2017zyq,DiValentino:2017iww,DiValentino:2017oaw,
	Bolejko:2017fos,DiValentino:2017rcr,Lancaster:2017ksf,Khosravi:2017hfi,Buen-Abad:2017gxg, DEramo:2018vss,Dutta:2018vmq,Banihashemi:2018has, belgacem2018,Pandey:2019plg,Agrawal:2019lmo,Agrawal:2019dlm, DiValentino:2019exe,
	Desmond:2019ygn, Pan:2019cot,  
  Vattis:2019efj,Poulin:2018cxd,lin2019,Li:2019san,Alexander:2019rsc,lin2019b,DiValentino:2019ffd,Archidiacono:2019wdp,Knox:2019rjx}, rather than a
manifestation of possible systematics in the local distance
ladder~\cite{Shi:1997aa,efs2014,Aubourg:2014yra,Odderskov:2014hqa,Macaulay:2018fxi,Aylor:2018drw,Taubenberger:2019qna,Kenworthy:2019qwq,Rameez:2019wdt}.

The spectacular success of the standard models of cosmology and particle
physics in describing all cosmological and particle physics observables,
however, makes the task of explaining the Hubble tension from \emph{new
  physics} (NP) rather nontrivial.  Particularly in this context, if the
CMB data are to be reinterpreted with NP, the peaks and troughs of the power
spectra must match data at least as well as the $\Lambda$CDM
parametrization of the big bang cosmology.  {The locations of acoustic
  peaks~\cite{sz1970c,Peebles1970} in CMB  data approximately correspond to
  the extrema of the cosine function characterizing the photon temperature
  transfer function,} $\cos(k\rast + \phi)$, where $k$ denotes  the
comoving wave number, $\rast$ is the comoving sound horizon at
recombination, and $\phi$ {is the phase shift with contribution
  ($\phi>0$) from
free-streaming neutrinos in  $\Lambda$CDM cosmology
\cite{Bashinsky:2003tk}. The peak
  positions correspond to the } wave numbers $\kp$, which satisfy  $\kp\rast = m\pi-\phi$, where $m\ge 1$ is an integer. The corresponding observed CMB peak multipoles $ (\ellp) $ are given by
\begin{equation}
\begin{split}
\ellp\ \approx \ \kp \DA \ = \ \left(  m\pi -\phi \right) \frac{\DA}{\rast} \; , \qquad \text{where} \\ 
 \DA = \int_0^{\zast} d \! z \frac{1}{H(z)} \; , \qquad \qquad  \rast=\int_{\zast}^{\infty} d \! z   \frac{c_s(z)}{H(z)} 	\; , 
\label{Eq:rs}
\end{split}
\end{equation}
{$c_s(z)$ is the speed of sound in the baryon-photon plasma, $ H(z) $ is the Hubble parameter, and
  $\DA$ is the comoving angular diameter distance to the redshift of
  recombination $\zast$. Finding a solution to the Hubble tension requires
keeping $\ellp$ fixed while increasing $H_0$. 

We see from
Eq.~\eqref{Eq:rs} that we can modify the late-time evolution of the Universe,
i.e., modify $H(z)$ for $z< \zast$, in such a way that $\DA$ remains
unchanged but $H_0\equiv H(0)$ is pushed higher, to reconcile CMB/BAO or \emph{acoustic} $H_0$
with local $H_0$~\cite{barreira,DiValentino:2016hlg, Qing-Guo:2016ykt,Kumar:2016zpg,Bolejko:2017fos, renk2017,DiValentino:2017zyq,DiValentino:2017rcr,Khosravi:2017hfi,DiValentino:2017iww,Dutta:2018vmq,belgacem2018,Banihashemi:2018has,Agrawal:2019dlm,DiValentino:2019exe,Desmond:2019ygn,Pan:2019cot,Agrawal:2019lmo,Li:2019san,DiValentino:2019ffd}. Since in these solutions the early
expansion history of the  Universe ($H(z)$ for $z> \zast$) is unchanged, $\rast$ remains
unaltered. Therefore, $\ellp$ remains unchanged from the observed
$\Lambda$CDM values.}    {A second class of proposals rely on altering
the cosmology before radiation domination, i.e., $H(z)$ for $z\gg \zast$. These
solutions } change $\rast$, while at the same time keeping $\rast/\DA$
fixed~\cite{umilta2015,Lesgourgues:2015wza,Alam:2016wpf,Ko:2016uft, Karwal:2016vyq, DiValentino:2017oaw,
  Lancaster:2017ksf,Buen-Abad:2017gxg, DEramo:2018vss, Vattis:2019efj, Pandey:2019plg,Poulin:2018cxd,lin2019,Alexander:2019rsc,lin2019b,Archidiacono:2019wdp}.    {\emph{All of the solutions} that have been proposed
so far to alleviate the Hubble tension fall into the above two classes and, in particular, keep the acoustic scale at recombination $\tast = \rast/\DA$
fixed} even after accommodating a larger Hubble constant. 

\begin{figure*}
	\centering
	\includegraphics[width=0.85\linewidth]{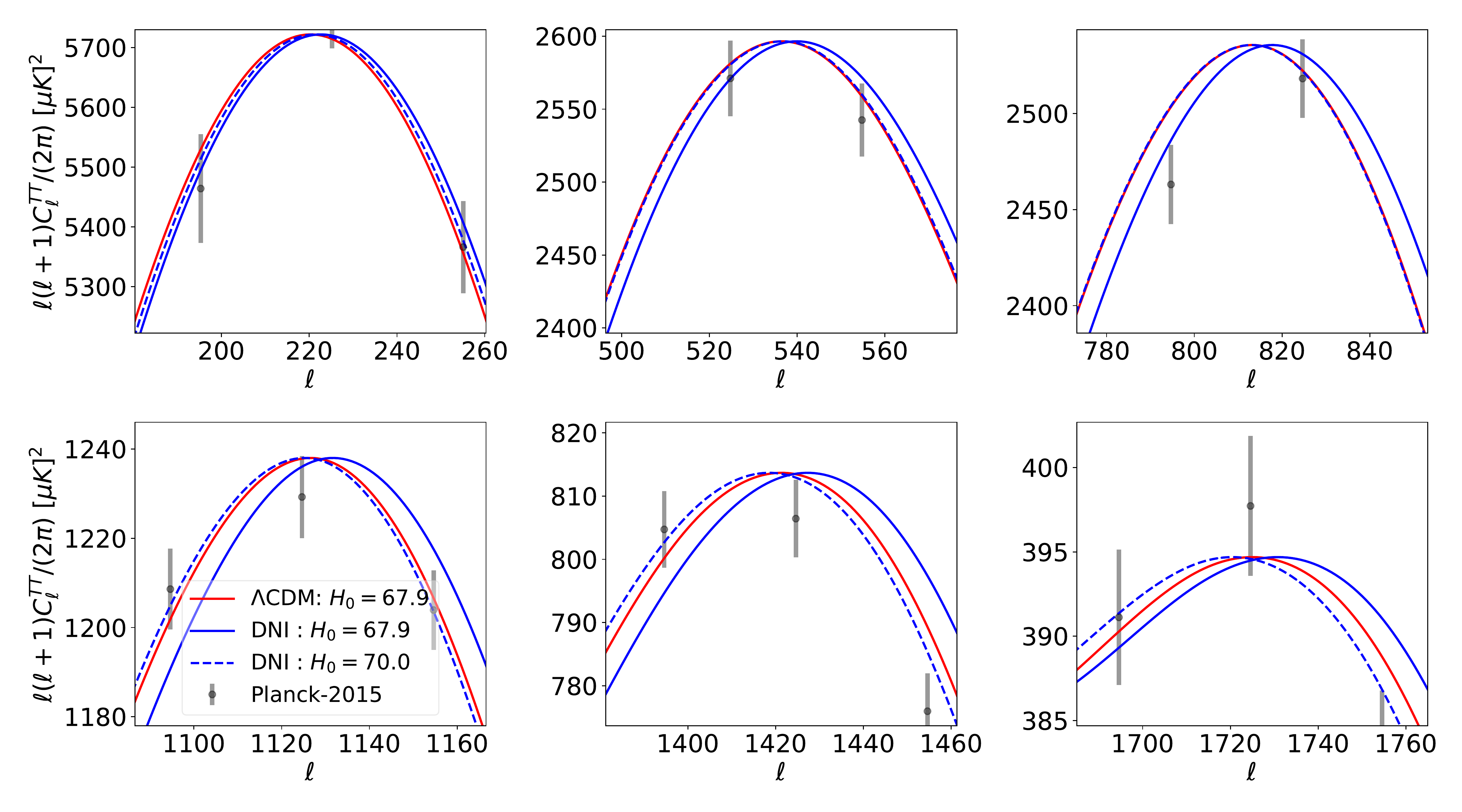}
	\caption{{CMB temperature (TT) power spectrum around first six acoustic peaks. The
			leftmost solid red line is the best-fit Planck \cite{Aghanim:2015xee} temperature power
			spectrum with a best-fit value of $H_0=67.9~{\rm
				kms^{-1}Mpc^{-1}}$. Introducing DNI, keeping all other cosmological parameters fixed,
			moves all peaks to the right/higher $\ell$ with larger shift for
			higher $\ell$ peaks (rightmost solid blue curves).
			However, DNI with higher $H_0$
			brings the peaks back to the
			original positions (dashed blue).  The amplitudes of DNI power
			spectra for each peak is adjusted so that the peak height
			is the same as the $\Lambda$CDM. Also shown as points with
			error bars is the binned
			Planck power spectrum.}}
	\label{fig:lcttee-hf}
\end{figure*}

\section{Undoing neutrino phase-shift}
In this paper, we propose a new class of solutions in which NP solves the Hubble tension by inducing changes in the phase shift $\phi$. These solutions are characterized by acoustic scales $\theta_{\ast}$ which are different from the acoustic scale in the $\Lambda$CDM model. To understand the nature of NP that can
accommodate a larger $H_0$, {let us consider a flat $\Lambda$CDM
cosmology, with the Hubble parameter given by $H(z)^2=H_0^2\left[\Om(1+z)^3
  + \Or(1+z)^4 + \left( 1- \Om - \Or\right) \right]$, where $\Omega_i$ are 
 the ratios of physical energy densities $ (\rho_i) $ to the critical energy
density
today and $i=m,r$ for total nonrelativistic
matter and total radiation, respectively. To separate out the effect of changing $H_0$, let us keep the
physical energy densities of matter and radiation, $\Om H_0^2$ and $\Or
H_0^2$, fixed.  A change
$H_0^2\rightarrow H_0^2 + \delta (H_0^2)$ then
implies $H(z)^2\rightarrow H(z)^2+\delta(
H_0^2)$.\footnote{{We note that including a curvature term, $\OK(1+z)^2$,
  will result in a redshift-dependent change in the Hubble parameter and
  therefore cannot compensate for a constant shift in $H(z)^2$.}} This constant shift in $H(z)$ is only important at low redshifts and becomes unimportant at high redshifts,
when $H(z)$ is much larger, and thus has negligible effect on $\rast$. Therefore, we
see from Eq. \eqref{Eq:rs} that 
increasing $H_0$ ($\delta (H_0^2)>0$) decreases $ \DA $ 
($\delta \DA<0$).}  If $\delta \DA$ is
to be compensated mostly from the shift in $\phi$ so that $\ellp$ remains unaltered, we get from
  Eq.~\eqref{Eq:rs} 
\begin{equation}
\frac{\delta \DA} {\DA}  \ - \ \frac{\delta \phi_m} {m\pi -\phi}  \ = 0 
\quad \Rightarrow \quad   
\delta \phi_m   \ \approx \  {m\pi } \frac{\delta \DA} {\DA}  \; , 
\end{equation}
where we have explicitly used the notation $\delta \phi_m$ to refer to {the
fact that 
the needed change in phase shift is different for different peaks. We have
also used the fact that $\phi \ll \pi$ in the approximate equality.}
Therefore, if NP needs to accommodate a larger $H_0$, it must induce a
\emph{negative} change in the phase shift that increases with $m$.  

Incredibly, \emph{undoing} the phase shift from free-streaming neutrinos in
the standard $\Lambda$CDM cosmology \cite{Bashinsky:2003tk} produces
{almost exactly the required effect (see Fig. \ref{fig:lcttee-hf}).}  Neutrinos
with NP interactions
scatter  and do not free stream, effectively generating a
negative phase shift with respect to $ \Lambda $CDM. Even though there exists a plethora of studies of cosmological
impacts from
nonstandard neutrinos interaction~\cite{Bell:2005dr,Mangano:2006mp,Serra:2009uu,Cyr-Racine:2013jua,Archidiacono:2013dua,Wilkinson:2014ksa,Boehm:2014vja,Bertoni:2014mva,Escudero:2015yka,2015JCAP...07..014F,Lancaster:2017ksf,Oldengott:2017fhy,Kreisch:2019yzn,2019arXiv190407810F,DiValentino:2017oaw,Ghosh:2017jdy}, as well as studies of phase shift in the context of varying relativistic degrees of freedom ($ N_\text{eff} $) on the phase shift~\cite{Follin:2015hya,Baumann:2015rya,Pan:2016zla,Baumann:2017lmt,Baumann:2017gkg,Choi:2018gho,Baumann:2019tdh}, a detailed study of the impact of new neutrino
interactions 
on 
the scale-dependent acoustic phase shift while keeping $ N_\text{eff} $ fixed at the standard value of $ 3.046 $ has not been performed yet.




\section{DNI : Dark Neutrino Interactions} In this work, we present a simple proof-of-principle model, namely, Dark Neutrino
Interactions (DNI), where a
component of dark matter interacts with neutrinos stopping them from free streaming. {The DNI undo the
phase shift induced by the free streaming neutrinos in the standard model
and thus push $H_0$ to higher values and yet are} safe from all
cosmological and particle physics bounds.  The necessary feature of this
model is a two component dark matter.  Only a small  fraction, $f  $, of the total dark
  matter (DM) energy density $ (\Omega_{ \rm DM}) $ is contributed by the component $ ({\rm namely,}~ \chi) $ that interacts
with neutrinos, the rest being the standard non-interacting cold dark
matter (CDM).   
\begin{equation}\label{eq:totdm}
\Omega_\chi \ = f \; \Omega_{ \rm DM} \;, \quad  \Omega_{ \rm CDM} = (1-f)  \; \Omega_{ \rm DM} \;.
\end{equation}
Note that having a small $f$ allows us to evade the
constraints typically obtained when all of the dark matter interacts with neutrinos~\cite{Mangano:2006mp, Serra:2009uu, Boehm:2014vja, Wilkinson:2014ksa, Bertoni:2014mva,Escudero:2015yka,Primulando:2017kxf}. 
The primary ingredients for our model are therefore $(i)$ an interacting dark matter component, $\chi$; $(ii)$
a messenger, $\psi$ (triplet in flavor); and $(iii)$ an electroweak (EW) gauge invariant effective operator involving the Higgs scalar $ H $ and the lepton doublet $ l $. After $ H $ acquires a nonzero vacuum expectation value ($v$), the effective operator gives marginal  interactions among neutrinos, messengers, and dark matter,
\begin{equation}
\mathcal{L} \ \supset \   \frac{1}{\Lambda} \  \left( H^\dag l \right) \left( \psi \chi \right)  \qquad \Rightarrow \qquad  
\     \ \frac{v}{\sqrt{2}\Lambda} \ \delta_{ij}  \ \nu_i \psi_j \chi \;,
\label{eq:interaction}
\end{equation}
where $ \Lambda $ is the scale of the effective operator  and  $i,j$ are flavor indices.  For a possible way to generate the interaction in Eq.~\eqref{eq:interaction} from a ultraviolet complete model using various symmetries see Ref.~\cite{Ghosh:2017jdy}.  By construction,  neutrinos remain massless and all three flavors interact with equal strength.

In this work, we focus on cases in which the mediators and dark matter
are nearly degenerate in mass. As shown in Ref.~\cite{Ghosh:2017jdy}, this allows the
scattering cross-section {($\sigma_{\chi\nu}$) between the dark
  matter and neutrinos to become independent of the neutrino temperature ($
  T_\nu $). The temperature independence of DNI enables neutrinos to
  decouple late, undoing the phase-shift from free streaming
  neutrinos for all the modes entering horizon until
  recombination.}  We can write the ``differential optical depth",
$\dot{\mu}\equiv \id \mu/\id \eta$, for neutrinos in the DNI model as
\begin{multline}\label{eq:mudot}
\dot{\mu} =   a n_{\chi} \sigma_{\chi\nu}   =   a \left( \frac{\rho_{\chi}}{m_{\chi}}\right)  \sigma_{\chi\nu} 
=  a f u   \rho_\text{dm}
\left(\sigma_\text{th} \over 100\text{GeV}\right),
\end{multline}
where $ a $ is the scale factor; $ \eta $ is the conformal time; $ \sigma_\text{th} = 6.65 \times
10^{-25} ~\text{cm}^2$ is the Thomson cross-section; $n_{\chi},
\rho_{\chi}, m_{\chi}$ denote the number density, the energy density, and
the mass of $\chi$, respectively; and
\begin{multline}\label{eq:uzero}
u \equiv  \frac{\sigma_{\chi\nu} }{  \sigma_{\text{Th}} }\times\frac{100
\text{GeV}}{ m_{\chi}}
 \simeq \left(5.5 \text{TeV} \over \Lambda\right)^4\left(1 \text{MeV} \over m_\chi\right)^3.
\end{multline}

The perturbation equations for $\nu$ and $\chi$ in DNI are coupled together~\cite{Wilkinson:2014ksa} similar to the
perturbations of the baryon-photon system. 

We use the notations and  perturbation variables in conformal Newtonian
gauge as defined in  Ref.~\cite{Ma:1995ey}, where
$ \delta_i $, $ \theta_i $ stand for over-density and divergence of fluid
velocity respectively for $ i $-th species. 
The total DM transfer functions are just the weighted sum of the
corresponding perturbation variables of the two dark
matter components,
\begin{align}
\delta_{\rm DM} &= f\delta_{\chi} + (1-f)\delta_{\rm CDM}\;,\nonumber\\
\theta_{\rm DM} &= f\theta_{\chi} + (1-f)\theta_{\rm CDM}\;.\label{Eq:f}
\end{align}
In the limit, $f\rightarrow 0$, we recover the standard $\Lambda$CDM
cosmology. In addition, for $f\ll 1$, the total dark matter transfer functions are
negligibly different from the CDM transfer functions and the modifications
to the dark matter power spectrum are of order $f^2$. Thus, the \emph{only}
difference from the $\Lambda$CDM cosmology, when  $f\ll 1$, comes from
modification of the free streaming of neutrinos. This regime is exactly what we
are interested in. 

The evolution equations of CDM are unchanged, while the equations of
interacting component of DM, $ \chi $, get additional terms due to interactions. The neutrino Boltzmann equations are also modified, picking up extra
interaction terms,
\begin{align}
\dot{\delta}_{\chi} &= -\theta_\chi + 3\dot{\phi}\;,\\
\dot{\theta}_\chi &= - {\dot{a}\over a}\theta_\chi + k^2\psi - \left(4\rho_\nu \over 3 \rho_\chi\right)\dmu(\theta_\chi-\theta_\nu)\;,\\
\dot{\delta}_\nu &= -{4 \over 3}\theta_\nu + 4\dot{\phi}\;,\\
\dot{\theta}_\nu &=k^2\left({1 \over 4}\delta_\nu - \sigma_\nu\right) + k^2\psi + \dmu(\theta_\chi-\theta_\nu)\;,\\
\dot{F}_{\nu l} &= {k \over (2l+1)}\left[lF_{\nu ~l-1}-(l+1)F_{\nu ~l+1}\right] - \dmu F_{\nu l}\;,
\end{align}
 where $ k $ is the comoving wave number; the dots represent derivatives with respect to the conformal time; $ \phi $ and $
 \psi $ are the Newtonian potentials and $ F_{\nu l} $, with $l \ge 2 $, is the $ l $-th multipole moment of neutrino distribution function.
The initial conditions are also modified as the initial anisotropic stress is zero for tightly coupled neutrinos.

\begin{figure}[t]
	\centering
	\includegraphics[width=\linewidth]{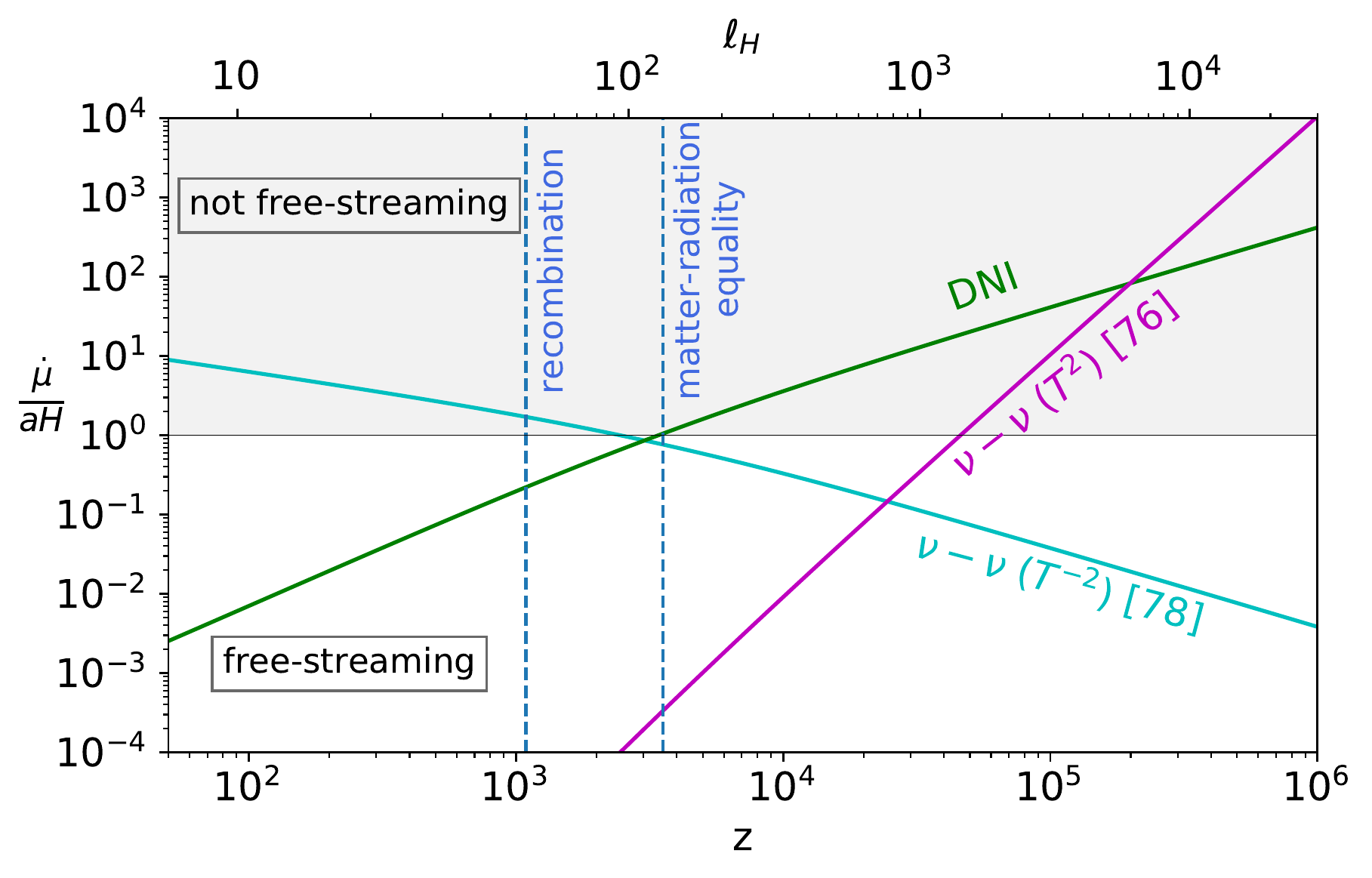}
	\caption{Comparison of \emph{optical depth} of neutrinos in DNI $ (fu = 0.034) $ with models of neutrino self-interaction~\cite{Oldengott:2017fhy} and~\cite{2019arXiv190407810F} having different temperature dependences.  
          The top axis shows the modes $ \ell_H $ which enter horizon at redshift $ z $.	}
	\label{fig:mfpcomp}
\end{figure}

{We plot the ratio of interaction rate to Hubble rate, $\dot{\mu}/(aH)$, in
Fig. \ref{fig:mfpcomp} for the current upper limits $ (fu=0.034) $ for our model derived in this work. For comparison, we also show cases with  neutrino self-interaction
models~\cite{Oldengott:2017fhy,2019arXiv190407810F} where crosssections vary as $T_\nu^2$ and  $T_\nu^{-2}$.
We see from Fig. \ref{fig:mfpcomp} that with the current upper bounds (fixed $ N_\text{eff} $) on neutrino
interactions, we can significantly modify the free streaming of neutrinos for all scales which enter horizon before recombination \emph{only in the
temperature independent case}.}

\section{Impact on Hubble tension} We have implemented the DNI cosmology in the publicly available code Cosmic Linear Anisotropy Solving
System ({CLASS})~\cite{2011JCAP...07..034B}. 
 In DNI cosmology,  the modes which enter horizon earlier
(higher $\ell$) get a larger phase shift (with respect to $\Lambda$CDM cosmology) compared to the modes which enter later as shown in Fig. \ref{fig:lcttee-hf} where we use $ f=10^{-3}, u =34 $.
This is because  the relative contribution of neutrinos (proportional to $ \rho_\nu /( \rho_r + \rho_m) $, where $ \rho_\nu $ is the neutrino energy density) to the metric perturbations
decreases with time as matter starts to dominate the energy density of the
Universe. This is almost exactly the $\ell$ dependence that we need to
alleviate the Hubble tension (Eq.~\eqref{Eq:rs}).
\begin{figure}[h]
	\centering
	\includegraphics[width=\linewidth]{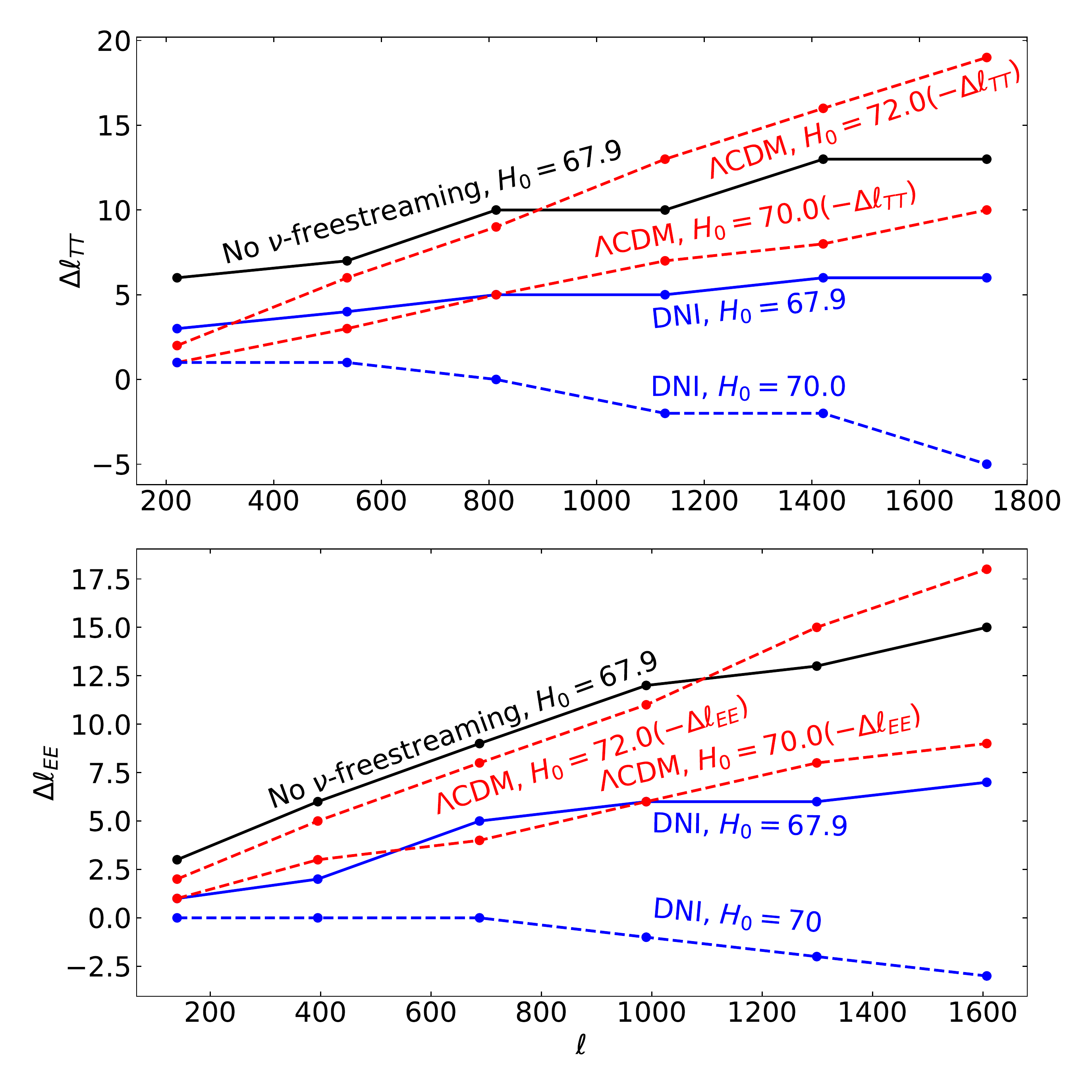}
	\caption{Shift of the position of peaks of CMB TT $
          (\Delta\ell_{TT}) $and EE $ (\Delta\ell_{EE}) $  spectrum in
          $\Lambda$CDM and DNI cosmologies $ (f = 10^{-3}, u=34) $ with respect to bestfit $ \Lambda $CDM model with $ H_0 = 67.9 ~\kpspm$.}
	\label{fig:lctt-ps-3}
\end{figure}
{We show this explicitly in Fig. \ref{fig:lctt-ps-3} where we plot
(negative of) the shift in peak positions for the CMB temperature and $E$-mode
polarization angular power spectra $ (\ell(\ell+1)C_\ell /(2\pi)) $ as we
change the Hubble constant in $\Lambda$CDM cosmology from the best fit
value while keeping other parameters ($\Omega_m H_0^2$, etc.) constant. 
For reference, we show the maximum effect we can get in the curve labelled ``No
$\nu$-freestreaming"  with
$\dot{\mu}/(aH) \ggg 1$.  We
see that the shift in $\ellp$ for DNI cosmology, with the current upper
bound in temperature independent interactions, is approximately of the same
size (but in the opposite direction) as $\Lambda$CDM cosmology with $H_0=70 {\rm km/s/Mpc}$. 
The scalings in $ \ell $ are also similar in both the cases. The small residual peak shifts at high multipoles are  within the errorbars.
Therefore,
we expect that the Hubble tension should reduce considerably in a DNI
cosmology. We verify this in the DNI curves with $H_0=70 {\rm km/s/Mpc}$, in
which the peak shifts are negligible compared to the best fit Planck
$\Lambda$CDM cosmology.  }


	\begin{figure*}[t]
		\centering
		\begin{subfigure}{0.32\linewidth}
		\centering
		\includegraphics[width=\linewidth]{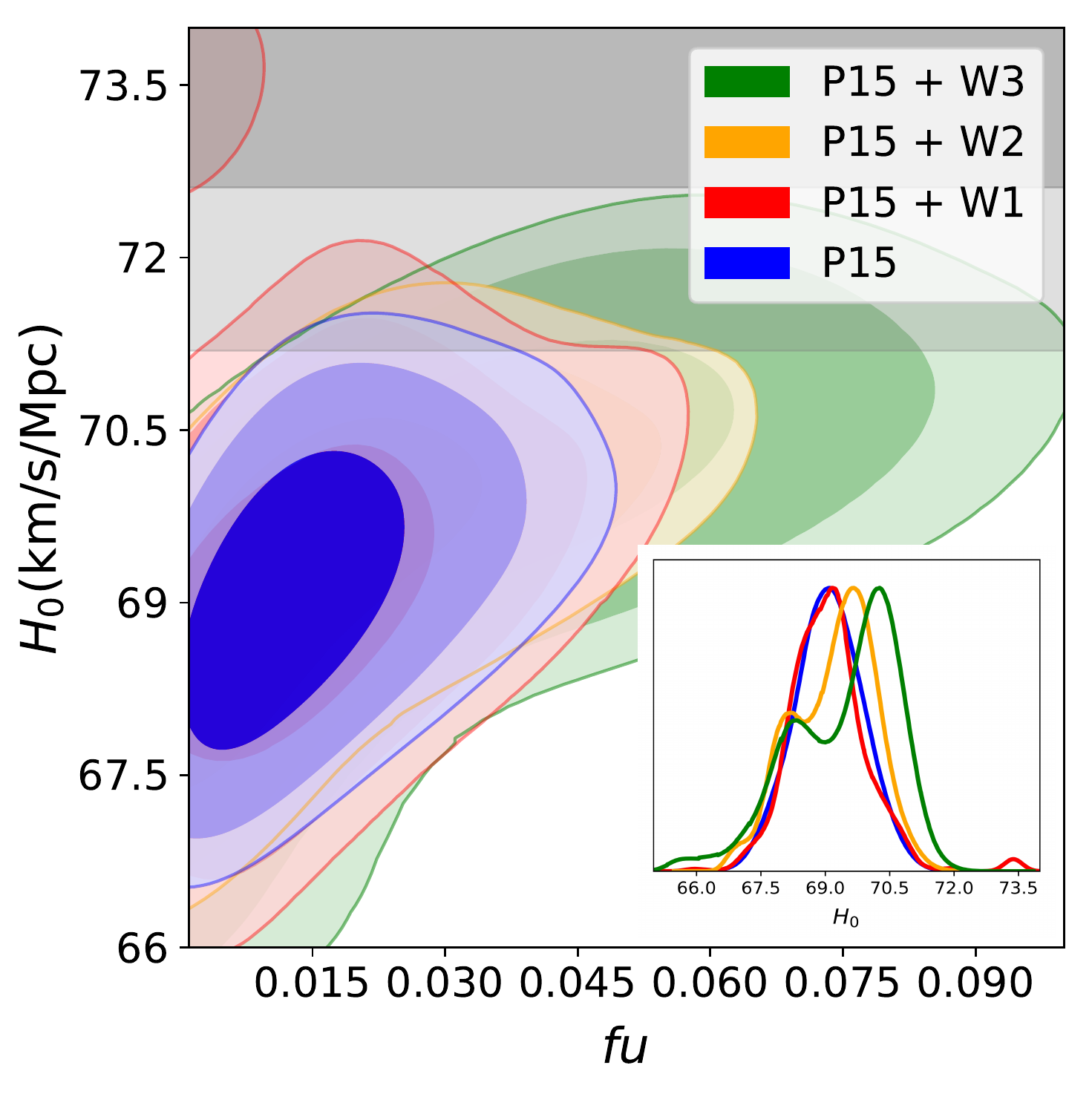}
		\label{fig:h0fu0}
	\end{subfigure}
	\begin{subfigure}{0.32\linewidth}
		\centering
		\includegraphics[width=\linewidth]{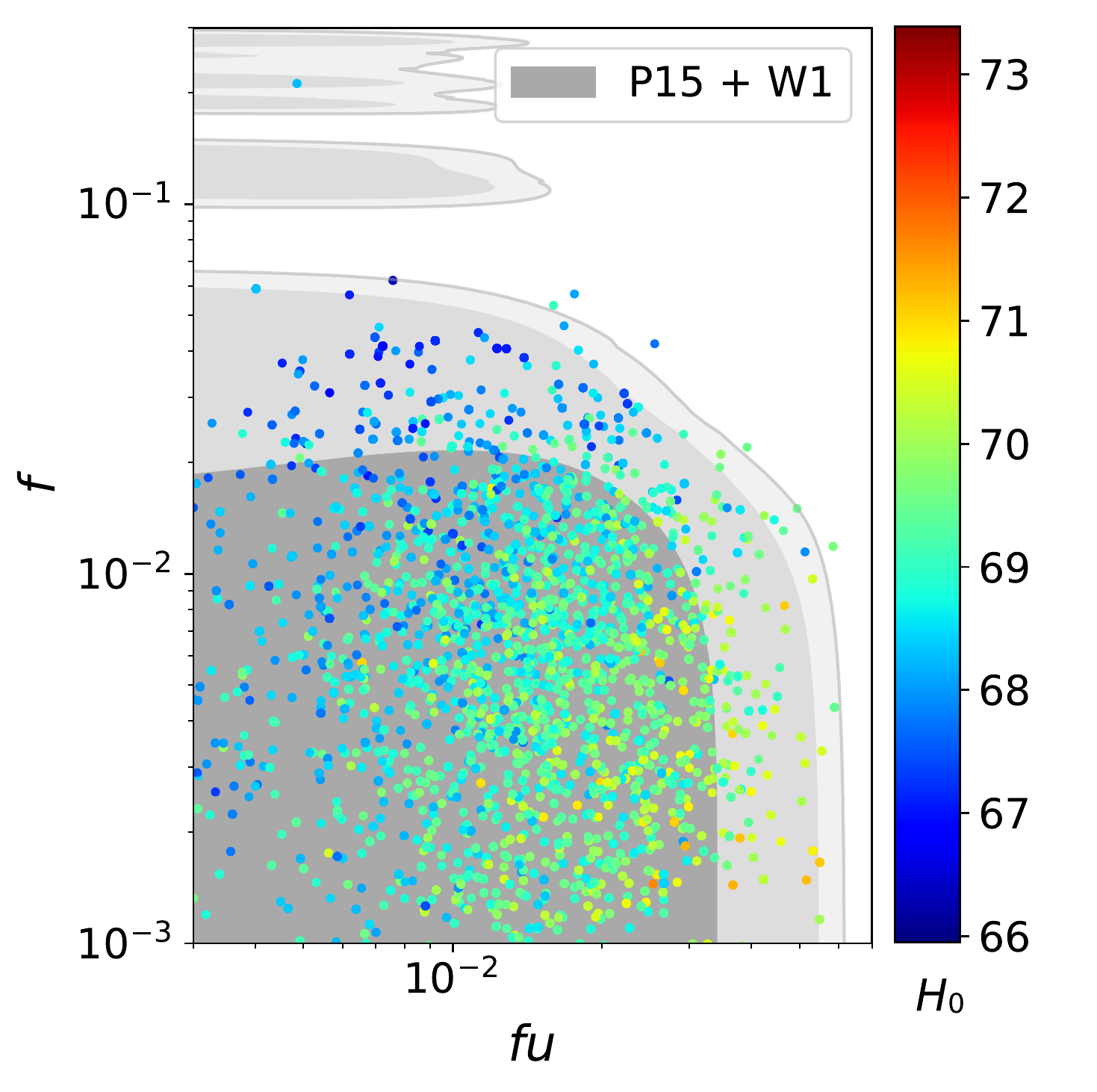}
		\label{fig:h0fu2}
	\end{subfigure}
	\begin{subfigure}{0.32\linewidth}
		\centering
		\includegraphics[height = 5.4 cm ,width=\linewidth]{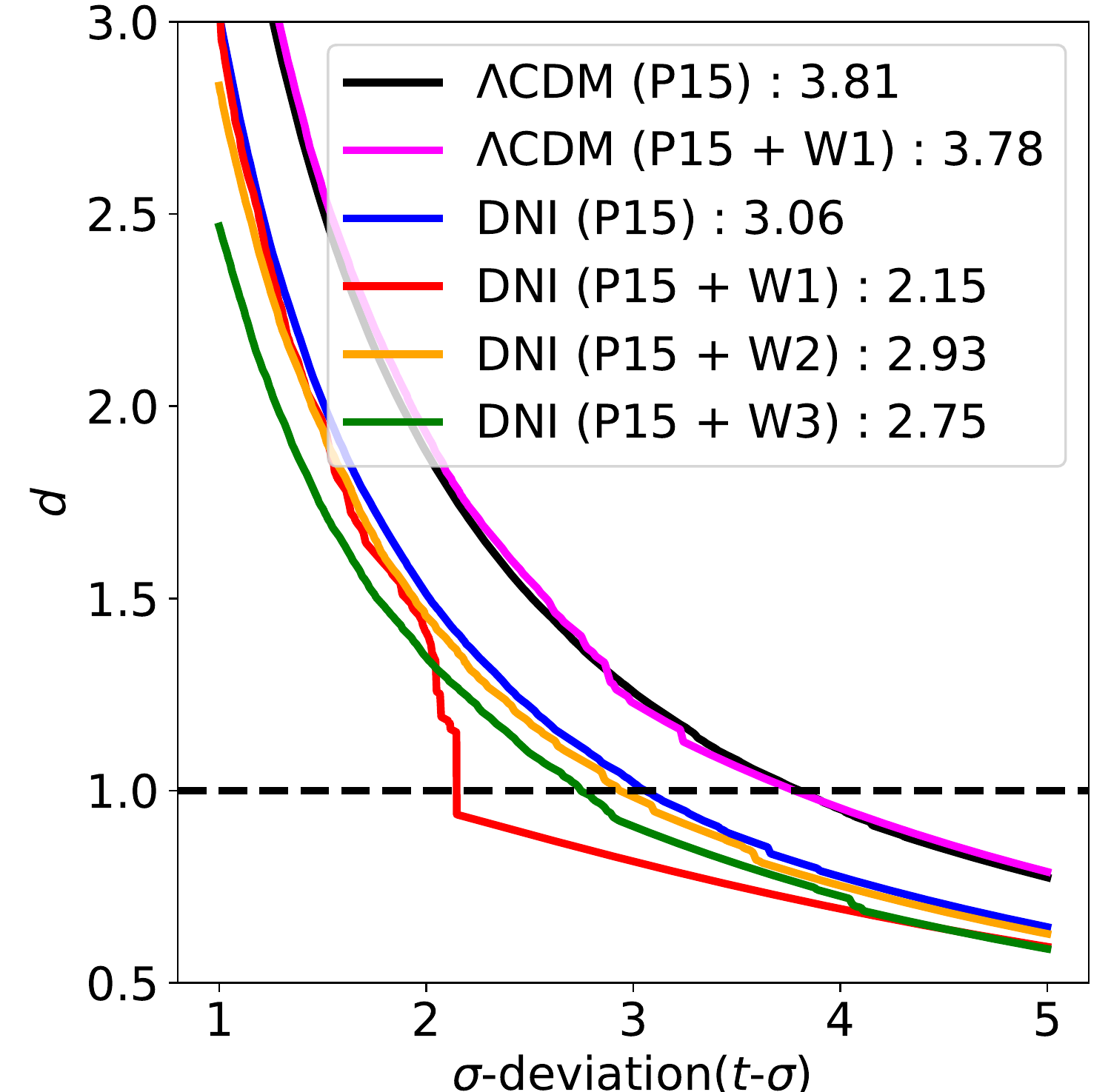}
		\label{fig:H-ten}
	\end{subfigure}
	\caption{The left panel shows $1\sigma,2\sigma,~\text{and}~3\sigma$ constraints in DNI and $H_0$
		for different data set combinations. The light and dark grey band shows $ 1\sigma $ and $ 2\sigma $ band for SH0ES~\cite{Riess:2019cxk} measurement respectively. The central panel shows the
		MCMC samples in the $f-fu$ plane. The right most panel shows
		calculation of Hubble tension (values given in the legend) taking into account
		non-Gaussianity of PDFs. The $ 2\sigma $ upper-limit from P15 is $ fu <0.034 $.}
	\label{fig:mcmc}
\end{figure*}

We perform a Markov-Chain Monte Carlo (MCMC) analysis of the DNI  model
using publicly available code \textsc{Monte Python}~\cite{Audren:2012wb}. In addition to the $ \Lambda $CDM parameters, we varied $ \omega_\chi = \Omega_{\chi} h^2$ and $ u $, where $h\equiv H_0/(100 ~{\rm kms^{-1}Mpc^{-1}})$ is the reduced Hubble constant. The fraction $ f $ was extracted as a derived parameter,  $ f = {\omega_\chi \over \omega_{\rm CDM} + \omega_{\chi}} $, where $ \omega_{ \rm CDM} = \Omega_{ \rm CDM} h^2 $. For all the parameters in the MCMC analysis
we satisfy the Gelman-Rubin convergence criterion $ R - 1 < 0.1 $.

For $ \omega_\chi $ and $ u $ we have used flat priors with no hard prior upper boundaries. The lower
	prior boundary was set to 0 for physicality.
We use the following cosmological data
sets: 
Planck CMB 2015 Low-$ \ell $ TEB, High $ \ell $ TT EE TE - Plik lite and CMB lensing T+P~\cite{Aghanim:2015xee} (named `P15') 
and  full shape of Galaxy power spectrum measured
by WiggleZ Dark Energy Survey~\cite{PhysRevD.86.103518}. We have checked that in DNI cosmology the difference when using lite vs full likelihood is insignificant. WiggleZ data are sensitive to the modification of matter power spectrum due to the small fraction of interacting dark matter and the strongly interacting neutrinos.
	The WiggleZ power spectrum goes upto $ k=0.5~\hmpci $. We have used
        different $ k $-cutoff of the full dataset for three separate
        analyses and label them W1, W2, W3 for cutoff  $\km = 0.12h, 0.2h,
        0.3\hmpci$ respectively.
        We used \textsc{CLASS} Halofit module
        \cite{Smith:2002dz} to incorporate non-linear modifications in the
        power spectrum, since the WiggleZ power spectrum (specifically W2 \& W3) goes
        	upto the k modes where these effects are important. Although the Halofit is tested tool for $ \Lambda $CDM cosmology, since DNI introduce very small changes in the matter power spectrum by construction, the use of Halofit is justified in this case.
        
We note that for the BAO data, it will be incorrect to use just the BAO scale
(or $\tast$)
extracted from the power spectrum (e.g., Ref.~\cite{alam2017}) assuming
$\Lambda$CDM cosmology, available as BAO likelihood modules in public MCMC codes, to constrain any new physics which
modifies the phase shift $\phi$ of the acoustic oscillations and allows
$\tast$ to vary from the $\Lambda$CDM value. This is the case for us and
also for any model with non-standard $N_{\rm
  eff}$, since any new free streaming relativistic species contributes to $\phi$ in a scale dependent manner.

The marginalized constraints from our MCMC analysis, with two extra parameters ($f$,$u$)  are shown in  Fig. \ref{fig:mcmc}.  The local measurement
from Ref.~\cite{Riess:2019cxk} of $ H_0 = 74.03 \pm  1.42
~\text{km} ~\text{s}^{-1} ~\text{Mpc}^{-1} $ is shown in gray horizontal
bands. There is a clear degeneracy between the
neutrino stopping power ($\propto fu$)  and $H_0$ which reduces the Hubble tension.
We see from the MCMC samples plotted in Fig. \ref{fig:mcmc} (centre) that
stronger neutrino interaction favours higher $ H_0 $. The 2D contours
  clearly indicates that data prefer small $ f ~(\lesssim 10^{-2}) $  and,
  as discussed above, for small $ f $ the results (limits) become
  independent of the value of $ f $ (Eq. \ref{Eq:f}). These are general features of DNI which are present in all the datasets we have analysed.

The 1-D probability distribution functions (PDF)
shown in inset of Fig.~\ref{fig:mcmc} (left) are
highly non-Gaussian. To quantify the tension between non-Gaussian PDFs, we
define a quantity $d=(H_1-H_2)/\sqrt{\sigma_1(t)^2+\sigma_2(t)^2}$, where
  $H_1,H_2$ are two $ H_0 $ measurements and
  $\sigma_1(t),\sigma_2(t)$ are the corresponding `$t$-$\sigma$' upper or lower
  limits.
  For a Gaussian PDF
  $\sigma(t)=t\sigma_G$, where $\sigma_G$ is the Gaussian $1$-$\sigma$
  error. We use Gaussian errorbar for the local $H_0$ measurement and plot
  the quantity $d$ in Fig. \ref{fig:mcmc} (right). The tension is then given by
  the value of $t$ where $d=1$. Our definition is equivalent to the
  usual definition of tension in the Gaussian case. We see that for
  $\Lambda$CDM the tension is at $3.8\sigma$,
  which reduces to $\lesssim 3\sigma$ in
  DNI cosmology. 
The small secondary peak for the `P15 + W1' dataset within the $ 3 $$ \sigma $ region results in a jump in $ d $ and reduction in tension to $ 2.1 $$ \sigma $. We note that the effect of neutrino interactions on the matter power spectrum is not just a phase shift in BAO. The matter power spectrum is also enhanced because the perturbations in neutrinos do not decay away completely on horizon entry in DNI cosmology and contribute to metric perturbations. Both the phase shift in the CMB and the modifications of the matter power spectrum help in reducing the tension to $ 2.1\sigma $.
  
  \setlength{\tabcolsep}{0.4em}
  \renewcommand{\arraystretch}{1.3}
  \begin{table*}
  	\centering
  	\begin{tabular}{l c|c c|c}
  		&\multicolumn{2}{c}{P15+W1+SH0ES} & \multicolumn{2}{c}{P15+W1}\\
  		\hline
  		&$ \Lambda $CDM &  DNI  & $ \Lambda $CDM &  DNI  \\
  		$H_0~(\kpspm)$(bf) & $68.89_{-0.59}^{+0.58}$~$ (68.86) $&$70.25_{-0.61}^{+0.63}$~$ (70.37) $&$68.01_{-0.6}^{+0.58}$~$ (68.08) $&$69.39_{-0.68}^{+0.69}$~$ (69.31) $\\
  		$ fu $ (bf) & $ 0 $ &$0.02321_{-0.012}^{+0.0065}(0.01874)$&$ 0 $& $0.01744_{-0.011}^{+0.0062}(0.01567)$\\
  		$100~\omega_{b }$ & $2.243_{-0.015}^{+0.015}$ & $2.251_{-0.015}^{+0.015}$& $2.226_{-0.016}^{+0.015}$& $2.238_{-0.015}^{+0.015}$\\
  		
  		$\omega_{\rm DM }$ & $0.1176_{-0.0013}^{+0.0013}$ &$0.1181_{-0.0013}^{+0.0013}$&$0.1194_{-0.0013}^{+0.0013}$& $0.1195_{-0.0013}^{+0.0013}$\\
  		$ln10^{10}A_{s }$ & $3.07_{-0.025}^{+0.024}$ & $3.005_{-0.026}^{+0.025}$ & $3.052_{-0.025}^{+0.017}$ & $2.998_{-0.027}^{+0.021}$\\
  		$n_{s }$ & $0.9709_{-0.0046}^{+0.0045}$ & $0.9492_{-0.0048}^{+0.0047}$ & $0.966_{-0.0045}^{+0.0043}$ & $0.9467_{-0.0051}^{+0.0044}$\\
  		$ \sigma_8 $ &$0.8283_{-0.009}^{+0.0088}$ & $0.831_{-0.0092}^{+0.0091}$ &$0.826_{-0.0087}^{+0.0072}$ &$0.8308_{-0.0087}^{+0.0076}$ \\
  		\rule[-2ex]{0pt}{6ex}
  		\makecell[l]{$ 100\tast $\\bf} &\makecell{$ 1.04201_{-0.00030}^{+0.00030}$ \\ $1.04205 $} & \makecell{$ 1.04643_{-0.00078}^{+0.00094}(+14.7\sigma)$ \\ $1.04614(+0.4\%) $} &\makecell{$ 1.04183_{-0.00029}^{+0.00031}$ \\ $1.04188 $} & \makecell{$1.04573_{-0.00087}^{+0.00125}(+13\sigma)$ \\ $1.04587 (+0.4\%)$}\\
  		$ \rast $(Mpc),bf &$ 145.07 $ & $ 144.93 ~(-0.1\%) $ &$ 144.81 $ &$ 144.52 ~(-0.2\%)$ \\
  		$ \DA $(Mpc),bf &$ 12.78 $ & $ 12.71 ~(-0.5\%)$&$ 12.75 $ & $ 12.68 ~(-0.6\%)$\\
  		$ \Delta\chi^2 $ & $ 0 $ & $ -9.08 $ & $ 0  $ & $ -2.42 $
  	\end{tabular}
  	\caption{Parameter table for different data-set combinations with
  		fixed  $f=10^{-3}$ for DNI. Best-fit values are indicated by `bf'. We also show the baryon density ($ \omega_b \equiv\Omega_b h^2 $),  total dark matter density ($ \omega_{\rm DM} \equiv\Omega_{ \rm DM} h^2 $) and the magnitude of matter power spectrum on $ 8h^{-1}~{\rm Mpc}$ scale $ (\sigma_8) $.} 
  	\label{tbl:chisq}
  \end{table*}

In Table \ref{tbl:chisq} we present results of a MCMC analysis of DNI cosmology for fixed $ f = 10^{-3} $, only varying $ u $ along with the $ \Lambda $CDM parameters\footnote{We varied the energy density of total DM $ \omega_{\rm DM} $ and set the CDM energy density $ \omega_{\rm CDM} = (1-f) \times \omega_{\rm DM} $ and interacting DM energy density $ \omega_\chi  = f \times \omega_{\rm DM}$.}, where we also include the local measurement of $ H_0 $ (SH0ES collaboration~\cite{Riess:2019cxk}). With respect to $ \Lambda $CDM, $ \chi^2 $ reduces by $ 9 $ in DNI with one extra parameter $ u $. The bestfit value of the Hubble constant turns out to be $ H_0 = 70.4 $.  Note that, as argued before, this increase in $ H_0 $ is associated with a decrease in $ D_A $ which in turn gets compensated mostly from a change in $ \phi $. Therefore, the bestfit for DNI cosmology is characterized by a $ \tast $ which is approximately $ 15\sigma $ away from that of $ \Lambda $CDM. There is however a small change in $ \rast $ which roughly compensates approximately $ 20\% $ change in $ D_A $.
Interestingly, DNI cosmology is a slightly better fit to the `P15 + W1'
datasets than the $ \Lambda $CDM cosmology and $ H_0 $ tension is reduced
to $ 2.9 \sigma $. Since $ f \ll 1$, the modification of neutrino
	free streaming alone results in a higher $ H_0 $ and the better fit
        of the data.  In particular, the modification of the dark matter
        power spectrum due to the effect of radiation pressure of
	neutrinos on dark matter is not important for $f\ll 1$.
	
\begin{figure*}[t!]
	\begin{subfigure}{0.43\linewidth}
		\centering
		\includegraphics[width=0.9\linewidth]{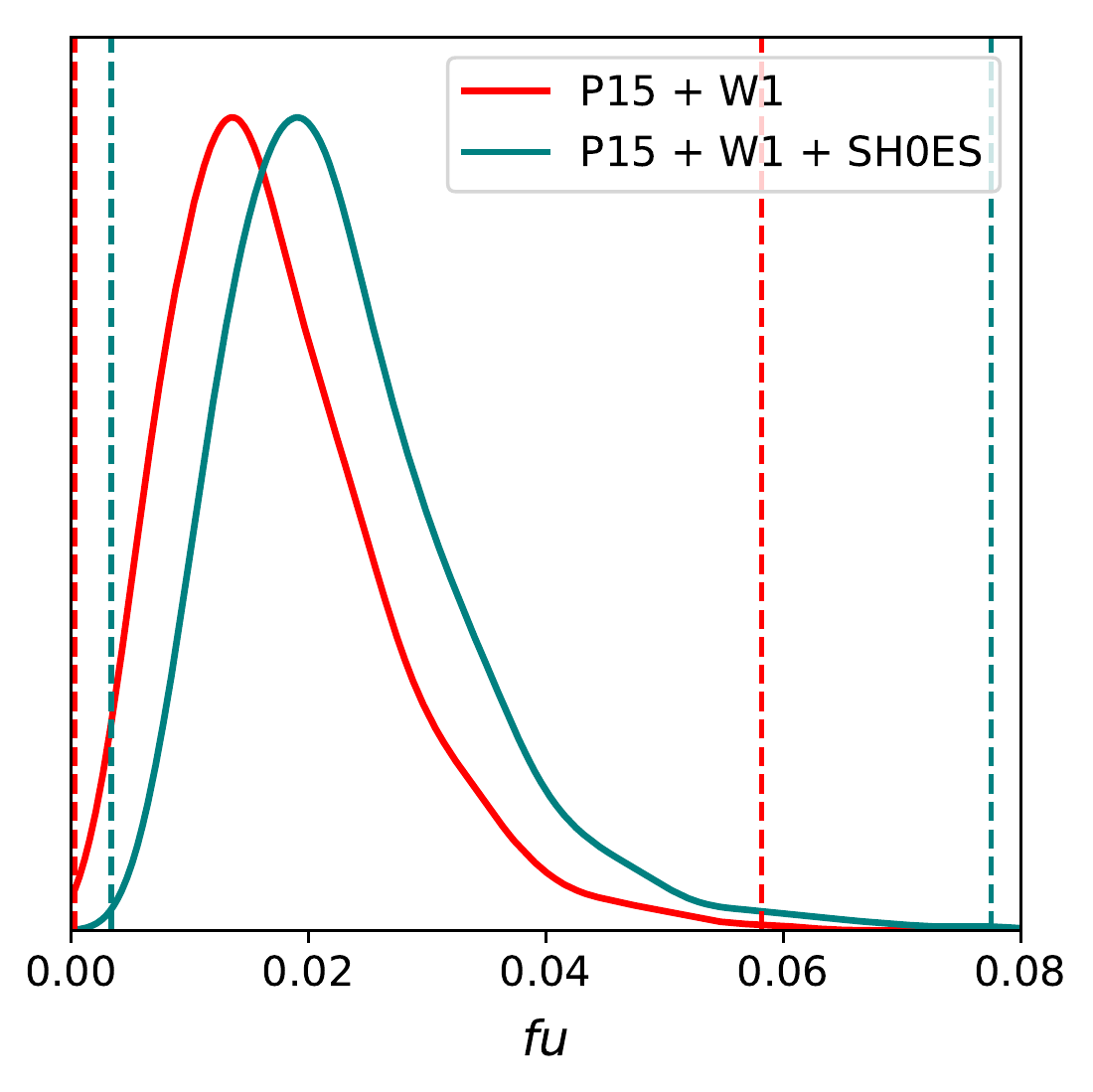}
	\end{subfigure}
    \begin{subfigure}{0.43\linewidth}
    	\centering
    	\includegraphics[width=\linewidth]{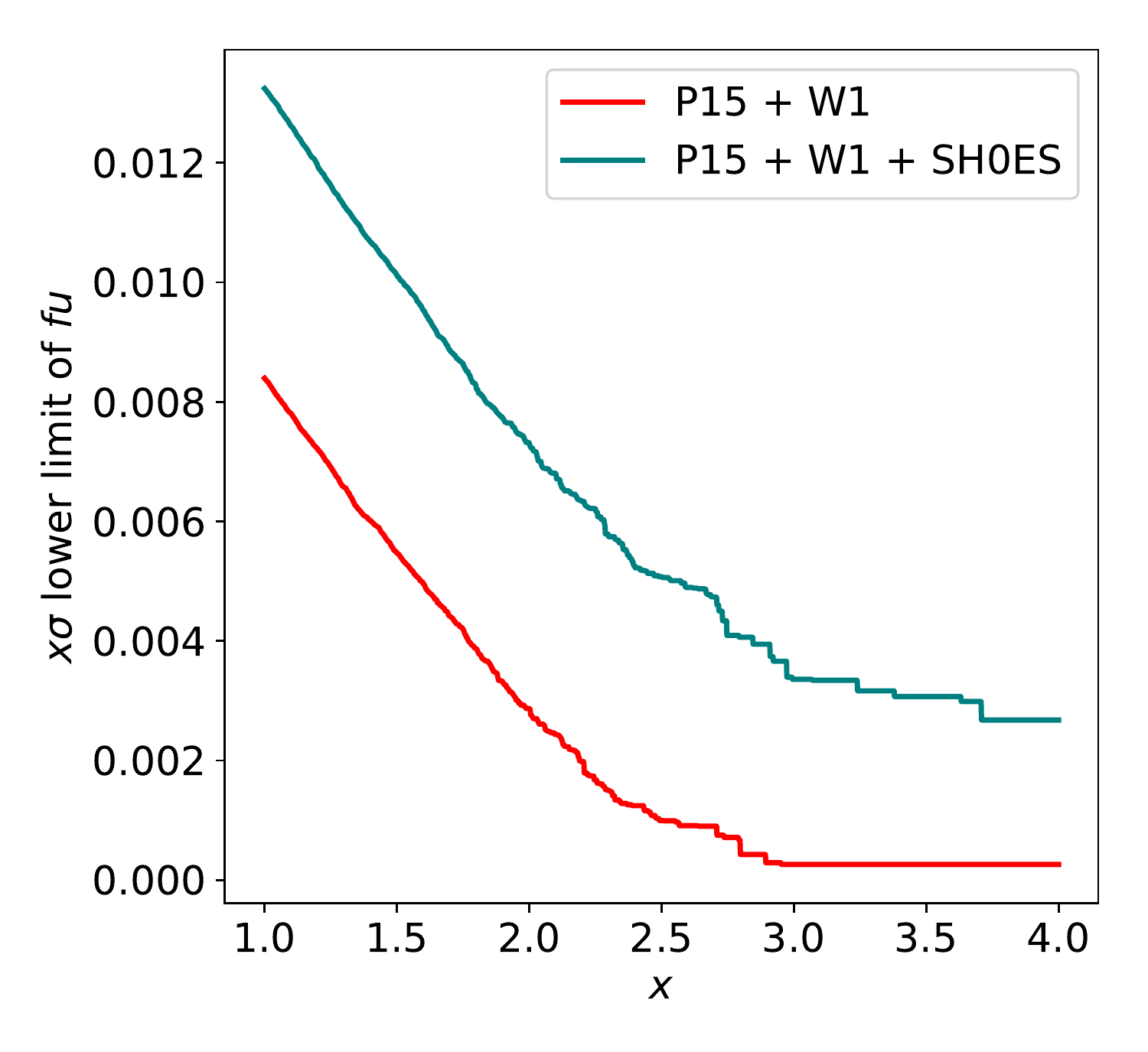}
    \end{subfigure}
	\caption{Left: 1-D posterior for $ fu $ for fixed $ f $ DNI
            cosmology (see Table.~\ref{tbl:chisq}). The dashed vertical
            lines marks the $ 3\,\sigma $ upper and lower limit for the
            corresponding dataset. Right: The $ x\,\sigma $ lower limit
            on $ fu $ plotted against $ x $.
          It can be seen from both the plot that $ fu = 0 $ is excluded at
          more than $ 3\,\sigma $ when we include SH0ES data.}
	\label{fig:fu-post}
\end{figure*}

We show the posterior distribution of $ fu $ for fixed $ f ~(=
  10^{-3})$ DNI cosmology  (see Table.~\ref{tbl:chisq}) in
  Fig. \ref{fig:fu-post} (left). DNI cosmology prefers non-zero dark  neutrino
  interactions at $\gtrsim 3\sigma$ in order to reconcile the local Hubble
  measurements with the CMB and high redshift $ (0.2 \lesssim z \lesssim 0.8) $ large scale structure data.
  The dashed lines show the $ 3\sigma $ upper and lower limits of $ fu $. Even without the SH0ES
  data a non-zero interaction is preferred, however, a zero interaction is
  not ruled out as the posterior extends to the $fu=0$ boundary. When we
  include SH0ES data,  the $fu=0$ point is outside the $3\sigma$ lower
  limit. This can be also be seen  in the right panel of
  Fig.~\ref{fig:fu-post} where we plot the $x\sigma$ lower limit vs
  $x$. This curve flattens out when the PDF starts falling sharply for the
  without-SH0ES case as we hit the boundary.   Thus
  we  conclude that non-zero interactions are demanded for reconciliation of
  SH0ES data with Planck and WiggleZ.
 For clarity we also show the posterior distributions of all the parameters in Fig.~\ref{fig:u1e-1p2-vs-u1e-1p3triangle} in Appendix~\ref{sec:app1}. The shift in $ \theta_\ast $ and $ H_0 $ is clearly visible in the figure. The other parameters having significant shift are $ n_\text{s} $ and $ A_\text{s} $.
  
  We see in Table.~\ref{tbl:chisq} that the bestfit $ fu\approx 2 \times 10^{-2} $ requires $ u \approx 20$. For $ m_\chi\lesssim1~\text{MeV} $ we find the scale of the effective operator to be $ \Lambda \gtrsim 2.5~\text{TeV} $ from Eq. \eqref{eq:uzero} . For this high a $ \Lambda $ we do not expect any significant constraint from particle physics.

\begin{figure}[h!]
	\centering
	\includegraphics[width=0.9\linewidth]{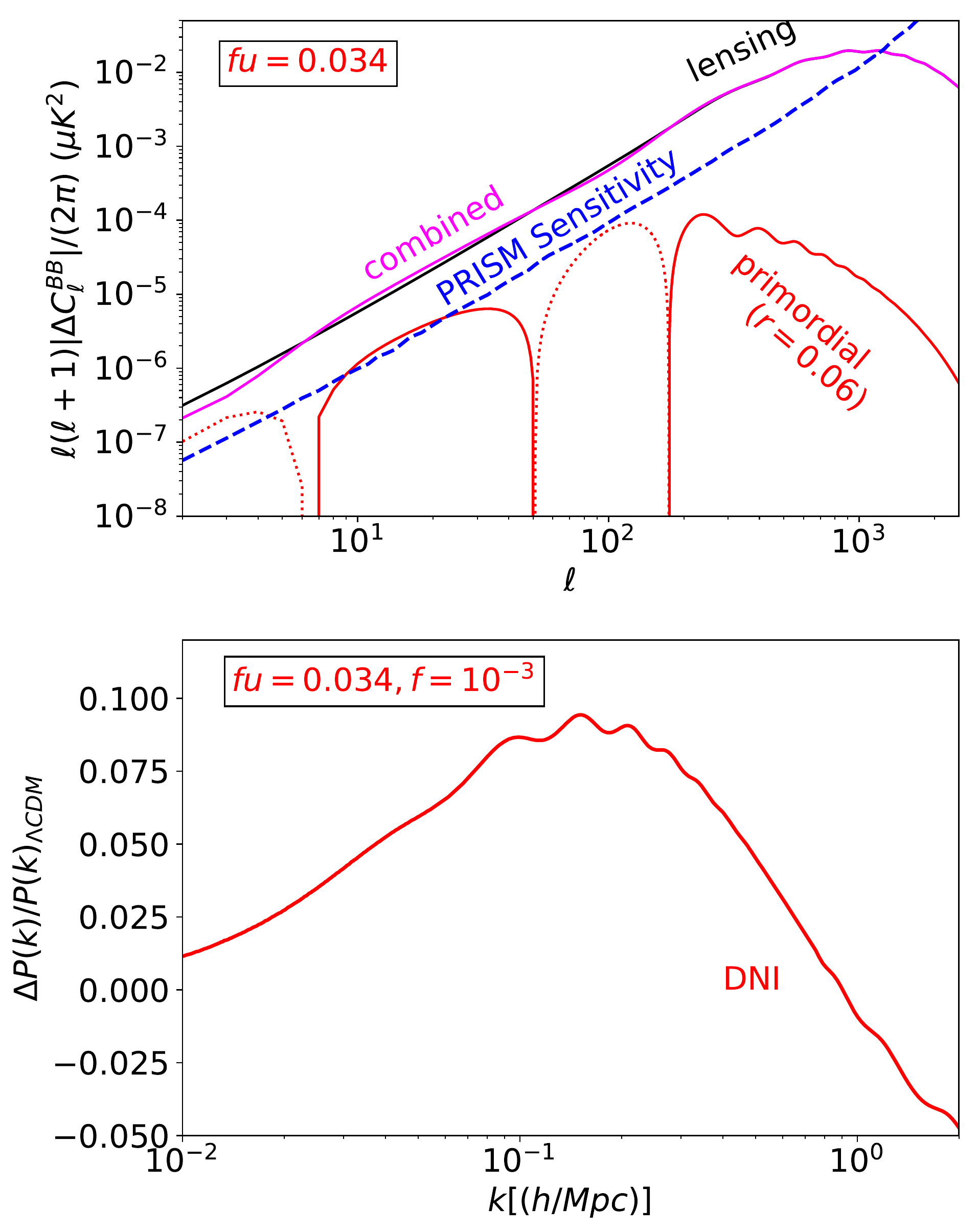}
	\caption{Absolute change in CMB B-modes [primordial (red),
			lensing (black) and combined (pink)] from $ \Lambda $CDM due to
			DNI is shown in the top panel. 	The solid (dotted) lines
			represent enhancement (suppression) of power. The blue dashed
			line shows sensitivity of PRISM~\cite{Andre:2013nfa}. The
			bottom panel shows the fractional change in matter power
			spectrum w.r.t $ \Lambda $CDM cosmology.
	}
	\label{fig:prismbb}
\end{figure}

The gravity of new neutrino
interactions 
modifies the $B$-mode CMB power spectrum \cite{Ghosh:2017jdy} and matter power spectrum as shown in Fig.~\ref{fig:prismbb} . 
We compare in Fig.~\ref{fig:prismbb} (top) the modification of $ B $-modes for tensor to scalar ratio $ r=0.06 $~\cite{Ade:2015fwj,Ade:2015xua,Ade:2018gkx} with the sensitivity of the proposed experiment Polarized Radiation Imaging
	and Spectroscopy Mission (PRISM)~\cite{Andre:2013nfa}. This effect, in principle, can be detected if $ r $ is close to the current upper limit~\cite{Ade:2015fwj,Ade:2015xua,Ade:2018gkx} with a future PRISM like experiment~\cite{Andre:2013nfa,core,pico,cmbs4,Hazumi2019}. 
	In DNI cosmology, the perturbations in neutrinos do not decay away and influence the matter perturbations gravitationally. This results in the enhancement of the matter power spectrum as shown in Fig~\ref{fig:prismbb} (bottom). Of course, there are shifts in the BAO peaks which are analogous to the shifts in the CMB acoustic peaks, as discussed before. The effects in the matter power spectrum are much stronger compared to the CMB B mode. The future Large Scale Structure surveys~\cite{Aghamousa:2016zmz,Amendola:2016saw,Ivezic:2008fe,2011arXiv1110.3193L} will greatly improve the accuracy of matter power spectrum measurements achieving approximately $ 1\% $ accuracy~\cite{2011arXiv1110.3193L} and will therefore be able probe the DNI cosmology.

\section{Conclusions} In this work we have proposed a qualitatively new framework that ameliorates the Hubble tension by primarily using 
the phase
shift in the acoustic oscillations of the primordial plasma. 
Amazingly, this framework undoes the neutrino induced phase-shift of $ \Lambda $CDM, generates the
shift in the acoustic peaks of CMB with the required scale dependence 
and pushes the \emph{acoustic}  $H_0$ higher
towards 
 the locally measured value of $H_0$. We note that the BAOs are the imprints of the same acoustic oscillations of the primordial plasma in the matter power spectrum~\cite{sz1970c,Eisenstein:1997ik} and, therefore, any phase shift or the lack of it would be carried over to the BAOs also. 
  Using the Planck CMB and WiggleZ (W1) datasets along with SH0ES, we find that new interactions of neutrinos are preferred and no interaction $ (fu=0) $ is disfavoured  at $ \gtrsim 3 \sigma $ (Fig.~5).
  We therefore might have found evidence of new interactions
of neutrinos in the Hubble tension. \emph{Our modified version of \textsc{CLASS} used in this paper is made publicly available at}
\href{https://github.com/subhajitghosh-phy/CLASS_DNI}{https://github.com/subhajitghosh-phy/CLASS\_DNI}.


%

\begin{acknowledgments}
	This work was supported by Science and Engineering Research Board (SERB),
Government of India Grant No. ECR/2015/000078 and No. ECR/2015/000196. This work was also
supported by Max-Planck-Gesellschaft funded Max Planck partner group
between Tata Institute of Fundamental Research, Mumbai and
Max-Planck-Institut f\"ur Astrophysik, Garching.  
The PDF plots and MCMC analysis were done using
public python package \textsc{GetDist}
\cite{Lewis:2002ah}. This work used computational facility
of Department of Theoretical Physics, Tata Institute of Fundamental Research. We thank Basudeb Dasgupta for useful comments. We acknowledge support of the Department of Atomic Energy, Government of India, under Project No. 12-R\&D-TFR-5.02-0200. 
\end{acknowledgments}

%


\bibliography{scalarvstensor}

\begin{thebibliography}{106}%
\makeatletter
\providecommand \@ifxundefined [1]{%
 \@ifx{#1\undefined}
}%
\providecommand \@ifnum [1]{%
 \ifnum #1\expandafter \@firstoftwo
 \else \expandafter \@secondoftwo
 \fi
}%
\providecommand \@ifx [1]{%
 \ifx #1\expandafter \@firstoftwo
 \else \expandafter \@secondoftwo
 \fi
}%
\providecommand \natexlab [1]{#1}%
\providecommand \enquote  [1]{``#1''}%
\providecommand \bibnamefont  [1]{#1}%
\providecommand \bibfnamefont [1]{#1}%
\providecommand \citenamefont [1]{#1}%
\providecommand \href@noop [0]{\@secondoftwo}%
\providecommand \href [0]{\begingroup \@sanitize@url \@href}%
\providecommand \@href[1]{\@@startlink{#1}\@@href}%
\providecommand \@@href[1]{\endgroup#1\@@endlink}%
\providecommand \@sanitize@url [0]{\catcode `\\12\catcode `\$12\catcode
  `\&12\catcode `\#12\catcode `\^12\catcode `\_12\catcode `\%12\relax}%
\providecommand \@@startlink[1]{}%
\providecommand \@@endlink[0]{}%
\providecommand \url  [0]{\begingroup\@sanitize@url \@url }%
\providecommand \@url [1]{\endgroup\@href {#1}{\urlprefix }}%
\providecommand \urlprefix  [0]{URL }%
\providecommand \Eprint [0]{\href }%
\providecommand \doibase [0]{http://dx.doi.org/}%
\providecommand \selectlanguage [0]{\@gobble}%
\providecommand \bibinfo  [0]{\@secondoftwo}%
\providecommand \bibfield  [0]{\@secondoftwo}%
\providecommand \translation [1]{[#1]}%
\providecommand \BibitemOpen [0]{}%
\providecommand \bibitemStop [0]{}%
\providecommand \bibitemNoStop [0]{.\EOS\space}%
\providecommand \EOS [0]{\spacefactor3000\relax}%
\providecommand \BibitemShut  [1]{\csname bibitem#1\endcsname}%
\let\auto@bib@innerbib\@empty
\bibitem [{\citenamefont {Ade}\ \emph {et~al.}(2016)\citenamefont {Ade} \emph
  {et~al.}}]{Ade:2015xua}%
  \BibitemOpen
  \bibfield  {author} {\bibinfo {author} {\bibfnamefont {P.~A.~R.}\
  \bibnamefont {Ade}} \emph {et~al.} (\bibinfo {collaboration} {Planck}),\
  }\bibfield  {title} {\enquote {\bibinfo {title} {{Planck 2015 results. XIII.
  Cosmological parameters}},}\ }\href {\doibase 10.1051/0004-6361/201525830}
  {\bibfield  {journal} {\bibinfo  {journal} {Astron. Astrophys.}\ }\textbf
  {\bibinfo {volume} {594}},\ \bibinfo {pages} {A13} (\bibinfo {year}
  {2016})},\ \Eprint {http://arxiv.org/abs/1502.01589} {arXiv:1502.01589
  [astro-ph.CO]} \BibitemShut {NoStop}%
\bibitem [{\citenamefont {Aghanim}\ \emph {et~al.}(2018)\citenamefont {Aghanim}
  \emph {et~al.}}]{Aghanim:2018eyx}%
  \BibitemOpen
  \bibfield  {author} {\bibinfo {author} {\bibfnamefont {N.}~\bibnamefont
  {Aghanim}} \emph {et~al.} (\bibinfo {collaboration} {Planck}),\ }\bibfield
  {title} {\enquote {\bibinfo {title} {{Planck 2018 results. VI. Cosmological
  parameters}},}\ }\href@noop {} {\  (\bibinfo {year} {2018})},\ \Eprint
  {http://arxiv.org/abs/1807.06209} {arXiv:1807.06209 [astro-ph.CO]}
  \BibitemShut {NoStop}%
\bibitem [{\citenamefont {Beutler}\ \emph {et~al.}(2011)\citenamefont
  {Beutler}, \citenamefont {Blake}, \citenamefont {Colless}, \citenamefont
  {Jones}, \citenamefont {Staveley-Smith}, \citenamefont {Campbell},
  \citenamefont {Parker}, \citenamefont {Saunders},\ and\ \citenamefont
  {Watson}}]{Beutler:2011hx}%
  \BibitemOpen
  \bibfield  {author} {\bibinfo {author} {\bibfnamefont {Florian}\ \bibnamefont
  {Beutler}}, \bibinfo {author} {\bibfnamefont {Chris}\ \bibnamefont {Blake}},
  \bibinfo {author} {\bibfnamefont {Matthew}\ \bibnamefont {Colless}}, \bibinfo
  {author} {\bibfnamefont {D.~Heath}\ \bibnamefont {Jones}}, \bibinfo {author}
  {\bibfnamefont {Lister}\ \bibnamefont {Staveley-Smith}}, \bibinfo {author}
  {\bibfnamefont {Lachlan}\ \bibnamefont {Campbell}}, \bibinfo {author}
  {\bibfnamefont {Quentin}\ \bibnamefont {Parker}}, \bibinfo {author}
  {\bibfnamefont {Will}\ \bibnamefont {Saunders}}, \ and\ \bibinfo {author}
  {\bibfnamefont {Fred}\ \bibnamefont {Watson}},\ }\bibfield  {title} {\enquote
  {\bibinfo {title} {{The 6dF Galaxy Survey: Baryon Acoustic Oscillations and
  the Local Hubble Constant}},}\ }\href {\doibase
  10.1111/j.1365-2966.2011.19250.x} {\bibfield  {journal} {\bibinfo  {journal}
  {Mon. Not. Roy. Astron. Soc.}\ }\textbf {\bibinfo {volume} {416}},\ \bibinfo
  {pages} {3017--3032} (\bibinfo {year} {2011})},\ \Eprint
  {http://arxiv.org/abs/1106.3366} {arXiv:1106.3366 [astro-ph.CO]} \BibitemShut
  {NoStop}%
\bibitem [{\citenamefont {Font-Ribera}\ \emph {et~al.}(2014)\citenamefont
  {Font-Ribera} \emph {et~al.}}]{Font-Ribera:2013wce}%
  \BibitemOpen
  \bibfield  {author} {\bibinfo {author} {\bibfnamefont {Andreu}\ \bibnamefont
  {Font-Ribera}} \emph {et~al.} (\bibinfo {collaboration} {BOSS}),\ }\bibfield
  {title} {\enquote {\bibinfo {title} {{Quasar-Lyman $\alpha$ Forest
  Cross-Correlation from BOSS DR11 : Baryon Acoustic Oscillations}},}\ }\href
  {\doibase 10.1088/1475-7516/2014/05/027} {\bibfield  {journal} {\bibinfo
  {journal} {JCAP}\ }\textbf {\bibinfo {volume} {1405}},\ \bibinfo {pages}
  {027} (\bibinfo {year} {2014})},\ \Eprint {http://arxiv.org/abs/1311.1767}
  {arXiv:1311.1767 [astro-ph.CO]} \BibitemShut {NoStop}%
\bibitem [{\citenamefont {Delubac}\ \emph {et~al.}(2015)\citenamefont {Delubac}
  \emph {et~al.}}]{Delubac:2014aqe}%
  \BibitemOpen
  \bibfield  {author} {\bibinfo {author} {\bibfnamefont {Timothée}\
  \bibnamefont {Delubac}} \emph {et~al.} (\bibinfo {collaboration} {BOSS}),\
  }\bibfield  {title} {\enquote {\bibinfo {title} {{Baryon acoustic
  oscillations in the Ly$\alpha$ forest of BOSS DR11 quasars}},}\ }\href
  {\doibase 10.1051/0004-6361/201423969} {\bibfield  {journal} {\bibinfo
  {journal} {Astron. Astrophys.}\ }\textbf {\bibinfo {volume} {574}},\ \bibinfo
  {pages} {A59} (\bibinfo {year} {2015})},\ \Eprint
  {http://arxiv.org/abs/1404.1801} {arXiv:1404.1801 [astro-ph.CO]} \BibitemShut
  {NoStop}%
\bibitem [{\citenamefont {Ross}\ \emph {et~al.}(2015)\citenamefont {Ross},
  \citenamefont {Samushia}, \citenamefont {Howlett}, \citenamefont {Percival},
  \citenamefont {Burden},\ and\ \citenamefont {Manera}}]{Ross:2014qpa}%
  \BibitemOpen
  \bibfield  {author} {\bibinfo {author} {\bibfnamefont {Ashley~J.}\
  \bibnamefont {Ross}}, \bibinfo {author} {\bibfnamefont {Lado}\ \bibnamefont
  {Samushia}}, \bibinfo {author} {\bibfnamefont {Cullan}\ \bibnamefont
  {Howlett}}, \bibinfo {author} {\bibfnamefont {Will~J.}\ \bibnamefont
  {Percival}}, \bibinfo {author} {\bibfnamefont {Angela}\ \bibnamefont
  {Burden}}, \ and\ \bibinfo {author} {\bibfnamefont {Marc}\ \bibnamefont
  {Manera}},\ }\bibfield  {title} {\enquote {\bibinfo {title} {{The clustering
  of the SDSS DR7 main Galaxy sample – I. A 4 per cent distance measure at $z
  = 0.15$}},}\ }\href {\doibase 10.1093/mnras/stv154} {\bibfield  {journal}
  {\bibinfo  {journal} {Mon. Not. Roy. Astron. Soc.}\ }\textbf {\bibinfo
  {volume} {449}},\ \bibinfo {pages} {835--847} (\bibinfo {year} {2015})},\
  \Eprint {http://arxiv.org/abs/1409.3242} {arXiv:1409.3242 [astro-ph.CO]}
  \BibitemShut {NoStop}%
\bibitem [{\citenamefont {Addison}\ \emph {et~al.}(2018)\citenamefont
  {Addison}, \citenamefont {Watts}, \citenamefont {Bennett}, \citenamefont
  {Halpern}, \citenamefont {Hinshaw},\ and\ \citenamefont
  {Weiland}}]{Addison:2017fdm}%
  \BibitemOpen
  \bibfield  {author} {\bibinfo {author} {\bibfnamefont {G.~E.}\ \bibnamefont
  {Addison}}, \bibinfo {author} {\bibfnamefont {D.~J.}\ \bibnamefont {Watts}},
  \bibinfo {author} {\bibfnamefont {C.~L.}\ \bibnamefont {Bennett}}, \bibinfo
  {author} {\bibfnamefont {M.}~\bibnamefont {Halpern}}, \bibinfo {author}
  {\bibfnamefont {G.}~\bibnamefont {Hinshaw}}, \ and\ \bibinfo {author}
  {\bibfnamefont {J.~L.}\ \bibnamefont {Weiland}},\ }\bibfield  {title}
  {\enquote {\bibinfo {title} {{Elucidating $\Lambda$CDM: Impact of Baryon
  Acoustic Oscillation Measurements on the Hubble Constant Discrepancy}},}\
  }\href {\doibase 10.3847/1538-4357/aaa1ed} {\bibfield  {journal} {\bibinfo
  {journal} {Astrophys. J.}\ }\textbf {\bibinfo {volume} {853}},\ \bibinfo
  {pages} {119} (\bibinfo {year} {2018})},\ \Eprint
  {http://arxiv.org/abs/1707.06547} {arXiv:1707.06547 [astro-ph.CO]}
  \BibitemShut {NoStop}%
\bibitem [{\citenamefont {Alam}\ \emph
  {et~al.}(2017{\natexlab{a}})\citenamefont {Alam} \emph {et~al.}}]{alam2017}%
  \BibitemOpen
  \bibfield  {author} {\bibinfo {author} {\bibfnamefont {Shadab}\ \bibnamefont
  {Alam}} \emph {et~al.} (\bibinfo {collaboration} {BOSS}),\ }\bibfield
  {title} {\enquote {\bibinfo {title} {{The clustering of galaxies in the
  completed SDSS-III Baryon Oscillation Spectroscopic Survey: cosmological
  analysis of the DR12 galaxy sample}},}\ }\href {\doibase
  10.1093/mnras/stx721} {\bibfield  {journal} {\bibinfo  {journal} {Mon. Not.
  Roy. Astron. Soc.}\ }\textbf {\bibinfo {volume} {470}},\ \bibinfo {pages}
  {2617--2652} (\bibinfo {year} {2017}{\natexlab{a}})},\ \Eprint
  {http://arxiv.org/abs/1607.03155} {arXiv:1607.03155 [astro-ph.CO]}
  \BibitemShut {NoStop}%
\bibitem [{\citenamefont {{Riess}}\ \emph {et~al.}(2016)\citenamefont
  {{Riess}}, \citenamefont {{Macri}}, \citenamefont {{Hoffmann}}, \citenamefont
  {{Scolnic}}, \citenamefont {{Casertano}}, \citenamefont {{Filippenko}},
  \citenamefont {{Tucker}}, \citenamefont {{Reid}}, \citenamefont {{Jones}},\
  and\ \citenamefont {{Silverman}}}]{riess2016}%
  \BibitemOpen
  \bibfield  {author} {\bibinfo {author} {\bibfnamefont {Adam~G.}\ \bibnamefont
  {{Riess}}}, \bibinfo {author} {\bibfnamefont {Lucas~M.}\ \bibnamefont
  {{Macri}}}, \bibinfo {author} {\bibfnamefont {Samantha~L.}\ \bibnamefont
  {{Hoffmann}}}, \bibinfo {author} {\bibfnamefont {Dan}\ \bibnamefont
  {{Scolnic}}}, \bibinfo {author} {\bibfnamefont {Stefano}\ \bibnamefont
  {{Casertano}}}, \bibinfo {author} {\bibfnamefont {Alexei~V.}\ \bibnamefont
  {{Filippenko}}}, \bibinfo {author} {\bibfnamefont {Brad~E.}\ \bibnamefont
  {{Tucker}}}, \bibinfo {author} {\bibfnamefont {Mark~J.}\ \bibnamefont
  {{Reid}}}, \bibinfo {author} {\bibfnamefont {David~O.}\ \bibnamefont
  {{Jones}}}, \ and\ \bibinfo {author} {\bibfnamefont {Jeffrey~M.}\
  \bibnamefont {{Silverman}}},\ }\bibfield  {title} {\enquote {\bibinfo {title}
  {{A 2.4\% Determination of the Local Value of the Hubble Constant}},}\ }\href
  {\doibase 10.3847/0004-637X/826/1/56} {\bibfield  {journal} {\bibinfo
  {journal} {\apj}\ }\textbf {\bibinfo {volume} {826}},\ \bibinfo {eid} {56}
  (\bibinfo {year} {2016})},\ \Eprint {http://arxiv.org/abs/1604.01424}
  {arXiv:1604.01424 [astro-ph.CO]} \BibitemShut {NoStop}%
\bibitem [{\citenamefont {Riess}\ \emph {et~al.}(2018)\citenamefont {Riess}
  \emph {et~al.}}]{riess2018}%
  \BibitemOpen
  \bibfield  {author} {\bibinfo {author} {\bibfnamefont {Adam~G.}\ \bibnamefont
  {Riess}} \emph {et~al.},\ }\bibfield  {title} {\enquote {\bibinfo {title}
  {{Milky Way Cepheid Standards for Measuring Cosmic Distances and Application
  to Gaia DR2: Implications for the Hubble Constant}},}\ }\href {\doibase
  10.3847/1538-4357/aac82e} {\bibfield  {journal} {\bibinfo  {journal}
  {Astrophys. J.}\ }\textbf {\bibinfo {volume} {861}},\ \bibinfo {pages} {126}
  (\bibinfo {year} {2018})},\ \Eprint {http://arxiv.org/abs/1804.10655}
  {arXiv:1804.10655 [astro-ph.CO]} \BibitemShut {NoStop}%
\bibitem [{\citenamefont {Riess}\ \emph {et~al.}(2019)\citenamefont {Riess},
  \citenamefont {Casertano}, \citenamefont {Yuan}, \citenamefont {Macri},\ and\
  \citenamefont {Scolnic}}]{Riess:2019cxk}%
  \BibitemOpen
  \bibfield  {author} {\bibinfo {author} {\bibfnamefont {Adam~G.}\ \bibnamefont
  {Riess}}, \bibinfo {author} {\bibfnamefont {Stefano}\ \bibnamefont
  {Casertano}}, \bibinfo {author} {\bibfnamefont {Wenlong}\ \bibnamefont
  {Yuan}}, \bibinfo {author} {\bibfnamefont {Lucas~M.}\ \bibnamefont {Macri}},
  \ and\ \bibinfo {author} {\bibfnamefont {Dan}\ \bibnamefont {Scolnic}},\
  }\bibfield  {title} {\enquote {\bibinfo {title} {{Large Magellanic Cloud
  Cepheid Standards Provide a 1\% Foundation for the Determination of the
  Hubble Constant and Stronger Evidence for Physics Beyond LambdaCDM}},}\
  }\href@noop {} {\  (\bibinfo {year} {2019})},\ \Eprint
  {http://arxiv.org/abs/1903.07603} {arXiv:1903.07603 [astro-ph.CO]}
  \BibitemShut {NoStop}%
\bibitem [{\citenamefont {Bonvin}\ \emph {et~al.}(2017)\citenamefont {Bonvin}
  \emph {et~al.}}]{h0licow2017}%
  \BibitemOpen
  \bibfield  {author} {\bibinfo {author} {\bibfnamefont {V.}~\bibnamefont
  {Bonvin}} \emph {et~al.},\ }\bibfield  {title} {\enquote {\bibinfo {title}
  {{H0LiCOW – V. New COSMOGRAIL time delays of HE 0435-1223: $H_0$ to 3.8 per
  cent precision from strong lensing in a flat $\Lambda$CDM model}},}\ }\href
  {\doibase 10.1093/mnras/stw3006} {\bibfield  {journal} {\bibinfo  {journal}
  {Mon. Not. Roy. Astron. Soc.}\ }\textbf {\bibinfo {volume} {465}},\ \bibinfo
  {pages} {4914--4930} (\bibinfo {year} {2017})},\ \Eprint
  {http://arxiv.org/abs/1607.01790} {arXiv:1607.01790 [astro-ph.CO]}
  \BibitemShut {NoStop}%
\bibitem [{\citenamefont {Birrer}\ \emph {et~al.}(2019)\citenamefont {Birrer}
  \emph {et~al.}}]{h0licow2019}%
  \BibitemOpen
  \bibfield  {author} {\bibinfo {author} {\bibfnamefont {S.}~\bibnamefont
  {Birrer}} \emph {et~al.},\ }\bibfield  {title} {\enquote {\bibinfo {title}
  {{H0LiCOW - IX. Cosmographic analysis of the doubly imaged quasar SDSS
  $1206+4332$ and a new measurement of the Hubble constant}},}\ }\href
  {\doibase 10.1093/mnras/stz200} {\bibfield  {journal} {\bibinfo  {journal}
  {Mon. Not. Roy. Astron. Soc.}\ }\textbf {\bibinfo {volume} {484}},\ \bibinfo
  {pages} {4726} (\bibinfo {year} {2019})},\ \Eprint
  {http://arxiv.org/abs/1809.01274} {arXiv:1809.01274 [astro-ph.CO]}
  \BibitemShut {NoStop}%
\bibitem [{\citenamefont {Freedman}\ \emph {et~al.}(2019)\citenamefont
  {Freedman} \emph {et~al.}}]{freedman2019}%
  \BibitemOpen
  \bibfield  {author} {\bibinfo {author} {\bibfnamefont {Wendy~L.}\
  \bibnamefont {Freedman}} \emph {et~al.},\ }\bibfield  {title} {\enquote
  {\bibinfo {title} {{The Carnegie-Chicago Hubble Program. VIII. An Independent
  Determination of the Hubble Constant Based on the Tip of the Red Giant
  Branch}},}\ }\href@noop {} {\  (\bibinfo {year} {2019})},\ \Eprint
  {http://arxiv.org/abs/1907.05922} {arXiv:1907.05922 [astro-ph.CO]}
  \BibitemShut {NoStop}%
\bibitem [{\citenamefont {{Yuan}}\ \emph {et~al.}(2019)\citenamefont {{Yuan}},
  \citenamefont {{Riess}}, \citenamefont {{Macri}}, \citenamefont
  {{Casertano}},\ and\ \citenamefont {{Scolnic}}}]{yuan2019}%
  \BibitemOpen
  \bibfield  {author} {\bibinfo {author} {\bibfnamefont {Wenlong}\ \bibnamefont
  {{Yuan}}}, \bibinfo {author} {\bibfnamefont {Adam~G.}\ \bibnamefont
  {{Riess}}}, \bibinfo {author} {\bibfnamefont {Lucas~M.}\ \bibnamefont
  {{Macri}}}, \bibinfo {author} {\bibfnamefont {Stefano}\ \bibnamefont
  {{Casertano}}}, \ and\ \bibinfo {author} {\bibfnamefont {Dan}\ \bibnamefont
  {{Scolnic}}},\ }\bibfield  {title} {\enquote {\bibinfo {title} {{Consistent
  Calibration of the Tip of the Red Giant Branch in the Large Magellanic Cloud
  on the Hubble Space Telescope Photometric System and Implications for the
  Determination of the Hubble Constant}},}\ }\href@noop {} {\bibfield
  {journal} {\bibinfo  {journal} {arXiv e-prints}\ ,\ \bibinfo {eid}
  {arXiv:1908.00993}} (\bibinfo {year} {2019})},\ \Eprint
  {http://arxiv.org/abs/1908.00993} {arXiv:1908.00993 [astro-ph.GA]}
  \BibitemShut {NoStop}%
\bibitem [{\citenamefont {Freedman}\ \emph {et~al.}(2020)\citenamefont
  {Freedman}, \citenamefont {Madore}, \citenamefont {Hoyt}, \citenamefont
  {Jang}, \citenamefont {Beaton}, \citenamefont {Lee}, \citenamefont {Monson},
  \citenamefont {Neeley},\ and\ \citenamefont {Rich}}]{Freedman:2020dne}%
  \BibitemOpen
  \bibfield  {author} {\bibinfo {author} {\bibfnamefont {Wendy~L.}\
  \bibnamefont {Freedman}}, \bibinfo {author} {\bibfnamefont {Barry~F.}\
  \bibnamefont {Madore}}, \bibinfo {author} {\bibfnamefont {Taylor}\
  \bibnamefont {Hoyt}}, \bibinfo {author} {\bibfnamefont {In~Sung}\
  \bibnamefont {Jang}}, \bibinfo {author} {\bibfnamefont {Rachael}\
  \bibnamefont {Beaton}}, \bibinfo {author} {\bibfnamefont {Myung~Gyoon}\
  \bibnamefont {Lee}}, \bibinfo {author} {\bibfnamefont {Andrew}\ \bibnamefont
  {Monson}}, \bibinfo {author} {\bibfnamefont {Jill}\ \bibnamefont {Neeley}}, \
  and\ \bibinfo {author} {\bibfnamefont {Jeffrey}\ \bibnamefont {Rich}},\
  }\bibfield  {title} {\enquote {\bibinfo {title} {{Calibration of the Tip of
  the Red Giant Branch (TRGB)}},}\ }\href {\doibase 10.3847/1538-4357/ab7339}
  {\  (\bibinfo {year} {2020}),\ 10.3847/1538-4357/ab7339},\ \Eprint
  {http://arxiv.org/abs/2002.01550} {arXiv:2002.01550 [astro-ph.GA]}
  \BibitemShut {NoStop}%
\bibitem [{\citenamefont {{Barreira}}\ \emph {et~al.}(2014)\citenamefont
  {{Barreira}}, \citenamefont {{Li}}, \citenamefont {{Baugh}},\ and\
  \citenamefont {{Pascoli}}}]{barreira}%
  \BibitemOpen
  \bibfield  {author} {\bibinfo {author} {\bibfnamefont {Alexandre}\
  \bibnamefont {{Barreira}}}, \bibinfo {author} {\bibfnamefont {Baojiu}\
  \bibnamefont {{Li}}}, \bibinfo {author} {\bibfnamefont {Carlton~M.}\
  \bibnamefont {{Baugh}}}, \ and\ \bibinfo {author} {\bibfnamefont {Silvia}\
  \bibnamefont {{Pascoli}}},\ }\bibfield  {title} {\enquote {\bibinfo {title}
  {{The observational status of Galileon gravity after Planck}},}\ }\href
  {\doibase 10.1088/1475-7516/2014/08/059} {\bibfield  {journal} {\bibinfo
  {journal} {\jcap}\ }\textbf {\bibinfo {volume} {2014}},\ \bibinfo {eid} {059}
  (\bibinfo {year} {2014})},\ \Eprint {http://arxiv.org/abs/1406.0485}
  {arXiv:1406.0485 [astro-ph.CO]} \BibitemShut {NoStop}%
\bibitem [{\citenamefont {{Umilt{\`a}}}\ \emph {et~al.}(2015)\citenamefont
  {{Umilt{\`a}}}, \citenamefont {{Ballardini}}, \citenamefont {{Finelli}},\
  and\ \citenamefont {{Paoletti}}}]{umilta2015}%
  \BibitemOpen
  \bibfield  {author} {\bibinfo {author} {\bibfnamefont {C.}~\bibnamefont
  {{Umilt{\`a}}}}, \bibinfo {author} {\bibfnamefont {M.}~\bibnamefont
  {{Ballardini}}}, \bibinfo {author} {\bibfnamefont {F.}~\bibnamefont
  {{Finelli}}}, \ and\ \bibinfo {author} {\bibfnamefont {D.}~\bibnamefont
  {{Paoletti}}},\ }\bibfield  {title} {\enquote {\bibinfo {title} {{CMB and BAO
  constraints for an induced gravity dark energy model with a quartic
  potential}},}\ }\href {\doibase 10.1088/1475-7516/2015/08/017} {\bibfield
  {journal} {\bibinfo  {journal} {\jcap}\ }\textbf {\bibinfo {volume} {2015}},\
  \bibinfo {eid} {017} (\bibinfo {year} {2015})},\ \Eprint
  {http://arxiv.org/abs/1507.00718} {arXiv:1507.00718 [astro-ph.CO]}
  \BibitemShut {NoStop}%
\bibitem [{\citenamefont {Lesgourgues}\ \emph {et~al.}(2016)\citenamefont
  {Lesgourgues}, \citenamefont {Marques-Tavares},\ and\ \citenamefont
  {Schmaltz}}]{Lesgourgues:2015wza}%
  \BibitemOpen
  \bibfield  {author} {\bibinfo {author} {\bibfnamefont {Julien}\ \bibnamefont
  {Lesgourgues}}, \bibinfo {author} {\bibfnamefont {Gustavo}\ \bibnamefont
  {Marques-Tavares}}, \ and\ \bibinfo {author} {\bibfnamefont {Martin}\
  \bibnamefont {Schmaltz}},\ }\bibfield  {title} {\enquote {\bibinfo {title}
  {{Evidence for dark matter interactions in cosmological precision data?}}}\
  }\href {\doibase 10.1088/1475-7516/2016/02/037} {\bibfield  {journal}
  {\bibinfo  {journal} {JCAP}\ }\textbf {\bibinfo {volume} {1602}},\ \bibinfo
  {pages} {037} (\bibinfo {year} {2016})},\ \Eprint
  {http://arxiv.org/abs/1507.04351} {arXiv:1507.04351 [astro-ph.CO]}
  \BibitemShut {NoStop}%
\bibitem [{\citenamefont {Alam}\ \emph
  {et~al.}(2017{\natexlab{b}})\citenamefont {Alam}, \citenamefont {Bag},\ and\
  \citenamefont {Sahni}}]{Alam:2016wpf}%
  \BibitemOpen
  \bibfield  {author} {\bibinfo {author} {\bibfnamefont {Ujjaini}\ \bibnamefont
  {Alam}}, \bibinfo {author} {\bibfnamefont {Satadru}\ \bibnamefont {Bag}}, \
  and\ \bibinfo {author} {\bibfnamefont {Varun}\ \bibnamefont {Sahni}},\
  }\bibfield  {title} {\enquote {\bibinfo {title} {{Constraining the Cosmology
  of the Phantom Brane using Distance Measures}},}\ }\href {\doibase
  10.1103/PhysRevD.95.023524} {\bibfield  {journal} {\bibinfo  {journal} {Phys.
  Rev.}\ }\textbf {\bibinfo {volume} {D95}},\ \bibinfo {pages} {023524}
  (\bibinfo {year} {2017}{\natexlab{b}})},\ \Eprint
  {http://arxiv.org/abs/1605.04707} {arXiv:1605.04707 [astro-ph.CO]}
  \BibitemShut {NoStop}%
\bibitem [{\citenamefont {Di~Valentino}\ \emph {et~al.}(2016)\citenamefont
  {Di~Valentino}, \citenamefont {Melchiorri},\ and\ \citenamefont
  {Silk}}]{DiValentino:2016hlg}%
  \BibitemOpen
  \bibfield  {author} {\bibinfo {author} {\bibfnamefont {Eleonora}\
  \bibnamefont {Di~Valentino}}, \bibinfo {author} {\bibfnamefont {Alessandro}\
  \bibnamefont {Melchiorri}}, \ and\ \bibinfo {author} {\bibfnamefont {Joseph}\
  \bibnamefont {Silk}},\ }\bibfield  {title} {\enquote {\bibinfo {title}
  {{Reconciling Planck with the local value of $H_0$ in extended parameter
  space}},}\ }\href {\doibase 10.1016/j.physletb.2016.08.043} {\bibfield
  {journal} {\bibinfo  {journal} {Phys. Lett.}\ }\textbf {\bibinfo {volume}
  {B761}},\ \bibinfo {pages} {242--246} (\bibinfo {year} {2016})},\ \Eprint
  {http://arxiv.org/abs/1606.00634} {arXiv:1606.00634 [astro-ph.CO]}
  \BibitemShut {NoStop}%
\bibitem [{\citenamefont {Huang}\ and\ \citenamefont
  {Wang}(2016)}]{Qing-Guo:2016ykt}%
  \BibitemOpen
  \bibfield  {author} {\bibinfo {author} {\bibfnamefont {Qing-Guo}\
  \bibnamefont {Huang}}\ and\ \bibinfo {author} {\bibfnamefont
  {Ke}~\bibnamefont {Wang}},\ }\bibfield  {title} {\enquote {\bibinfo {title}
  {{How the dark energy can reconcile Planck with local determination of the
  Hubble constant}},}\ }\href {\doibase 10.1140/epjc/s10052-016-4352-x}
  {\bibfield  {journal} {\bibinfo  {journal} {Eur. Phys. J.}\ }\textbf
  {\bibinfo {volume} {C76}},\ \bibinfo {pages} {506} (\bibinfo {year}
  {2016})},\ \Eprint {http://arxiv.org/abs/1606.05965} {arXiv:1606.05965
  [astro-ph.CO]} \BibitemShut {NoStop}%
\bibitem [{\citenamefont {Ko}\ and\ \citenamefont {Tang}(2016)}]{Ko:2016uft}%
  \BibitemOpen
  \bibfield  {author} {\bibinfo {author} {\bibfnamefont {P.}~\bibnamefont
  {Ko}}\ and\ \bibinfo {author} {\bibfnamefont {Yong}\ \bibnamefont {Tang}},\
  }\bibfield  {title} {\enquote {\bibinfo {title} {{Light dark photon and
  fermionic dark radiation for the Hubble constant and the structure
  formation}},}\ }\href {\doibase 10.1016/j.physletb.2016.10.001} {\bibfield
  {journal} {\bibinfo  {journal} {Phys. Lett.}\ }\textbf {\bibinfo {volume}
  {B762}},\ \bibinfo {pages} {462--466} (\bibinfo {year} {2016})},\ \Eprint
  {http://arxiv.org/abs/1608.01083} {arXiv:1608.01083 [hep-ph]} \BibitemShut
  {NoStop}%
\bibitem [{\citenamefont {Karwal}\ and\ \citenamefont
  {Kamionkowski}(2016)}]{Karwal:2016vyq}%
  \BibitemOpen
  \bibfield  {author} {\bibinfo {author} {\bibfnamefont {Tanvi}\ \bibnamefont
  {Karwal}}\ and\ \bibinfo {author} {\bibfnamefont {Marc}\ \bibnamefont
  {Kamionkowski}},\ }\bibfield  {title} {\enquote {\bibinfo {title} {{Dark
  energy at early times, the Hubble parameter, and the string axiverse}},}\
  }\href {\doibase 10.1103/PhysRevD.94.103523} {\bibfield  {journal} {\bibinfo
  {journal} {Phys. Rev.}\ }\textbf {\bibinfo {volume} {D94}},\ \bibinfo {pages}
  {103523} (\bibinfo {year} {2016})},\ \Eprint
  {http://arxiv.org/abs/1608.01309} {arXiv:1608.01309 [astro-ph.CO]}
  \BibitemShut {NoStop}%
\bibitem [{\citenamefont {Kumar}\ and\ \citenamefont
  {Nunes}(2016)}]{Kumar:2016zpg}%
  \BibitemOpen
  \bibfield  {author} {\bibinfo {author} {\bibfnamefont {Suresh}\ \bibnamefont
  {Kumar}}\ and\ \bibinfo {author} {\bibfnamefont {Rafael~C.}\ \bibnamefont
  {Nunes}},\ }\bibfield  {title} {\enquote {\bibinfo {title} {{Probing the
  interaction between dark matter and dark energy in the presence of massive
  neutrinos}},}\ }\href {\doibase 10.1103/PhysRevD.94.123511} {\bibfield
  {journal} {\bibinfo  {journal} {Phys. Rev.}\ }\textbf {\bibinfo {volume}
  {D94}},\ \bibinfo {pages} {123511} (\bibinfo {year} {2016})},\ \Eprint
  {http://arxiv.org/abs/1608.02454} {arXiv:1608.02454 [astro-ph.CO]}
  \BibitemShut {NoStop}%
\bibitem [{\citenamefont {{Renk}}\ \emph {et~al.}(2017)\citenamefont {{Renk}},
  \citenamefont {{Zumalac{\'a}rregui}}, \citenamefont {{Montanari}},\ and\
  \citenamefont {{Barreira}}}]{renk2017}%
  \BibitemOpen
  \bibfield  {author} {\bibinfo {author} {\bibfnamefont {Janina}\ \bibnamefont
  {{Renk}}}, \bibinfo {author} {\bibfnamefont {Miguel}\ \bibnamefont
  {{Zumalac{\'a}rregui}}}, \bibinfo {author} {\bibfnamefont {Francesco}\
  \bibnamefont {{Montanari}}}, \ and\ \bibinfo {author} {\bibfnamefont
  {Alexandre}\ \bibnamefont {{Barreira}}},\ }\bibfield  {title} {\enquote
  {\bibinfo {title} {{Galileon gravity in light of ISW, CMB, BAO and H$_{0}$
  data}},}\ }\href {\doibase 10.1088/1475-7516/2017/10/020} {\bibfield
  {journal} {\bibinfo  {journal} {\jcap}\ }\textbf {\bibinfo {volume} {2017}},\
  \bibinfo {eid} {020} (\bibinfo {year} {2017})},\ \Eprint
  {http://arxiv.org/abs/1707.02263} {arXiv:1707.02263 [astro-ph.CO]}
  \BibitemShut {NoStop}%
\bibitem [{\citenamefont {Di~Valentino}\ \emph
  {et~al.}(2017{\natexlab{a}})\citenamefont {Di~Valentino}, \citenamefont
  {Melchiorri}, \citenamefont {Linder},\ and\ \citenamefont
  {Silk}}]{DiValentino:2017zyq}%
  \BibitemOpen
  \bibfield  {author} {\bibinfo {author} {\bibfnamefont {Eleonora}\
  \bibnamefont {Di~Valentino}}, \bibinfo {author} {\bibfnamefont {Alessandro}\
  \bibnamefont {Melchiorri}}, \bibinfo {author} {\bibfnamefont {Eric~V.}\
  \bibnamefont {Linder}}, \ and\ \bibinfo {author} {\bibfnamefont {Joseph}\
  \bibnamefont {Silk}},\ }\bibfield  {title} {\enquote {\bibinfo {title}
  {{Constraining Dark Energy Dynamics in Extended Parameter Space}},}\ }\href
  {\doibase 10.1103/PhysRevD.96.023523} {\bibfield  {journal} {\bibinfo
  {journal} {Phys. Rev.}\ }\textbf {\bibinfo {volume} {D96}},\ \bibinfo {pages}
  {023523} (\bibinfo {year} {2017}{\natexlab{a}})},\ \Eprint
  {http://arxiv.org/abs/1704.00762} {arXiv:1704.00762 [astro-ph.CO]}
  \BibitemShut {NoStop}%
\bibitem [{\citenamefont {Di~Valentino}\ \emph
  {et~al.}(2017{\natexlab{b}})\citenamefont {Di~Valentino}, \citenamefont
  {Melchiorri},\ and\ \citenamefont {Mena}}]{DiValentino:2017iww}%
  \BibitemOpen
  \bibfield  {author} {\bibinfo {author} {\bibfnamefont {Eleonora}\
  \bibnamefont {Di~Valentino}}, \bibinfo {author} {\bibfnamefont {Alessandro}\
  \bibnamefont {Melchiorri}}, \ and\ \bibinfo {author} {\bibfnamefont {Olga}\
  \bibnamefont {Mena}},\ }\bibfield  {title} {\enquote {\bibinfo {title} {{Can
  interacting dark energy solve the $H_0$ tension?}}}\ }\href {\doibase
  10.1103/PhysRevD.96.043503} {\bibfield  {journal} {\bibinfo  {journal} {Phys.
  Rev.}\ }\textbf {\bibinfo {volume} {D96}},\ \bibinfo {pages} {043503}
  (\bibinfo {year} {2017}{\natexlab{b}})},\ \Eprint
  {http://arxiv.org/abs/1704.08342} {arXiv:1704.08342 [astro-ph.CO]}
  \BibitemShut {NoStop}%
\bibitem [{\citenamefont {Di~Valentino}\ \emph
  {et~al.}(2018{\natexlab{a}})\citenamefont {Di~Valentino}, \citenamefont
  {Bøehm}, \citenamefont {Hivon},\ and\ \citenamefont
  {Bouchet}}]{DiValentino:2017oaw}%
  \BibitemOpen
  \bibfield  {author} {\bibinfo {author} {\bibfnamefont {Eleonora}\
  \bibnamefont {Di~Valentino}}, \bibinfo {author} {\bibfnamefont {Céline}\
  \bibnamefont {Bøehm}}, \bibinfo {author} {\bibfnamefont {Eric}\ \bibnamefont
  {Hivon}}, \ and\ \bibinfo {author} {\bibfnamefont {François~R.}\
  \bibnamefont {Bouchet}},\ }\bibfield  {title} {\enquote {\bibinfo {title}
  {{Reducing the $H_0$ and $\sigma_8$ tensions with Dark Matter-neutrino
  interactions}},}\ }\href {\doibase 10.1103/PhysRevD.97.043513} {\bibfield
  {journal} {\bibinfo  {journal} {Phys. Rev.}\ }\textbf {\bibinfo {volume}
  {D97}},\ \bibinfo {pages} {043513} (\bibinfo {year} {2018}{\natexlab{a}})},\
  \Eprint {http://arxiv.org/abs/1710.02559} {arXiv:1710.02559 [astro-ph.CO]}
  \BibitemShut {NoStop}%
\bibitem [{\citenamefont {Bolejko}(2018)}]{Bolejko:2017fos}%
  \BibitemOpen
  \bibfield  {author} {\bibinfo {author} {\bibfnamefont {Krzysztof}\
  \bibnamefont {Bolejko}},\ }\bibfield  {title} {\enquote {\bibinfo {title}
  {{Emerging spatial curvature can resolve the tension between high-redshift
  CMB and low-redshift distance ladder measurements of the Hubble constant}},}\
  }\href {\doibase 10.1103/PhysRevD.97.103529} {\bibfield  {journal} {\bibinfo
  {journal} {Phys. Rev.}\ }\textbf {\bibinfo {volume} {D97}},\ \bibinfo {pages}
  {103529} (\bibinfo {year} {2018})},\ \Eprint
  {http://arxiv.org/abs/1712.02967} {arXiv:1712.02967 [astro-ph.CO]}
  \BibitemShut {NoStop}%
\bibitem [{\citenamefont {Di~Valentino}\ \emph
  {et~al.}(2018{\natexlab{b}})\citenamefont {Di~Valentino}, \citenamefont
  {Linder},\ and\ \citenamefont {Melchiorri}}]{DiValentino:2017rcr}%
  \BibitemOpen
  \bibfield  {author} {\bibinfo {author} {\bibfnamefont {Eleonora}\
  \bibnamefont {Di~Valentino}}, \bibinfo {author} {\bibfnamefont {Eric~V.}\
  \bibnamefont {Linder}}, \ and\ \bibinfo {author} {\bibfnamefont {Alessandro}\
  \bibnamefont {Melchiorri}},\ }\bibfield  {title} {\enquote {\bibinfo {title}
  {{Vacuum phase transition solves the $H_0$ tension}},}\ }\href {\doibase
  10.1103/PhysRevD.97.043528} {\bibfield  {journal} {\bibinfo  {journal} {Phys.
  Rev.}\ }\textbf {\bibinfo {volume} {D97}},\ \bibinfo {pages} {043528}
  (\bibinfo {year} {2018}{\natexlab{b}})},\ \Eprint
  {http://arxiv.org/abs/1710.02153} {arXiv:1710.02153 [astro-ph.CO]}
  \BibitemShut {NoStop}%
\bibitem [{\citenamefont {Lancaster}\ \emph {et~al.}(2017)\citenamefont
  {Lancaster}, \citenamefont {Cyr-Racine}, \citenamefont {Knox},\ and\
  \citenamefont {Pan}}]{Lancaster:2017ksf}%
  \BibitemOpen
  \bibfield  {author} {\bibinfo {author} {\bibfnamefont {Lachlan}\ \bibnamefont
  {Lancaster}}, \bibinfo {author} {\bibfnamefont {Francis-Yan}\ \bibnamefont
  {Cyr-Racine}}, \bibinfo {author} {\bibfnamefont {Lloyd}\ \bibnamefont
  {Knox}}, \ and\ \bibinfo {author} {\bibfnamefont {Zhen}\ \bibnamefont
  {Pan}},\ }\bibfield  {title} {\enquote {\bibinfo {title} {{A tale of two
  modes: Neutrino free-streaming in the early universe}},}\ }\href {\doibase
  10.1088/1475-7516/2017/07/033} {\bibfield  {journal} {\bibinfo  {journal}
  {JCAP}\ }\textbf {\bibinfo {volume} {1707}},\ \bibinfo {pages} {033}
  (\bibinfo {year} {2017})},\ \Eprint {http://arxiv.org/abs/1704.06657}
  {arXiv:1704.06657 [astro-ph.CO]} \BibitemShut {NoStop}%
\bibitem [{\citenamefont {Khosravi}\ \emph {et~al.}(2019)\citenamefont
  {Khosravi}, \citenamefont {Baghram}, \citenamefont {Afshordi},\ and\
  \citenamefont {Altamirano}}]{Khosravi:2017hfi}%
  \BibitemOpen
  \bibfield  {author} {\bibinfo {author} {\bibfnamefont {Nima}\ \bibnamefont
  {Khosravi}}, \bibinfo {author} {\bibfnamefont {Shant}\ \bibnamefont
  {Baghram}}, \bibinfo {author} {\bibfnamefont {Niayesh}\ \bibnamefont
  {Afshordi}}, \ and\ \bibinfo {author} {\bibfnamefont {Natacha}\ \bibnamefont
  {Altamirano}},\ }\bibfield  {title} {\enquote {\bibinfo {title} {{$H_0$
  tension as a hint for a transition in gravitational theory}},}\ }\href
  {\doibase 10.1103/PhysRevD.99.103526} {\bibfield  {journal} {\bibinfo
  {journal} {Phys. Rev.}\ }\textbf {\bibinfo {volume} {D99}},\ \bibinfo {pages}
  {103526} (\bibinfo {year} {2019})},\ \Eprint
  {http://arxiv.org/abs/1710.09366} {arXiv:1710.09366 [astro-ph.CO]}
  \BibitemShut {NoStop}%
\bibitem [{\citenamefont {Buen-Abad}\ \emph {et~al.}(2018)\citenamefont
  {Buen-Abad}, \citenamefont {Schmaltz}, \citenamefont {Lesgourgues},\ and\
  \citenamefont {Brinckmann}}]{Buen-Abad:2017gxg}%
  \BibitemOpen
  \bibfield  {author} {\bibinfo {author} {\bibfnamefont {Manuel~A.}\
  \bibnamefont {Buen-Abad}}, \bibinfo {author} {\bibfnamefont {Martin}\
  \bibnamefont {Schmaltz}}, \bibinfo {author} {\bibfnamefont {Julien}\
  \bibnamefont {Lesgourgues}}, \ and\ \bibinfo {author} {\bibfnamefont {Thejs}\
  \bibnamefont {Brinckmann}},\ }\bibfield  {title} {\enquote {\bibinfo {title}
  {{Interacting Dark Sector and Precision Cosmology}},}\ }\href {\doibase
  10.1088/1475-7516/2018/01/008} {\bibfield  {journal} {\bibinfo  {journal}
  {JCAP}\ }\textbf {\bibinfo {volume} {1801}},\ \bibinfo {pages} {008}
  (\bibinfo {year} {2018})},\ \Eprint {http://arxiv.org/abs/1708.09406}
  {arXiv:1708.09406 [astro-ph.CO]} \BibitemShut {NoStop}%
\bibitem [{\citenamefont {D'Eramo}\ \emph {et~al.}(2018)\citenamefont
  {D'Eramo}, \citenamefont {Ferreira}, \citenamefont {Notari},\ and\
  \citenamefont {Bernal}}]{DEramo:2018vss}%
  \BibitemOpen
  \bibfield  {author} {\bibinfo {author} {\bibfnamefont {Francesco}\
  \bibnamefont {D'Eramo}}, \bibinfo {author} {\bibfnamefont {Ricardo~Z.}\
  \bibnamefont {Ferreira}}, \bibinfo {author} {\bibfnamefont {Alessio}\
  \bibnamefont {Notari}}, \ and\ \bibinfo {author} {\bibfnamefont {José~Luis}\
  \bibnamefont {Bernal}},\ }\bibfield  {title} {\enquote {\bibinfo {title}
  {{Hot Axions and the $H_0$ tension}},}\ }\href {\doibase
  10.1088/1475-7516/2018/11/014} {\bibfield  {journal} {\bibinfo  {journal}
  {JCAP}\ }\textbf {\bibinfo {volume} {1811}},\ \bibinfo {pages} {014}
  (\bibinfo {year} {2018})},\ \Eprint {http://arxiv.org/abs/1808.07430}
  {arXiv:1808.07430 [hep-ph]} \BibitemShut {NoStop}%
\bibitem [{\citenamefont {Dutta}\ \emph {et~al.}(2018)\citenamefont {Dutta},
  \citenamefont {Ruchika}, \citenamefont {Roy}, \citenamefont {Sen},\ and\
  \citenamefont {Sheikh-Jabbari}}]{Dutta:2018vmq}%
  \BibitemOpen
  \bibfield  {author} {\bibinfo {author} {\bibfnamefont {Koushik}\ \bibnamefont
  {Dutta}}, \bibinfo {author} {\bibnamefont {Ruchika}}, \bibinfo {author}
  {\bibfnamefont {Anirban}\ \bibnamefont {Roy}}, \bibinfo {author}
  {\bibfnamefont {Anjan~A.}\ \bibnamefont {Sen}}, \ and\ \bibinfo {author}
  {\bibfnamefont {M.~M.}\ \bibnamefont {Sheikh-Jabbari}},\ }\bibfield  {title}
  {\enquote {\bibinfo {title} {{Beyond $\Lambda$CDM with Low and High Redshift
  Data: Implications for Dark Energy}},}\ }\href@noop {} {\  (\bibinfo {year}
  {2018})},\ \Eprint {http://arxiv.org/abs/1808.06623} {arXiv:1808.06623
  [astro-ph.CO]} \BibitemShut {NoStop}%
\bibitem [{\citenamefont {Banihashemi}\ \emph {et~al.}(2019)\citenamefont
  {Banihashemi}, \citenamefont {Khosravi},\ and\ \citenamefont
  {Shirazi}}]{Banihashemi:2018has}%
  \BibitemOpen
  \bibfield  {author} {\bibinfo {author} {\bibfnamefont {Abdolali}\
  \bibnamefont {Banihashemi}}, \bibinfo {author} {\bibfnamefont {Nima}\
  \bibnamefont {Khosravi}}, \ and\ \bibinfo {author} {\bibfnamefont {Amir~H.}\
  \bibnamefont {Shirazi}},\ }\bibfield  {title} {\enquote {\bibinfo {title}
  {{Ginzburg-Landau Theory of Dark Energy: A Framework to Study Both Temporal
  and Spatial Cosmological Tensions Simultaneously}},}\ }\href {\doibase
  10.1103/PhysRevD.99.083509} {\bibfield  {journal} {\bibinfo  {journal} {Phys.
  Rev.}\ }\textbf {\bibinfo {volume} {D99}},\ \bibinfo {pages} {083509}
  (\bibinfo {year} {2019})},\ \Eprint {http://arxiv.org/abs/1810.11007}
  {arXiv:1810.11007 [astro-ph.CO]} \BibitemShut {NoStop}%
\bibitem [{\citenamefont {{Belgacem}}\ \emph {et~al.}(2018)\citenamefont
  {{Belgacem}}, \citenamefont {{Dirian}}, \citenamefont {{Foffa}},\ and\
  \citenamefont {{Maggiore}}}]{belgacem2018}%
  \BibitemOpen
  \bibfield  {author} {\bibinfo {author} {\bibfnamefont {Enis}\ \bibnamefont
  {{Belgacem}}}, \bibinfo {author} {\bibfnamefont {Yves}\ \bibnamefont
  {{Dirian}}}, \bibinfo {author} {\bibfnamefont {Stefano}\ \bibnamefont
  {{Foffa}}}, \ and\ \bibinfo {author} {\bibfnamefont {Michele}\ \bibnamefont
  {{Maggiore}}},\ }\bibfield  {title} {\enquote {\bibinfo {title} {{Nonlocal
  gravity. Conceptual aspects and cosmological predictions}},}\ }\href
  {\doibase 10.1088/1475-7516/2018/03/002} {\bibfield  {journal} {\bibinfo
  {journal} {\jcap}\ }\textbf {\bibinfo {volume} {2018}},\ \bibinfo {eid} {002}
  (\bibinfo {year} {2018})},\ \Eprint {http://arxiv.org/abs/1712.07066}
  {arXiv:1712.07066 [hep-th]} \BibitemShut {NoStop}%
\bibitem [{\citenamefont {Pandey}\ \emph {et~al.}(2019)\citenamefont {Pandey},
  \citenamefont {Karwal},\ and\ \citenamefont {Das}}]{Pandey:2019plg}%
  \BibitemOpen
  \bibfield  {author} {\bibinfo {author} {\bibfnamefont {Kanhaiya~L.}\
  \bibnamefont {Pandey}}, \bibinfo {author} {\bibfnamefont {Tanvi}\
  \bibnamefont {Karwal}}, \ and\ \bibinfo {author} {\bibfnamefont {Subinoy}\
  \bibnamefont {Das}},\ }\bibfield  {title} {\enquote {\bibinfo {title}
  {{Alleviating the $H_0$ and $\sigma_8$ anomalies with a decaying dark matter
  model}},}\ }\href@noop {} {\  (\bibinfo {year} {2019})},\ \Eprint
  {http://arxiv.org/abs/1902.10636} {arXiv:1902.10636 [astro-ph.CO]}
  \BibitemShut {NoStop}%
\bibitem [{\citenamefont {Agrawal}\ \emph
  {et~al.}(2019{\natexlab{a}})\citenamefont {Agrawal}, \citenamefont
  {Cyr-Racine}, \citenamefont {Pinner},\ and\ \citenamefont
  {Randall}}]{Agrawal:2019lmo}%
  \BibitemOpen
  \bibfield  {author} {\bibinfo {author} {\bibfnamefont {Prateek}\ \bibnamefont
  {Agrawal}}, \bibinfo {author} {\bibfnamefont {Francis-Yan}\ \bibnamefont
  {Cyr-Racine}}, \bibinfo {author} {\bibfnamefont {David}\ \bibnamefont
  {Pinner}}, \ and\ \bibinfo {author} {\bibfnamefont {Lisa}\ \bibnamefont
  {Randall}},\ }\bibfield  {title} {\enquote {\bibinfo {title} {{Rock 'n' Roll
  Solutions to the Hubble Tension}},}\ }\href@noop {} {\  (\bibinfo {year}
  {2019}{\natexlab{a}})},\ \Eprint {http://arxiv.org/abs/1904.01016}
  {arXiv:1904.01016 [astro-ph.CO]} \BibitemShut {NoStop}%
\bibitem [{\citenamefont {Agrawal}\ \emph
  {et~al.}(2019{\natexlab{b}})\citenamefont {Agrawal}, \citenamefont {Obied},\
  and\ \citenamefont {Vafa}}]{Agrawal:2019dlm}%
  \BibitemOpen
  \bibfield  {author} {\bibinfo {author} {\bibfnamefont {Prateek}\ \bibnamefont
  {Agrawal}}, \bibinfo {author} {\bibfnamefont {Georges}\ \bibnamefont
  {Obied}}, \ and\ \bibinfo {author} {\bibfnamefont {Cumrun}\ \bibnamefont
  {Vafa}},\ }\bibfield  {title} {\enquote {\bibinfo {title} {{$H_0$ Tension,
  Swampland Conjectures and the Epoch of Fading Dark Matter}},}\ }\href@noop {}
  {\  (\bibinfo {year} {2019}{\natexlab{b}})},\ \Eprint
  {http://arxiv.org/abs/1906.08261} {arXiv:1906.08261 [astro-ph.CO]}
  \BibitemShut {NoStop}%
\bibitem [{\citenamefont {Di~Valentino}\ \emph
  {et~al.}(2019{\natexlab{a}})\citenamefont {Di~Valentino}, \citenamefont
  {Ferreira}, \citenamefont {Visinelli},\ and\ \citenamefont
  {Danielsson}}]{DiValentino:2019exe}%
  \BibitemOpen
  \bibfield  {author} {\bibinfo {author} {\bibfnamefont {Eleonora}\
  \bibnamefont {Di~Valentino}}, \bibinfo {author} {\bibfnamefont {Ricardo~Z.}\
  \bibnamefont {Ferreira}}, \bibinfo {author} {\bibfnamefont {Luca}\
  \bibnamefont {Visinelli}}, \ and\ \bibinfo {author} {\bibfnamefont {Ulf}\
  \bibnamefont {Danielsson}},\ }\bibfield  {title} {\enquote {\bibinfo {title}
  {{Late time transitions in the quintessence field and the $H_0$ tension}},}\
  }\href@noop {} {\  (\bibinfo {year} {2019}{\natexlab{a}})},\ \Eprint
  {http://arxiv.org/abs/1906.11255} {arXiv:1906.11255 [astro-ph.CO]}
  \BibitemShut {NoStop}%
\bibitem [{\citenamefont {Desmond}\ \emph {et~al.}(2019)\citenamefont
  {Desmond}, \citenamefont {Jain},\ and\ \citenamefont
  {Sakstein}}]{Desmond:2019ygn}%
  \BibitemOpen
  \bibfield  {author} {\bibinfo {author} {\bibfnamefont {Harry}\ \bibnamefont
  {Desmond}}, \bibinfo {author} {\bibfnamefont {Bhuvnesh}\ \bibnamefont
  {Jain}}, \ and\ \bibinfo {author} {\bibfnamefont {Jeremy}\ \bibnamefont
  {Sakstein}},\ }\bibfield  {title} {\enquote {\bibinfo {title} {{A local
  resolution of the Hubble tension: The impact of screened fifth forces on the
  cosmic distance ladder}},}\ }\href@noop {} {\  (\bibinfo {year} {2019})},\
  \Eprint {http://arxiv.org/abs/1907.03778} {arXiv:1907.03778 [astro-ph.CO]}
  \BibitemShut {NoStop}%
\bibitem [{\citenamefont {Pan}\ \emph {et~al.}(2019)\citenamefont {Pan},
  \citenamefont {Yang}, \citenamefont {Di~Valentino}, \citenamefont
  {Shafieloo},\ and\ \citenamefont {Chakraborty}}]{Pan:2019cot}%
  \BibitemOpen
  \bibfield  {author} {\bibinfo {author} {\bibfnamefont {Supriya}\ \bibnamefont
  {Pan}}, \bibinfo {author} {\bibfnamefont {Weiqiang}\ \bibnamefont {Yang}},
  \bibinfo {author} {\bibfnamefont {Eleonora}\ \bibnamefont {Di~Valentino}},
  \bibinfo {author} {\bibfnamefont {Arman}\ \bibnamefont {Shafieloo}}, \ and\
  \bibinfo {author} {\bibfnamefont {Subenoy}\ \bibnamefont {Chakraborty}},\
  }\bibfield  {title} {\enquote {\bibinfo {title} {{Reconciling $H_0$ tension
  in a six parameter space?}}}\ }\href@noop {} {\  (\bibinfo {year} {2019})},\
  \Eprint {http://arxiv.org/abs/1907.12551} {arXiv:1907.12551 [astro-ph.CO]}
  \BibitemShut {NoStop}%
\bibitem [{\citenamefont {Vattis}\ \emph {et~al.}(2019)\citenamefont {Vattis},
  \citenamefont {Koushiappas},\ and\ \citenamefont {Loeb}}]{Vattis:2019efj}%
  \BibitemOpen
  \bibfield  {author} {\bibinfo {author} {\bibfnamefont {Kyriakos}\
  \bibnamefont {Vattis}}, \bibinfo {author} {\bibfnamefont {Savvas~M.}\
  \bibnamefont {Koushiappas}}, \ and\ \bibinfo {author} {\bibfnamefont
  {Abraham}\ \bibnamefont {Loeb}},\ }\bibfield  {title} {\enquote {\bibinfo
  {title} {{Dark matter decaying in the late Universe can relieve the H0
  tension}},}\ }\href {\doibase 10.1103/PhysRevD.99.121302} {\bibfield
  {journal} {\bibinfo  {journal} {Phys. Rev.}\ }\textbf {\bibinfo {volume}
  {D99}},\ \bibinfo {pages} {121302} (\bibinfo {year} {2019})},\ \Eprint
  {http://arxiv.org/abs/1903.06220} {arXiv:1903.06220 [astro-ph.CO]}
  \BibitemShut {NoStop}%
\bibitem [{\citenamefont {Poulin}\ \emph {et~al.}(2019)\citenamefont {Poulin},
  \citenamefont {Smith}, \citenamefont {Karwal},\ and\ \citenamefont
  {Kamionkowski}}]{Poulin:2018cxd}%
  \BibitemOpen
  \bibfield  {author} {\bibinfo {author} {\bibfnamefont {Vivian}\ \bibnamefont
  {Poulin}}, \bibinfo {author} {\bibfnamefont {Tristan~L.}\ \bibnamefont
  {Smith}}, \bibinfo {author} {\bibfnamefont {Tanvi}\ \bibnamefont {Karwal}}, \
  and\ \bibinfo {author} {\bibfnamefont {Marc}\ \bibnamefont {Kamionkowski}},\
  }\bibfield  {title} {\enquote {\bibinfo {title} {{Early Dark Energy Can
  Resolve The Hubble Tension}},}\ }\href {\doibase
  10.1103/PhysRevLett.122.221301} {\bibfield  {journal} {\bibinfo  {journal}
  {Phys. Rev. Lett.}\ }\textbf {\bibinfo {volume} {122}},\ \bibinfo {pages}
  {221301} (\bibinfo {year} {2019})},\ \Eprint
  {http://arxiv.org/abs/1811.04083} {arXiv:1811.04083 [astro-ph.CO]}
  \BibitemShut {NoStop}%
\bibitem [{\citenamefont {{Lin}}\ \emph
  {et~al.}(2019{\natexlab{a}})\citenamefont {{Lin}}, \citenamefont {{Raveri}},\
  and\ \citenamefont {{Hu}}}]{lin2019}%
  \BibitemOpen
  \bibfield  {author} {\bibinfo {author} {\bibfnamefont {Meng-Xiang}\
  \bibnamefont {{Lin}}}, \bibinfo {author} {\bibfnamefont {Marco}\ \bibnamefont
  {{Raveri}}}, \ and\ \bibinfo {author} {\bibfnamefont {Wayne}\ \bibnamefont
  {{Hu}}},\ }\bibfield  {title} {\enquote {\bibinfo {title} {{Phenomenology of
  modified gravity at recombination}},}\ }\href {\doibase
  10.1103/PhysRevD.99.043514} {\bibfield  {journal} {\bibinfo  {journal}
  {\prd}\ }\textbf {\bibinfo {volume} {99}},\ \bibinfo {eid} {043514} (\bibinfo
  {year} {2019}{\natexlab{a}})},\ \Eprint {http://arxiv.org/abs/1810.02333}
  {arXiv:1810.02333 [astro-ph.CO]} \BibitemShut {NoStop}%
\bibitem [{\citenamefont {Li}\ \emph {et~al.}(2019)\citenamefont {Li},
  \citenamefont {Shafieloo}, \citenamefont {Sahni},\ and\ \citenamefont
  {Starobinsky}}]{Li:2019san}%
  \BibitemOpen
  \bibfield  {author} {\bibinfo {author} {\bibfnamefont {Xiao-Lei}\
  \bibnamefont {Li}}, \bibinfo {author} {\bibfnamefont {Arman}\ \bibnamefont
  {Shafieloo}}, \bibinfo {author} {\bibfnamefont {Varun}\ \bibnamefont
  {Sahni}}, \ and\ \bibinfo {author} {\bibfnamefont {Alexei~A.}\ \bibnamefont
  {Starobinsky}},\ }\bibfield  {title} {\enquote {\bibinfo {title} {{Revisiting
  Metastable Dark Energy and Tensions in the Estimation of Cosmological
  Parameters}},}\ }\href@noop {} {\  (\bibinfo {year} {2019})},\ \Eprint
  {http://arxiv.org/abs/1904.03790} {arXiv:1904.03790 [astro-ph.CO]}
  \BibitemShut {NoStop}%
\bibitem [{\citenamefont {Alexander}\ and\ \citenamefont
  {McDonough}(2019)}]{Alexander:2019rsc}%
  \BibitemOpen
  \bibfield  {author} {\bibinfo {author} {\bibfnamefont {Stephon}\ \bibnamefont
  {Alexander}}\ and\ \bibinfo {author} {\bibfnamefont {Evan}\ \bibnamefont
  {McDonough}},\ }\bibfield  {title} {\enquote {\bibinfo {title}
  {{Axion-Dilaton Destabilization and the Hubble Tension}},}\ }\href {\doibase
  10.1016/j.physletb.2019.134830} {\  (\bibinfo {year} {2019}),\
  10.1016/j.physletb.2019.134830},\ \Eprint {http://arxiv.org/abs/1904.08912}
  {arXiv:1904.08912 [astro-ph.CO]} \BibitemShut {NoStop}%
\bibitem [{\citenamefont {{Lin}}\ \emph
  {et~al.}(2019{\natexlab{b}})\citenamefont {{Lin}}, \citenamefont
  {{Benevento}}, \citenamefont {{Hu}},\ and\ \citenamefont
  {{Raveri}}}]{lin2019b}%
  \BibitemOpen
  \bibfield  {author} {\bibinfo {author} {\bibfnamefont {Meng-Xiang}\
  \bibnamefont {{Lin}}}, \bibinfo {author} {\bibfnamefont {Giampaolo}\
  \bibnamefont {{Benevento}}}, \bibinfo {author} {\bibfnamefont {Wayne}\
  \bibnamefont {{Hu}}}, \ and\ \bibinfo {author} {\bibfnamefont {Marco}\
  \bibnamefont {{Raveri}}},\ }\bibfield  {title} {\enquote {\bibinfo {title}
  {{Acoustic Dark Energy: Potential Conversion of the Hubble Tension}},}\
  }\href@noop {} {\bibfield  {journal} {\bibinfo  {journal} {arXiv e-prints}\
  ,\ \bibinfo {eid} {arXiv:1905.12618}} (\bibinfo {year}
  {2019}{\natexlab{b}})},\ \Eprint {http://arxiv.org/abs/1905.12618}
  {arXiv:1905.12618 [astro-ph.CO]} \BibitemShut {NoStop}%
\bibitem [{\citenamefont {Di~Valentino}\ \emph
  {et~al.}(2019{\natexlab{b}})\citenamefont {Di~Valentino}, \citenamefont
  {Melchiorri}, \citenamefont {Mena},\ and\ \citenamefont
  {Vagnozzi}}]{DiValentino:2019ffd}%
  \BibitemOpen
  \bibfield  {author} {\bibinfo {author} {\bibfnamefont {Eleonora}\
  \bibnamefont {Di~Valentino}}, \bibinfo {author} {\bibfnamefont {Alessandro}\
  \bibnamefont {Melchiorri}}, \bibinfo {author} {\bibfnamefont {Olga}\
  \bibnamefont {Mena}}, \ and\ \bibinfo {author} {\bibfnamefont {Sunny}\
  \bibnamefont {Vagnozzi}},\ }\bibfield  {title} {\enquote {\bibinfo {title}
  {{Interacting dark energy after the latest Planck, DES, and $H_0$
  measurements: an excellent solution to the $H_0$ and cosmic shear
  tensions}},}\ }\href@noop {} {\  (\bibinfo {year} {2019}{\natexlab{b}})},\
  \Eprint {http://arxiv.org/abs/1908.04281} {arXiv:1908.04281 [astro-ph.CO]}
  \BibitemShut {NoStop}%
\bibitem [{\citenamefont {Archidiacono}\ \emph {et~al.}(2019)\citenamefont
  {Archidiacono}, \citenamefont {Hooper}, \citenamefont {Murgia}, \citenamefont
  {Bohr}, \citenamefont {Lesgourgues},\ and\ \citenamefont
  {Viel}}]{Archidiacono:2019wdp}%
  \BibitemOpen
  \bibfield  {author} {\bibinfo {author} {\bibfnamefont {Maria}\ \bibnamefont
  {Archidiacono}}, \bibinfo {author} {\bibfnamefont {Deanna~C.}\ \bibnamefont
  {Hooper}}, \bibinfo {author} {\bibfnamefont {Riccardo}\ \bibnamefont
  {Murgia}}, \bibinfo {author} {\bibfnamefont {Sebastian}\ \bibnamefont
  {Bohr}}, \bibinfo {author} {\bibfnamefont {Julien}\ \bibnamefont
  {Lesgourgues}}, \ and\ \bibinfo {author} {\bibfnamefont {Matteo}\
  \bibnamefont {Viel}},\ }\bibfield  {title} {\enquote {\bibinfo {title}
  {{Constraining Dark Matter -- Dark Radiation interactions with CMB, BAO, and
  Lyman-$\alpha$}},}\ }\href@noop {} {\  (\bibinfo {year} {2019})},\ \Eprint
  {http://arxiv.org/abs/1907.01496} {arXiv:1907.01496 [astro-ph.CO]}
  \BibitemShut {NoStop}%
\bibitem [{\citenamefont {Knox}\ and\ \citenamefont
  {Millea}(2019)}]{Knox:2019rjx}%
  \BibitemOpen
  \bibfield  {author} {\bibinfo {author} {\bibfnamefont {Lloyd}\ \bibnamefont
  {Knox}}\ and\ \bibinfo {author} {\bibfnamefont {Marius}\ \bibnamefont
  {Millea}},\ }\bibfield  {title} {\enquote {\bibinfo {title} {{The Hubble
  Hunter's Guide}},}\ }\href@noop {} {\  (\bibinfo {year} {2019})},\ \Eprint
  {http://arxiv.org/abs/1908.03663} {arXiv:1908.03663 [astro-ph.CO]}
  \BibitemShut {NoStop}%
\bibitem [{\citenamefont {Shi}\ and\ \citenamefont
  {Turner}(1998)}]{Shi:1997aa}%
  \BibitemOpen
  \bibfield  {author} {\bibinfo {author} {\bibfnamefont {Xiang-Dong}\
  \bibnamefont {Shi}}\ and\ \bibinfo {author} {\bibfnamefont {Michael~S.}\
  \bibnamefont {Turner}},\ }\bibfield  {title} {\enquote {\bibinfo {title}
  {{Expectations for the difference between local and global measurements of
  the Hubble constant}},}\ }\href {\doibase 10.1086/305169} {\bibfield
  {journal} {\bibinfo  {journal} {Astrophys. J.}\ }\textbf {\bibinfo {volume}
  {493}},\ \bibinfo {pages} {519} (\bibinfo {year} {1998})},\ \Eprint
  {http://arxiv.org/abs/astro-ph/9707101} {arXiv:astro-ph/9707101 [astro-ph]}
  \BibitemShut {NoStop}%
\bibitem [{\citenamefont {{Efstathiou}}(2014)}]{efs2014}%
  \BibitemOpen
  \bibfield  {author} {\bibinfo {author} {\bibfnamefont {George}\ \bibnamefont
  {{Efstathiou}}},\ }\bibfield  {title} {\enquote {\bibinfo {title} {{H$_{0}$
  revisited}},}\ }\href {\doibase 10.1093/mnras/stu278} {\bibfield  {journal}
  {\bibinfo  {journal} {\mnras}\ }\textbf {\bibinfo {volume} {440}},\ \bibinfo
  {pages} {1138--1152} (\bibinfo {year} {2014})},\ \Eprint
  {http://arxiv.org/abs/1311.3461} {arXiv:1311.3461 [astro-ph.CO]} \BibitemShut
  {NoStop}%
\bibitem [{\citenamefont {Aubourg}\ \emph {et~al.}(2015)\citenamefont {Aubourg}
  \emph {et~al.}}]{Aubourg:2014yra}%
  \BibitemOpen
  \bibfield  {author} {\bibinfo {author} {\bibfnamefont {Éric}\ \bibnamefont
  {Aubourg}} \emph {et~al.},\ }\bibfield  {title} {\enquote {\bibinfo {title}
  {{Cosmological implications of baryon acoustic oscillation measurements}},}\
  }\href {\doibase 10.1103/PhysRevD.92.123516} {\bibfield  {journal} {\bibinfo
  {journal} {Phys. Rev.}\ }\textbf {\bibinfo {volume} {D92}},\ \bibinfo {pages}
  {123516} (\bibinfo {year} {2015})},\ \Eprint {http://arxiv.org/abs/1411.1074}
  {arXiv:1411.1074 [astro-ph.CO]} \BibitemShut {NoStop}%
\bibitem [{\citenamefont {Odderskov}\ \emph {et~al.}(2014)\citenamefont
  {Odderskov}, \citenamefont {Hannestad},\ and\ \citenamefont
  {Haugbølle}}]{Odderskov:2014hqa}%
  \BibitemOpen
  \bibfield  {author} {\bibinfo {author} {\bibfnamefont {Io}~\bibnamefont
  {Odderskov}}, \bibinfo {author} {\bibfnamefont {Steen}\ \bibnamefont
  {Hannestad}}, \ and\ \bibinfo {author} {\bibfnamefont {Troels}\ \bibnamefont
  {Haugbølle}},\ }\bibfield  {title} {\enquote {\bibinfo {title} {{On the
  local variation of the Hubble constant}},}\ }\href {\doibase
  10.1088/1475-7516/2014/10/028} {\bibfield  {journal} {\bibinfo  {journal}
  {JCAP}\ }\textbf {\bibinfo {volume} {1410}},\ \bibinfo {pages} {028}
  (\bibinfo {year} {2014})},\ \Eprint {http://arxiv.org/abs/1407.7364}
  {arXiv:1407.7364 [astro-ph.CO]} \BibitemShut {NoStop}%
\bibitem [{\citenamefont {Macaulay}\ \emph {et~al.}(2019)\citenamefont
  {Macaulay} \emph {et~al.}}]{Macaulay:2018fxi}%
  \BibitemOpen
  \bibfield  {author} {\bibinfo {author} {\bibfnamefont {E.}~\bibnamefont
  {Macaulay}} \emph {et~al.} (\bibinfo {collaboration} {DES}),\ }\bibfield
  {title} {\enquote {\bibinfo {title} {{First Cosmological Results using Type
  Ia Supernovae from the Dark Energy Survey: Measurement of the Hubble
  Constant}},}\ }\href {\doibase 10.1093/mnras/stz978} {\bibfield  {journal}
  {\bibinfo  {journal} {Mon. Not. Roy. Astron. Soc.}\ }\textbf {\bibinfo
  {volume} {486}},\ \bibinfo {pages} {2184--2196} (\bibinfo {year} {2019})},\
  \Eprint {http://arxiv.org/abs/1811.02376} {arXiv:1811.02376 [astro-ph.CO]}
  \BibitemShut {NoStop}%
\bibitem [{\citenamefont {Aylor}\ \emph {et~al.}(2019)\citenamefont {Aylor},
  \citenamefont {Joy}, \citenamefont {Knox}, \citenamefont {Millea},
  \citenamefont {Raghunathan},\ and\ \citenamefont {Wu}}]{Aylor:2018drw}%
  \BibitemOpen
  \bibfield  {author} {\bibinfo {author} {\bibfnamefont {Kevin}\ \bibnamefont
  {Aylor}}, \bibinfo {author} {\bibfnamefont {MacKenzie}\ \bibnamefont {Joy}},
  \bibinfo {author} {\bibfnamefont {Lloyd}\ \bibnamefont {Knox}}, \bibinfo
  {author} {\bibfnamefont {Marius}\ \bibnamefont {Millea}}, \bibinfo {author}
  {\bibfnamefont {Srinivasan}\ \bibnamefont {Raghunathan}}, \ and\ \bibinfo
  {author} {\bibfnamefont {W.~L.~Kimmy}\ \bibnamefont {Wu}},\ }\bibfield
  {title} {\enquote {\bibinfo {title} {{Sounds Discordant: Classical Distance
  Ladder \& $\Lambda$CDM -based Determinations of the Cosmological Sound
  Horizon}},}\ }\href {\doibase 10.3847/1538-4357/ab0898} {\bibfield  {journal}
  {\bibinfo  {journal} {Astrophys. J.}\ }\textbf {\bibinfo {volume} {874}},\
  \bibinfo {pages} {4} (\bibinfo {year} {2019})},\ \Eprint
  {http://arxiv.org/abs/1811.00537} {arXiv:1811.00537 [astro-ph.CO]}
  \BibitemShut {NoStop}%
\bibitem [{\citenamefont {Taubenberger}\ \emph {et~al.}(2019)\citenamefont
  {Taubenberger}, \citenamefont {Suyu}, \citenamefont {Komatsu}, \citenamefont
  {Jee}, \citenamefont {Birrer}, \citenamefont {Bonvin}, \citenamefont
  {Courbin}, \citenamefont {Rusu}, \citenamefont {Shajib},\ and\ \citenamefont
  {Wong}}]{Taubenberger:2019qna}%
  \BibitemOpen
  \bibfield  {author} {\bibinfo {author} {\bibfnamefont {S.}~\bibnamefont
  {Taubenberger}}, \bibinfo {author} {\bibfnamefont {S.~H.}\ \bibnamefont
  {Suyu}}, \bibinfo {author} {\bibfnamefont {E.}~\bibnamefont {Komatsu}},
  \bibinfo {author} {\bibfnamefont {I.}~\bibnamefont {Jee}}, \bibinfo {author}
  {\bibfnamefont {S.}~\bibnamefont {Birrer}}, \bibinfo {author} {\bibfnamefont
  {V.}~\bibnamefont {Bonvin}}, \bibinfo {author} {\bibfnamefont
  {F.}~\bibnamefont {Courbin}}, \bibinfo {author} {\bibfnamefont {C.~E.}\
  \bibnamefont {Rusu}}, \bibinfo {author} {\bibfnamefont {A.~J.}\ \bibnamefont
  {Shajib}}, \ and\ \bibinfo {author} {\bibfnamefont {K.~C.}\ \bibnamefont
  {Wong}},\ }\bibfield  {title} {\enquote {\bibinfo {title} {{The Hubble
  Constant determined through an inverse distance ladder including quasar time
  delays and Type Ia supernovae}},}\ }\href {\doibase
  10.1051/0004-6361/201935980} {\bibfield  {journal} {\bibinfo  {journal}
  {Astron. Astrophys.}\ }\textbf {\bibinfo {volume} {628}},\ \bibinfo {pages}
  {L7} (\bibinfo {year} {2019})},\ \Eprint {http://arxiv.org/abs/1905.12496}
  {arXiv:1905.12496 [astro-ph.CO]} \BibitemShut {NoStop}%
\bibitem [{\citenamefont {Kenworthy}\ \emph {et~al.}(2019)\citenamefont
  {Kenworthy}, \citenamefont {Scolnic},\ and\ \citenamefont
  {Riess}}]{Kenworthy:2019qwq}%
  \BibitemOpen
  \bibfield  {author} {\bibinfo {author} {\bibfnamefont {W.~D'Arcy}\
  \bibnamefont {Kenworthy}}, \bibinfo {author} {\bibfnamefont {Dan}\
  \bibnamefont {Scolnic}}, \ and\ \bibinfo {author} {\bibfnamefont {Adam}\
  \bibnamefont {Riess}},\ }\bibfield  {title} {\enquote {\bibinfo {title} {{The
  Local Perspective on the Hubble Tension: Local Structure Does Not Impact
  Measurement of the Hubble Constant}},}\ }\href {\doibase
  10.3847/1538-4357/ab0ebf} {\bibfield  {journal} {\bibinfo  {journal}
  {Astrophys. J.}\ }\textbf {\bibinfo {volume} {875}},\ \bibinfo {pages} {145}
  (\bibinfo {year} {2019})},\ \Eprint {http://arxiv.org/abs/1901.08681}
  {arXiv:1901.08681 [astro-ph.CO]} \BibitemShut {NoStop}%
\bibitem [{\citenamefont {Rameez}\ and\ \citenamefont
  {Sarkar}(2019)}]{Rameez:2019wdt}%
  \BibitemOpen
  \bibfield  {author} {\bibinfo {author} {\bibfnamefont {Mohamed}\ \bibnamefont
  {Rameez}}\ and\ \bibinfo {author} {\bibfnamefont {Subir}\ \bibnamefont
  {Sarkar}},\ }\bibfield  {title} {\enquote {\bibinfo {title} {{Is there really
  a `Hubble tension'?}}}\ }\href@noop {} {\  (\bibinfo {year} {2019})},\
  \Eprint {http://arxiv.org/abs/1911.06456} {arXiv:1911.06456 [astro-ph.CO]}
  \BibitemShut {NoStop}%
\bibitem [{\citenamefont {{Sunyaev}}\ and\ \citenamefont
  {{Zeldovich}}(1970)}]{sz1970c}%
  \BibitemOpen
  \bibfield  {author} {\bibinfo {author} {\bibfnamefont {R.~A.}\ \bibnamefont
  {{Sunyaev}}}\ and\ \bibinfo {author} {\bibfnamefont {Y.~B.}\ \bibnamefont
  {{Zeldovich}}},\ }\bibfield  {title} {\enquote {\bibinfo {title}
  {{Small-Scale Fluctuations of Relic Radiation}},}\ }\href {\doibase
  10.1007/BF00653471} {\bibfield  {journal} {\bibinfo  {journal} {\apss}\
  }\textbf {\bibinfo {volume} {7}},\ \bibinfo {pages} {3--19} (\bibinfo {year}
  {1970})}\BibitemShut {NoStop}%
\bibitem [{\citenamefont {{Peebles}}\ and\ \citenamefont
  {{Yu}}(1970)}]{Peebles1970}%
  \BibitemOpen
  \bibfield  {author} {\bibinfo {author} {\bibfnamefont {P.~J.~E.}\
  \bibnamefont {{Peebles}}}\ and\ \bibinfo {author} {\bibfnamefont {J.~T.}\
  \bibnamefont {{Yu}}},\ }\bibfield  {title} {\enquote {\bibinfo {title}
  {{Primeval Adiabatic Perturbation in an Expanding Universe}},}\ }\href
  {\doibase 10.1086/150713} {\bibfield  {journal} {\bibinfo  {journal} {\apj}\
  }\textbf {\bibinfo {volume} {162}},\ \bibinfo {pages} {815} (\bibinfo {year}
  {1970})}\BibitemShut {NoStop}%
\bibitem [{\citenamefont {Bashinsky}\ and\ \citenamefont
  {Seljak}(2004)}]{Bashinsky:2003tk}%
  \BibitemOpen
  \bibfield  {author} {\bibinfo {author} {\bibfnamefont {Sergei}\ \bibnamefont
  {Bashinsky}}\ and\ \bibinfo {author} {\bibfnamefont {Uros}\ \bibnamefont
  {Seljak}},\ }\bibfield  {title} {\enquote {\bibinfo {title} {{Neutrino
  perturbations in CMB anisotropy and matter clustering}},}\ }\href {\doibase
  10.1103/PhysRevD.69.083002} {\bibfield  {journal} {\bibinfo  {journal} {Phys.
  Rev.}\ }\textbf {\bibinfo {volume} {D69}},\ \bibinfo {pages} {083002}
  (\bibinfo {year} {2004})},\ \Eprint {http://arxiv.org/abs/astro-ph/0310198}
  {arXiv:astro-ph/0310198 [astro-ph]} \BibitemShut {NoStop}%
\bibitem [{\citenamefont {Aghanim}\ \emph {et~al.}(2016)\citenamefont {Aghanim}
  \emph {et~al.}}]{Aghanim:2015xee}%
  \BibitemOpen
  \bibfield  {author} {\bibinfo {author} {\bibfnamefont {N.}~\bibnamefont
  {Aghanim}} \emph {et~al.} (\bibinfo {collaboration} {Planck}),\ }\bibfield
  {title} {\enquote {\bibinfo {title} {{Planck 2015 results. XI. CMB power
  spectra, likelihoods, and robustness of parameters}},}\ }\href {\doibase
  10.1051/0004-6361/201526926} {\bibfield  {journal} {\bibinfo  {journal}
  {Astron. Astrophys.}\ }\textbf {\bibinfo {volume} {594}},\ \bibinfo {pages}
  {A11} (\bibinfo {year} {2016})},\ \Eprint {http://arxiv.org/abs/1507.02704}
  {arXiv:1507.02704 [astro-ph.CO]} \BibitemShut {NoStop}%
\bibitem [{\citenamefont {Bell}\ \emph {et~al.}(2006)\citenamefont {Bell},
  \citenamefont {Pierpaoli},\ and\ \citenamefont {Sigurdson}}]{Bell:2005dr}%
  \BibitemOpen
  \bibfield  {author} {\bibinfo {author} {\bibfnamefont {Nicole~F.}\
  \bibnamefont {Bell}}, \bibinfo {author} {\bibfnamefont {Elena}\ \bibnamefont
  {Pierpaoli}}, \ and\ \bibinfo {author} {\bibfnamefont {Kris}\ \bibnamefont
  {Sigurdson}},\ }\bibfield  {title} {\enquote {\bibinfo {title} {{Cosmological
  signatures of interacting neutrinos}},}\ }\href {\doibase
  10.1103/PhysRevD.73.063523} {\bibfield  {journal} {\bibinfo  {journal} {Phys.
  Rev.}\ }\textbf {\bibinfo {volume} {D73}},\ \bibinfo {pages} {063523}
  (\bibinfo {year} {2006})},\ \Eprint {http://arxiv.org/abs/astro-ph/0511410}
  {arXiv:astro-ph/0511410 [astro-ph]} \BibitemShut {NoStop}%
\bibitem [{\citenamefont {Mangano}\ \emph {et~al.}(2006)\citenamefont
  {Mangano}, \citenamefont {Melchiorri}, \citenamefont {Serra}, \citenamefont
  {Cooray},\ and\ \citenamefont {Kamionkowski}}]{Mangano:2006mp}%
  \BibitemOpen
  \bibfield  {author} {\bibinfo {author} {\bibfnamefont {Gianpiero}\
  \bibnamefont {Mangano}}, \bibinfo {author} {\bibfnamefont {Alessandro}\
  \bibnamefont {Melchiorri}}, \bibinfo {author} {\bibfnamefont {Paolo}\
  \bibnamefont {Serra}}, \bibinfo {author} {\bibfnamefont {Asantha}\
  \bibnamefont {Cooray}}, \ and\ \bibinfo {author} {\bibfnamefont {Marc}\
  \bibnamefont {Kamionkowski}},\ }\bibfield  {title} {\enquote {\bibinfo
  {title} {{Cosmological bounds on dark matter-neutrino interactions}},}\
  }\href {\doibase 10.1103/PhysRevD.74.043517} {\bibfield  {journal} {\bibinfo
  {journal} {Phys. Rev.}\ }\textbf {\bibinfo {volume} {D74}},\ \bibinfo {pages}
  {043517} (\bibinfo {year} {2006})},\ \Eprint
  {http://arxiv.org/abs/astro-ph/0606190} {arXiv:astro-ph/0606190 [astro-ph]}
  \BibitemShut {NoStop}%
\bibitem [{\citenamefont {Serra}\ \emph {et~al.}(2010)\citenamefont {Serra},
  \citenamefont {Zalamea}, \citenamefont {Cooray}, \citenamefont {Mangano},\
  and\ \citenamefont {Melchiorri}}]{Serra:2009uu}%
  \BibitemOpen
  \bibfield  {author} {\bibinfo {author} {\bibfnamefont {Paolo}\ \bibnamefont
  {Serra}}, \bibinfo {author} {\bibfnamefont {Federico}\ \bibnamefont
  {Zalamea}}, \bibinfo {author} {\bibfnamefont {Asantha}\ \bibnamefont
  {Cooray}}, \bibinfo {author} {\bibfnamefont {Gianpiero}\ \bibnamefont
  {Mangano}}, \ and\ \bibinfo {author} {\bibfnamefont {Alessandro}\
  \bibnamefont {Melchiorri}},\ }\bibfield  {title} {\enquote {\bibinfo {title}
  {{Constraints on neutrino -- dark matter interactions from cosmic microwave
  background and large scale structure data}},}\ }\href {\doibase
  10.1103/PhysRevD.81.043507} {\bibfield  {journal} {\bibinfo  {journal} {Phys.
  Rev.}\ }\textbf {\bibinfo {volume} {D81}},\ \bibinfo {pages} {043507}
  (\bibinfo {year} {2010})},\ \Eprint {http://arxiv.org/abs/0911.4411}
  {arXiv:0911.4411 [astro-ph.CO]} \BibitemShut {NoStop}%
\bibitem [{\citenamefont {Cyr-Racine}\ and\ \citenamefont
  {Sigurdson}(2014)}]{Cyr-Racine:2013jua}%
  \BibitemOpen
  \bibfield  {author} {\bibinfo {author} {\bibfnamefont {Francis-Yan}\
  \bibnamefont {Cyr-Racine}}\ and\ \bibinfo {author} {\bibfnamefont {Kris}\
  \bibnamefont {Sigurdson}},\ }\bibfield  {title} {\enquote {\bibinfo {title}
  {{Limits on Neutrino-Neutrino Scattering in the Early Universe}},}\ }\href
  {\doibase 10.1103/PhysRevD.90.123533} {\bibfield  {journal} {\bibinfo
  {journal} {Phys. Rev.}\ }\textbf {\bibinfo {volume} {D90}},\ \bibinfo {pages}
  {123533} (\bibinfo {year} {2014})},\ \Eprint {http://arxiv.org/abs/1306.1536}
  {arXiv:1306.1536 [astro-ph.CO]} \BibitemShut {NoStop}%
\bibitem [{\citenamefont {Archidiacono}\ and\ \citenamefont
  {Hannestad}(2014)}]{Archidiacono:2013dua}%
  \BibitemOpen
  \bibfield  {author} {\bibinfo {author} {\bibfnamefont {Maria}\ \bibnamefont
  {Archidiacono}}\ and\ \bibinfo {author} {\bibfnamefont {Steen}\ \bibnamefont
  {Hannestad}},\ }\bibfield  {title} {\enquote {\bibinfo {title} {{Updated
  constraints on non-standard neutrino interactions from Planck}},}\ }\href
  {\doibase 10.1088/1475-7516/2014/07/046} {\bibfield  {journal} {\bibinfo
  {journal} {JCAP}\ }\textbf {\bibinfo {volume} {1407}},\ \bibinfo {pages}
  {046} (\bibinfo {year} {2014})},\ \Eprint {http://arxiv.org/abs/1311.3873}
  {arXiv:1311.3873 [astro-ph.CO]} \BibitemShut {NoStop}%
\bibitem [{\citenamefont {Wilkinson}\ \emph {et~al.}(2014)\citenamefont
  {Wilkinson}, \citenamefont {Boehm},\ and\ \citenamefont
  {Lesgourgues}}]{Wilkinson:2014ksa}%
  \BibitemOpen
  \bibfield  {author} {\bibinfo {author} {\bibfnamefont {Ryan~J.}\ \bibnamefont
  {Wilkinson}}, \bibinfo {author} {\bibfnamefont {Celine}\ \bibnamefont
  {Boehm}}, \ and\ \bibinfo {author} {\bibfnamefont {Julien}\ \bibnamefont
  {Lesgourgues}},\ }\bibfield  {title} {\enquote {\bibinfo {title}
  {{Constraining Dark Matter-Neutrino Interactions using the CMB and
  Large-Scale Structure}},}\ }\href {\doibase 10.1088/1475-7516/2014/05/011}
  {\bibfield  {journal} {\bibinfo  {journal} {JCAP}\ }\textbf {\bibinfo
  {volume} {1405}},\ \bibinfo {pages} {011} (\bibinfo {year} {2014})},\ \Eprint
  {http://arxiv.org/abs/1401.7597} {arXiv:1401.7597 [astro-ph.CO]} \BibitemShut
  {NoStop}%
\bibitem [{\citenamefont {Boehm}\ \emph {et~al.}(2014)\citenamefont {Boehm},
  \citenamefont {Schewtschenko}, \citenamefont {Wilkinson}, \citenamefont
  {Baugh},\ and\ \citenamefont {Pascoli}}]{Boehm:2014vja}%
  \BibitemOpen
  \bibfield  {author} {\bibinfo {author} {\bibfnamefont {C.}~\bibnamefont
  {Boehm}}, \bibinfo {author} {\bibfnamefont {J.~A.}\ \bibnamefont
  {Schewtschenko}}, \bibinfo {author} {\bibfnamefont {R.~J.}\ \bibnamefont
  {Wilkinson}}, \bibinfo {author} {\bibfnamefont {C.~M.}\ \bibnamefont
  {Baugh}}, \ and\ \bibinfo {author} {\bibfnamefont {S.}~\bibnamefont
  {Pascoli}},\ }\bibfield  {title} {\enquote {\bibinfo {title} {{Using the
  Milky Way satellites to study interactions between cold dark matter and
  radiation}},}\ }\href {\doibase 10.1093/mnrasl/slu115} {\bibfield  {journal}
  {\bibinfo  {journal} {Mon. Not. Roy. Astron. Soc.}\ }\textbf {\bibinfo
  {volume} {445}},\ \bibinfo {pages} {L31--L35} (\bibinfo {year} {2014})},\
  \Eprint {http://arxiv.org/abs/1404.7012} {arXiv:1404.7012 [astro-ph.CO]}
  \BibitemShut {NoStop}%
\bibitem [{\citenamefont {Bertoni}\ \emph {et~al.}(2015)\citenamefont
  {Bertoni}, \citenamefont {Ipek}, \citenamefont {McKeen},\ and\ \citenamefont
  {Nelson}}]{Bertoni:2014mva}%
  \BibitemOpen
  \bibfield  {author} {\bibinfo {author} {\bibfnamefont {Bridget}\ \bibnamefont
  {Bertoni}}, \bibinfo {author} {\bibfnamefont {Seyda}\ \bibnamefont {Ipek}},
  \bibinfo {author} {\bibfnamefont {David}\ \bibnamefont {McKeen}}, \ and\
  \bibinfo {author} {\bibfnamefont {Ann~E.}\ \bibnamefont {Nelson}},\
  }\bibfield  {title} {\enquote {\bibinfo {title} {{Constraints and
  consequences of reducing small scale structure via large dark matter-neutrino
  interactions}},}\ }\href {\doibase 10.1007/JHEP04(2015)170} {\bibfield
  {journal} {\bibinfo  {journal} {JHEP}\ }\textbf {\bibinfo {volume} {04}},\
  \bibinfo {pages} {170} (\bibinfo {year} {2015})},\ \Eprint
  {http://arxiv.org/abs/1412.3113} {arXiv:1412.3113 [hep-ph]} \BibitemShut
  {NoStop}%
\bibitem [{\citenamefont {Escudero}\ \emph {et~al.}(2015)\citenamefont
  {Escudero}, \citenamefont {Mena}, \citenamefont {Vincent}, \citenamefont
  {Wilkinson},\ and\ \citenamefont {Bœhm}}]{Escudero:2015yka}%
  \BibitemOpen
  \bibfield  {author} {\bibinfo {author} {\bibfnamefont {Miguel}\ \bibnamefont
  {Escudero}}, \bibinfo {author} {\bibfnamefont {Olga}\ \bibnamefont {Mena}},
  \bibinfo {author} {\bibfnamefont {Aaron~C.}\ \bibnamefont {Vincent}},
  \bibinfo {author} {\bibfnamefont {Ryan~J.}\ \bibnamefont {Wilkinson}}, \ and\
  \bibinfo {author} {\bibfnamefont {Céline}\ \bibnamefont {Bœhm}},\
  }\bibfield  {title} {\enquote {\bibinfo {title} {{Exploring dark matter
  microphysics with galaxy surveys}},}\ }\href {\doibase
  10.1088/1475-7516/2015/9/034, 10.1088/1475-7516/2015/09/034} {\bibfield
  {journal} {\bibinfo  {journal} {JCAP}\ }\textbf {\bibinfo {volume} {1509}},\
  \bibinfo {pages} {034} (\bibinfo {year} {2015})},\ \Eprint
  {http://arxiv.org/abs/1505.06735} {arXiv:1505.06735 [astro-ph.CO]}
  \BibitemShut {NoStop}%
\bibitem [{\citenamefont {{Forastieri}}\ \emph {et~al.}(2015)\citenamefont
  {{Forastieri}}, \citenamefont {{Lattanzi}},\ and\ \citenamefont
  {{Natoli}}}]{2015JCAP...07..014F}%
  \BibitemOpen
  \bibfield  {author} {\bibinfo {author} {\bibfnamefont {Francesco}\
  \bibnamefont {{Forastieri}}}, \bibinfo {author} {\bibfnamefont
  {Massimiliano}\ \bibnamefont {{Lattanzi}}}, \ and\ \bibinfo {author}
  {\bibfnamefont {Paolo}\ \bibnamefont {{Natoli}}},\ }\bibfield  {title}
  {\enquote {\bibinfo {title} {{Constraints on secret neutrino interactions
  after Planck}},}\ }\href {\doibase 10.1088/1475-7516/2015/07/014} {\bibfield
  {journal} {\bibinfo  {journal} {Journal of Cosmology and Astro-Particle
  Physics}\ }\textbf {\bibinfo {volume} {2015}},\ \bibinfo {eid} {014}
  (\bibinfo {year} {2015})},\ \Eprint {http://arxiv.org/abs/1504.04999}
  {arXiv:1504.04999 [astro-ph.CO]} \BibitemShut {NoStop}%
\bibitem [{\citenamefont {Oldengott}\ \emph {et~al.}(2017)\citenamefont
  {Oldengott}, \citenamefont {Tram}, \citenamefont {Rampf},\ and\ \citenamefont
  {Wong}}]{Oldengott:2017fhy}%
  \BibitemOpen
  \bibfield  {author} {\bibinfo {author} {\bibfnamefont {Isabel~M.}\
  \bibnamefont {Oldengott}}, \bibinfo {author} {\bibfnamefont {Thomas}\
  \bibnamefont {Tram}}, \bibinfo {author} {\bibfnamefont {Cornelius}\
  \bibnamefont {Rampf}}, \ and\ \bibinfo {author} {\bibfnamefont {Yvonne
  Y.~Y.}\ \bibnamefont {Wong}},\ }\bibfield  {title} {\enquote {\bibinfo
  {title} {{Interacting neutrinos in cosmology: exact description and
  constraints}},}\ }\href {\doibase 10.1088/1475-7516/2017/11/027} {\bibfield
  {journal} {\bibinfo  {journal} {JCAP}\ }\textbf {\bibinfo {volume} {1711}},\
  \bibinfo {pages} {027} (\bibinfo {year} {2017})},\ \Eprint
  {http://arxiv.org/abs/1706.02123} {arXiv:1706.02123 [astro-ph.CO]}
  \BibitemShut {NoStop}%
\bibitem [{\citenamefont {Kreisch}\ \emph {et~al.}(2019)\citenamefont
  {Kreisch}, \citenamefont {Cyr-Racine},\ and\ \citenamefont
  {Doré}}]{Kreisch:2019yzn}%
  \BibitemOpen
  \bibfield  {author} {\bibinfo {author} {\bibfnamefont {Christina~D.}\
  \bibnamefont {Kreisch}}, \bibinfo {author} {\bibfnamefont {Francis-Yan}\
  \bibnamefont {Cyr-Racine}}, \ and\ \bibinfo {author} {\bibfnamefont
  {Olivier}\ \bibnamefont {Doré}},\ }\bibfield  {title} {\enquote {\bibinfo
  {title} {{The Neutrino Puzzle: Anomalies, Interactions, and Cosmological
  Tensions}},}\ }\href@noop {} {\  (\bibinfo {year} {2019})},\ \Eprint
  {http://arxiv.org/abs/1902.00534} {arXiv:1902.00534 [astro-ph.CO]}
  \BibitemShut {NoStop}%
\bibitem [{\citenamefont {{Forastieri}}\ \emph {et~al.}(2019)\citenamefont
  {{Forastieri}}, \citenamefont {{Lattanzi}},\ and\ \citenamefont
  {{Natoli}}}]{2019arXiv190407810F}%
  \BibitemOpen
  \bibfield  {author} {\bibinfo {author} {\bibfnamefont {F.}~\bibnamefont
  {{Forastieri}}}, \bibinfo {author} {\bibfnamefont {M.}~\bibnamefont
  {{Lattanzi}}}, \ and\ \bibinfo {author} {\bibfnamefont {P.}~\bibnamefont
  {{Natoli}}},\ }\bibfield  {title} {\enquote {\bibinfo {title} {{Cosmological
  constraints on neutrino self-interactions with a light mediator}},}\
  }\href@noop {} {\bibfield  {journal} {\bibinfo  {journal} {arXiv e-prints}\
  ,\ \bibinfo {eid} {arXiv:1904.07810}} (\bibinfo {year} {2019})},\ \Eprint
  {http://arxiv.org/abs/1904.07810} {arXiv:1904.07810 [astro-ph.CO]}
  \BibitemShut {NoStop}%
\bibitem [{\citenamefont {Ghosh}\ \emph {et~al.}(2018)\citenamefont {Ghosh},
  \citenamefont {Khatri},\ and\ \citenamefont {Roy}}]{Ghosh:2017jdy}%
  \BibitemOpen
  \bibfield  {author} {\bibinfo {author} {\bibfnamefont {Subhajit}\
  \bibnamefont {Ghosh}}, \bibinfo {author} {\bibfnamefont {Rishi}\ \bibnamefont
  {Khatri}}, \ and\ \bibinfo {author} {\bibfnamefont {Tuhin~S.}\ \bibnamefont
  {Roy}},\ }\bibfield  {title} {\enquote {\bibinfo {title} {{Dark neutrino
  interactions make gravitational waves blue}},}\ }\href {\doibase
  10.1103/PhysRevD.97.063529} {\bibfield  {journal} {\bibinfo  {journal} {Phys.
  Rev.}\ }\textbf {\bibinfo {volume} {D97}},\ \bibinfo {pages} {063529}
  (\bibinfo {year} {2018})},\ \Eprint {http://arxiv.org/abs/1711.09929}
  {arXiv:1711.09929 [astro-ph.CO]} \BibitemShut {NoStop}%
\bibitem [{\citenamefont {Follin}\ \emph {et~al.}(2015)\citenamefont {Follin},
  \citenamefont {Knox}, \citenamefont {Millea},\ and\ \citenamefont
  {Pan}}]{Follin:2015hya}%
  \BibitemOpen
  \bibfield  {author} {\bibinfo {author} {\bibfnamefont {Brent}\ \bibnamefont
  {Follin}}, \bibinfo {author} {\bibfnamefont {Lloyd}\ \bibnamefont {Knox}},
  \bibinfo {author} {\bibfnamefont {Marius}\ \bibnamefont {Millea}}, \ and\
  \bibinfo {author} {\bibfnamefont {Zhen}\ \bibnamefont {Pan}},\ }\bibfield
  {title} {\enquote {\bibinfo {title} {{First Detection of the Acoustic
  Oscillation Phase Shift Expected from the Cosmic Neutrino Background}},}\
  }\href {\doibase 10.1103/PhysRevLett.115.091301} {\bibfield  {journal}
  {\bibinfo  {journal} {Phys. Rev. Lett.}\ }\textbf {\bibinfo {volume} {115}},\
  \bibinfo {pages} {091301} (\bibinfo {year} {2015})},\ \Eprint
  {http://arxiv.org/abs/1503.07863} {arXiv:1503.07863 [astro-ph.CO]}
  \BibitemShut {NoStop}%
\bibitem [{\citenamefont {Baumann}\ \emph {et~al.}(2016)\citenamefont
  {Baumann}, \citenamefont {Green}, \citenamefont {Meyers},\ and\ \citenamefont
  {Wallisch}}]{Baumann:2015rya}%
  \BibitemOpen
  \bibfield  {author} {\bibinfo {author} {\bibfnamefont {Daniel}\ \bibnamefont
  {Baumann}}, \bibinfo {author} {\bibfnamefont {Daniel}\ \bibnamefont {Green}},
  \bibinfo {author} {\bibfnamefont {Joel}\ \bibnamefont {Meyers}}, \ and\
  \bibinfo {author} {\bibfnamefont {Benjamin}\ \bibnamefont {Wallisch}},\
  }\bibfield  {title} {\enquote {\bibinfo {title} {{Phases of New Physics in
  the CMB}},}\ }\href {\doibase 10.1088/1475-7516/2016/01/007} {\bibfield
  {journal} {\bibinfo  {journal} {JCAP}\ }\textbf {\bibinfo {volume} {1601}},\
  \bibinfo {pages} {007} (\bibinfo {year} {2016})},\ \Eprint
  {http://arxiv.org/abs/1508.06342} {arXiv:1508.06342 [astro-ph.CO]}
  \BibitemShut {NoStop}%
\bibitem [{\citenamefont {Pan}\ \emph {et~al.}(2016)\citenamefont {Pan},
  \citenamefont {Knox}, \citenamefont {Mulroe},\ and\ \citenamefont
  {Narimani}}]{Pan:2016zla}%
  \BibitemOpen
  \bibfield  {author} {\bibinfo {author} {\bibfnamefont {Zhen}\ \bibnamefont
  {Pan}}, \bibinfo {author} {\bibfnamefont {Lloyd}\ \bibnamefont {Knox}},
  \bibinfo {author} {\bibfnamefont {Brigid}\ \bibnamefont {Mulroe}}, \ and\
  \bibinfo {author} {\bibfnamefont {Ali}\ \bibnamefont {Narimani}},\ }\bibfield
   {title} {\enquote {\bibinfo {title} {{Cosmic Microwave Background Acoustic
  Peak Locations}},}\ }\href {\doibase 10.1093/mnras/stw833} {\bibfield
  {journal} {\bibinfo  {journal} {Mon. Not. Roy. Astron. Soc.}\ }\textbf
  {\bibinfo {volume} {459}},\ \bibinfo {pages} {2513--2524} (\bibinfo {year}
  {2016})},\ \Eprint {http://arxiv.org/abs/1603.03091} {arXiv:1603.03091
  [astro-ph.CO]} \BibitemShut {NoStop}%
\bibitem [{\citenamefont {Baumann}\ \emph {et~al.}(2017)\citenamefont
  {Baumann}, \citenamefont {Green},\ and\ \citenamefont
  {Zaldarriaga}}]{Baumann:2017lmt}%
  \BibitemOpen
  \bibfield  {author} {\bibinfo {author} {\bibfnamefont {Daniel}\ \bibnamefont
  {Baumann}}, \bibinfo {author} {\bibfnamefont {Daniel}\ \bibnamefont {Green}},
  \ and\ \bibinfo {author} {\bibfnamefont {Matias}\ \bibnamefont
  {Zaldarriaga}},\ }\bibfield  {title} {\enquote {\bibinfo {title} {{Phases of
  New Physics in the BAO Spectrum}},}\ }\href {\doibase
  10.1088/1475-7516/2017/11/007} {\bibfield  {journal} {\bibinfo  {journal}
  {JCAP}\ }\textbf {\bibinfo {volume} {1711}},\ \bibinfo {pages} {007}
  (\bibinfo {year} {2017})},\ \Eprint {http://arxiv.org/abs/1703.00894}
  {arXiv:1703.00894 [astro-ph.CO]} \BibitemShut {NoStop}%
\bibitem [{\citenamefont {Baumann}\ \emph {et~al.}(2018)\citenamefont
  {Baumann}, \citenamefont {Green},\ and\ \citenamefont
  {Wallisch}}]{Baumann:2017gkg}%
  \BibitemOpen
  \bibfield  {author} {\bibinfo {author} {\bibfnamefont {Daniel}\ \bibnamefont
  {Baumann}}, \bibinfo {author} {\bibfnamefont {Daniel}\ \bibnamefont {Green}},
  \ and\ \bibinfo {author} {\bibfnamefont {Benjamin}\ \bibnamefont
  {Wallisch}},\ }\bibfield  {title} {\enquote {\bibinfo {title} {{Searching for
  light relics with large-scale structure}},}\ }\href {\doibase
  10.1088/1475-7516/2018/08/029} {\bibfield  {journal} {\bibinfo  {journal}
  {JCAP}\ }\textbf {\bibinfo {volume} {1808}},\ \bibinfo {pages} {029}
  (\bibinfo {year} {2018})},\ \Eprint {http://arxiv.org/abs/1712.08067}
  {arXiv:1712.08067 [astro-ph.CO]} \BibitemShut {NoStop}%
\bibitem [{\citenamefont {Choi}\ \emph {et~al.}(2018)\citenamefont {Choi},
  \citenamefont {Chiang},\ and\ \citenamefont {LoVerde}}]{Choi:2018gho}%
  \BibitemOpen
  \bibfield  {author} {\bibinfo {author} {\bibfnamefont {Gongjun}\ \bibnamefont
  {Choi}}, \bibinfo {author} {\bibfnamefont {Chi-Ting}\ \bibnamefont {Chiang}},
  \ and\ \bibinfo {author} {\bibfnamefont {Marilena}\ \bibnamefont {LoVerde}},\
  }\bibfield  {title} {\enquote {\bibinfo {title} {{Probing Decoupling in Dark
  Sectors with the Cosmic Microwave Background}},}\ }\href {\doibase
  10.1088/1475-7516/2018/06/044} {\bibfield  {journal} {\bibinfo  {journal}
  {JCAP}\ }\textbf {\bibinfo {volume} {1806}},\ \bibinfo {pages} {044}
  (\bibinfo {year} {2018})},\ \Eprint {http://arxiv.org/abs/1804.10180}
  {arXiv:1804.10180 [astro-ph.CO]} \BibitemShut {NoStop}%
\bibitem [{\citenamefont {Baumann}\ \emph {et~al.}(2019)\citenamefont
  {Baumann}, \citenamefont {Beutler}, \citenamefont {Flauger}, \citenamefont
  {Green}, \citenamefont {Vargas-Magaña}, \citenamefont {Slosar},
  \citenamefont {Wallisch},\ and\ \citenamefont {Yèche}}]{Baumann:2019tdh}%
  \BibitemOpen
  \bibfield  {author} {\bibinfo {author} {\bibfnamefont {Daniel}\ \bibnamefont
  {Baumann}}, \bibinfo {author} {\bibfnamefont {Florian}\ \bibnamefont
  {Beutler}}, \bibinfo {author} {\bibfnamefont {Raphael}\ \bibnamefont
  {Flauger}}, \bibinfo {author} {\bibfnamefont {Daniel}\ \bibnamefont {Green}},
  \bibinfo {author} {\bibfnamefont {Mariana}\ \bibnamefont {Vargas-Magaña}},
  \bibinfo {author} {\bibfnamefont {Anže}\ \bibnamefont {Slosar}}, \bibinfo
  {author} {\bibfnamefont {Benjamin}\ \bibnamefont {Wallisch}}, \ and\ \bibinfo
  {author} {\bibfnamefont {Christophe}\ \bibnamefont {Yèche}},\ }\bibfield
  {title} {\enquote {\bibinfo {title} {{First constraint on the
  neutrino-induced phase shift in the spectrum of baryon acoustic
  oscillations}},}\ }\href {\doibase 10.1038/s41567-019-0435-6} {\bibfield
  {journal} {\bibinfo  {journal} {Nature Phys.}\ }\textbf {\bibinfo {volume}
  {15}},\ \bibinfo {pages} {465--469} (\bibinfo {year} {2019})},\ \Eprint
  {http://arxiv.org/abs/1803.10741} {arXiv:1803.10741 [astro-ph.CO]}
  \BibitemShut {NoStop}%
\bibitem [{\citenamefont {Primulando}\ and\ \citenamefont
  {Uttayarat}(2018)}]{Primulando:2017kxf}%
  \BibitemOpen
  \bibfield  {author} {\bibinfo {author} {\bibfnamefont {Reinard}\ \bibnamefont
  {Primulando}}\ and\ \bibinfo {author} {\bibfnamefont {Patipan}\ \bibnamefont
  {Uttayarat}},\ }\bibfield  {title} {\enquote {\bibinfo {title} {{Dark
  Matter-Neutrino Interaction in Light of Collider and Neutrino Telescope
  Data}},}\ }\href {\doibase 10.1007/JHEP06(2018)026} {\bibfield  {journal}
  {\bibinfo  {journal} {JHEP}\ }\textbf {\bibinfo {volume} {06}},\ \bibinfo
  {pages} {026} (\bibinfo {year} {2018})},\ \Eprint
  {http://arxiv.org/abs/1710.08567} {arXiv:1710.08567 [hep-ph]} \BibitemShut
  {NoStop}%
\bibitem [{\citenamefont {Ma}\ and\ \citenamefont
  {Bertschinger}(1995)}]{Ma:1995ey}%
  \BibitemOpen
  \bibfield  {author} {\bibinfo {author} {\bibfnamefont {Chung-Pei}\
  \bibnamefont {Ma}}\ and\ \bibinfo {author} {\bibfnamefont {Edmund}\
  \bibnamefont {Bertschinger}},\ }\bibfield  {title} {\enquote {\bibinfo
  {title} {{Cosmological perturbation theory in the synchronous and conformal
  Newtonian gauges}},}\ }\href {\doibase 10.1086/176550} {\bibfield  {journal}
  {\bibinfo  {journal} {Astrophys. J.}\ }\textbf {\bibinfo {volume} {455}},\
  \bibinfo {pages} {7--25} (\bibinfo {year} {1995})},\ \Eprint
  {http://arxiv.org/abs/astro-ph/9506072} {arXiv:astro-ph/9506072 [astro-ph]}
  \BibitemShut {NoStop}%
\bibitem [{\citenamefont {{Blas}}\ \emph {et~al.}(2011)\citenamefont {{Blas}},
  \citenamefont {{Lesgourgues}},\ and\ \citenamefont
  {{Tram}}}]{2011JCAP...07..034B}%
  \BibitemOpen
  \bibfield  {author} {\bibinfo {author} {\bibfnamefont {Diego}\ \bibnamefont
  {{Blas}}}, \bibinfo {author} {\bibfnamefont {Julien}\ \bibnamefont
  {{Lesgourgues}}}, \ and\ \bibinfo {author} {\bibfnamefont {Thomas}\
  \bibnamefont {{Tram}}},\ }\bibfield  {title} {\enquote {\bibinfo {title}
  {{The Cosmic Linear Anisotropy Solving System (CLASS). Part II: Approximation
  schemes}},}\ }\href {\doibase 10.1088/1475-7516/2011/07/034} {\bibfield
  {journal} {\bibinfo  {journal} {Journal of Cosmology and Astro-Particle
  Physics}\ }\textbf {\bibinfo {volume} {2011}},\ \bibinfo {eid} {034}
  (\bibinfo {year} {2011})},\ \Eprint {http://arxiv.org/abs/1104.2933}
  {arXiv:1104.2933 [astro-ph.CO]} \BibitemShut {NoStop}%
\bibitem [{\citenamefont {Audren}\ \emph {et~al.}(2013)\citenamefont {Audren},
  \citenamefont {Lesgourgues}, \citenamefont {Benabed},\ and\ \citenamefont
  {Prunet}}]{Audren:2012wb}%
  \BibitemOpen
  \bibfield  {author} {\bibinfo {author} {\bibfnamefont {Benjamin}\
  \bibnamefont {Audren}}, \bibinfo {author} {\bibfnamefont {Julien}\
  \bibnamefont {Lesgourgues}}, \bibinfo {author} {\bibfnamefont {Karim}\
  \bibnamefont {Benabed}}, \ and\ \bibinfo {author} {\bibfnamefont {Simon}\
  \bibnamefont {Prunet}},\ }\bibfield  {title} {\enquote {\bibinfo {title}
  {{Conservative Constraints on Early Cosmology: an illustration of the Monte
  Python cosmological parameter inference code}},}\ }\href {\doibase
  10.1088/1475-7516/2013/02/001} {\bibfield  {journal} {\bibinfo  {journal}
  {JCAP}\ }\textbf {\bibinfo {volume} {1302}},\ \bibinfo {pages} {001}
  (\bibinfo {year} {2013})},\ \Eprint {http://arxiv.org/abs/1210.7183}
  {arXiv:1210.7183 [astro-ph.CO]} \BibitemShut {NoStop}%
\bibitem [{\citenamefont {Parkinson}\ \emph {et~al.}(2012)\citenamefont
  {Parkinson} \emph {et~al.}}]{PhysRevD.86.103518}%
  \BibitemOpen
  \bibfield  {author} {\bibinfo {author} {\bibfnamefont {David}\ \bibnamefont
  {Parkinson}} \emph {et~al.},\ }\bibfield  {title} {\enquote {\bibinfo {title}
  {{The WiggleZ Dark Energy Survey: Final data release and cosmological
  results}},}\ }\href {\doibase 10.1103/PhysRevD.86.103518} {\bibfield
  {journal} {\bibinfo  {journal} {Phys. Rev.}\ }\textbf {\bibinfo {volume}
  {D86}},\ \bibinfo {pages} {103518} (\bibinfo {year} {2012})},\ \Eprint
  {http://arxiv.org/abs/1210.2130} {arXiv:1210.2130 [astro-ph.CO]} \BibitemShut
  {NoStop}%
\bibitem [{\citenamefont {Smith}\ \emph {et~al.}(2003)\citenamefont {Smith},
  \citenamefont {Peacock}, \citenamefont {Jenkins}, \citenamefont {White},
  \citenamefont {Frenk}, \citenamefont {Pearce}, \citenamefont {Thomas},
  \citenamefont {Efstathiou},\ and\ \citenamefont {Couchmann}}]{Smith:2002dz}%
  \BibitemOpen
  \bibfield  {author} {\bibinfo {author} {\bibfnamefont {R.~E.}\ \bibnamefont
  {Smith}}, \bibinfo {author} {\bibfnamefont {J.~A.}\ \bibnamefont {Peacock}},
  \bibinfo {author} {\bibfnamefont {A.}~\bibnamefont {Jenkins}}, \bibinfo
  {author} {\bibfnamefont {S.~D.~M.}\ \bibnamefont {White}}, \bibinfo {author}
  {\bibfnamefont {C.~S.}\ \bibnamefont {Frenk}}, \bibinfo {author}
  {\bibfnamefont {F.~R.}\ \bibnamefont {Pearce}}, \bibinfo {author}
  {\bibfnamefont {P.~A.}\ \bibnamefont {Thomas}}, \bibinfo {author}
  {\bibfnamefont {G.}~\bibnamefont {Efstathiou}}, \ and\ \bibinfo {author}
  {\bibfnamefont {H.~M.~P.}\ \bibnamefont {Couchmann}} (\bibinfo
  {collaboration} {VIRGO Consortium}),\ }\bibfield  {title} {\enquote {\bibinfo
  {title} {{Stable clustering, the halo model and nonlinear cosmological power
  spectra}},}\ }\href {\doibase 10.1046/j.1365-8711.2003.06503.x} {\bibfield
  {journal} {\bibinfo  {journal} {Mon. Not. Roy. Astron. Soc.}\ }\textbf
  {\bibinfo {volume} {341}},\ \bibinfo {pages} {1311} (\bibinfo {year}
  {2003})},\ \Eprint {http://arxiv.org/abs/astro-ph/0207664}
  {arXiv:astro-ph/0207664 [astro-ph]} \BibitemShut {NoStop}%
\bibitem [{\citenamefont {Andre}\ \emph {et~al.}(2014)\citenamefont {Andre}
  \emph {et~al.}}]{Andre:2013nfa}%
  \BibitemOpen
  \bibfield  {author} {\bibinfo {author} {\bibfnamefont {Philippe}\
  \bibnamefont {Andre}} \emph {et~al.} (\bibinfo {collaboration} {PRISM}),\
  }\bibfield  {title} {\enquote {\bibinfo {title} {{PRISM (Polarized Radiation
  Imaging and Spectroscopy Mission): An Extended White Paper}},}\ }\href
  {\doibase 10.1088/1475-7516/2014/02/006} {\bibfield  {journal} {\bibinfo
  {journal} {JCAP}\ }\textbf {\bibinfo {volume} {1402}},\ \bibinfo {pages}
  {006} (\bibinfo {year} {2014})},\ \Eprint {http://arxiv.org/abs/1310.1554}
  {arXiv:1310.1554 [astro-ph.CO]} \BibitemShut {NoStop}%
\bibitem [{\citenamefont {Ade}\ \emph {et~al.}(2015)\citenamefont {Ade} \emph
  {et~al.}}]{Ade:2015fwj}%
  \BibitemOpen
  \bibfield  {author} {\bibinfo {author} {\bibfnamefont {P.~A.~R.}\
  \bibnamefont {Ade}} \emph {et~al.} (\bibinfo {collaboration} {BICEP2, Keck
  Array}),\ }\bibfield  {title} {\enquote {\bibinfo {title} {{BICEP2 / Keck
  Array V: Measurements of B-mode Polarization at Degree Angular Scales and 150
  GHz by the Keck Array}},}\ }\href {\doibase 10.1088/0004-637X/811/2/126}
  {\bibfield  {journal} {\bibinfo  {journal} {Astrophys. J.}\ }\textbf
  {\bibinfo {volume} {811}},\ \bibinfo {pages} {126} (\bibinfo {year}
  {2015})},\ \Eprint {http://arxiv.org/abs/1502.00643} {arXiv:1502.00643
  [astro-ph.CO]} \BibitemShut {NoStop}%
\bibitem [{\citenamefont {Ade}\ \emph {et~al.}(2018)\citenamefont {Ade} \emph
  {et~al.}}]{Ade:2018gkx}%
  \BibitemOpen
  \bibfield  {author} {\bibinfo {author} {\bibfnamefont {P.~A.~R.}\
  \bibnamefont {Ade}} \emph {et~al.} (\bibinfo {collaboration} {BICEP2, Keck
  Array}),\ }\bibfield  {title} {\enquote {\bibinfo {title} {{BICEP2 / Keck
  Array x: Constraints on Primordial Gravitational Waves using Planck, WMAP,
  and New BICEP2/Keck Observations through the 2015 Season}},}\ }\href
  {\doibase 10.1103/PhysRevLett.121.221301} {\bibfield  {journal} {\bibinfo
  {journal} {Phys. Rev. Lett.}\ }\textbf {\bibinfo {volume} {121}},\ \bibinfo
  {pages} {221301} (\bibinfo {year} {2018})},\ \Eprint
  {http://arxiv.org/abs/1810.05216} {arXiv:1810.05216 [astro-ph.CO]}
  \BibitemShut {NoStop}%
\bibitem [{\citenamefont {Finelli}\ \emph {et~al.}(2018)\citenamefont {Finelli}
  \emph {et~al.}}]{core}%
  \BibitemOpen
  \bibfield  {author} {\bibinfo {author} {\bibfnamefont {Fabio}\ \bibnamefont
  {Finelli}} \emph {et~al.} (\bibinfo {collaboration} {CORE}),\ }\bibfield
  {title} {\enquote {\bibinfo {title} {{Exploring cosmic origins with CORE:
  Inflation}},}\ }\href {\doibase 10.1088/1475-7516/2018/04/016} {\bibfield
  {journal} {\bibinfo  {journal} {JCAP}\ }\textbf {\bibinfo {volume} {1804}},\
  \bibinfo {pages} {016} (\bibinfo {year} {2018})},\ \Eprint
  {http://arxiv.org/abs/1612.08270} {arXiv:1612.08270 [astro-ph.CO]}
  \BibitemShut {NoStop}%
\bibitem [{\citenamefont {Alvarez}\ \emph {et~al.}(2019)\citenamefont {Alvarez}
  \emph {et~al.}}]{pico}%
  \BibitemOpen
  \bibfield  {author} {\bibinfo {author} {\bibfnamefont {Marcelo}\ \bibnamefont
  {Alvarez}} \emph {et~al.},\ }\bibfield  {title} {\enquote {\bibinfo {title}
  {{PICO: Probe of Inflation and Cosmic Origins}},}\ }\href@noop {} {\
  (\bibinfo {year} {2019})},\ \Eprint {http://arxiv.org/abs/1908.07495}
  {arXiv:1908.07495 [astro-ph.IM]} \BibitemShut {NoStop}%
\bibitem [{\citenamefont {Abazajian}\ \emph {et~al.}(2016)\citenamefont
  {Abazajian} \emph {et~al.}}]{cmbs4}%
  \BibitemOpen
  \bibfield  {author} {\bibinfo {author} {\bibfnamefont {Kevork~N.}\
  \bibnamefont {Abazajian}} \emph {et~al.} (\bibinfo {collaboration}
  {CMB-S4}),\ }\bibfield  {title} {\enquote {\bibinfo {title} {{CMB-S4 Science
  Book, First Edition}},}\ }\href@noop {} {\  (\bibinfo {year} {2016})},\
  \Eprint {http://arxiv.org/abs/1610.02743} {arXiv:1610.02743 [astro-ph.CO]}
  \BibitemShut {NoStop}%
\bibitem [{\citenamefont {Hazumi}\ \emph {et~al.}(2019)\citenamefont {Hazumi}
  \emph {et~al.}}]{Hazumi2019}%
  \BibitemOpen
  \bibfield  {author} {\bibinfo {author} {\bibfnamefont {M.}~\bibnamefont
  {Hazumi}} \emph {et~al.},\ }\bibfield  {title} {\enquote {\bibinfo {title}
  {{LiteBIRD: A Satellite for the Studies of B-Mode Polarization and Inflation
  from Cosmic Background Radiation Detection}},}\ }\href {\doibase
  10.1007/s10909-019-02150-5} {\bibfield  {journal} {\bibinfo  {journal} {J.
  Low. Temp. Phys.}\ }\textbf {\bibinfo {volume} {194}},\ \bibinfo {pages}
  {443--452} (\bibinfo {year} {2019})}\BibitemShut {NoStop}%
\bibitem [{\citenamefont {Aghamousa}\ \emph {et~al.}(2016)\citenamefont
  {Aghamousa} \emph {et~al.}}]{Aghamousa:2016zmz}%
  \BibitemOpen
  \bibfield  {author} {\bibinfo {author} {\bibfnamefont {Amir}\ \bibnamefont
  {Aghamousa}} \emph {et~al.} (\bibinfo {collaboration} {DESI}),\ }\bibfield
  {title} {\enquote {\bibinfo {title} {{The DESI Experiment Part I:
  Science,Targeting, and Survey Design}},}\ }\href@noop {} {\  (\bibinfo {year}
  {2016})},\ \Eprint {http://arxiv.org/abs/1611.00036} {arXiv:1611.00036
  [astro-ph.IM]} \BibitemShut {NoStop}%
\bibitem [{\citenamefont {Amendola}\ \emph {et~al.}(2018)\citenamefont
  {Amendola} \emph {et~al.}}]{Amendola:2016saw}%
  \BibitemOpen
  \bibfield  {author} {\bibinfo {author} {\bibfnamefont {Luca}\ \bibnamefont
  {Amendola}} \emph {et~al.},\ }\bibfield  {title} {\enquote {\bibinfo {title}
  {{Cosmology and fundamental physics with the Euclid satellite}},}\ }\href
  {\doibase 10.1007/s41114-017-0010-3} {\bibfield  {journal} {\bibinfo
  {journal} {Living Rev. Rel.}\ }\textbf {\bibinfo {volume} {21}},\ \bibinfo
  {pages} {2} (\bibinfo {year} {2018})},\ \Eprint
  {http://arxiv.org/abs/1606.00180} {arXiv:1606.00180 [astro-ph.CO]}
  \BibitemShut {NoStop}%
\bibitem [{\citenamefont {Ivezić}\ \emph {et~al.}(2019)\citenamefont {Ivezić}
  \emph {et~al.}}]{Ivezic:2008fe}%
  \BibitemOpen
  \bibfield  {author} {\bibinfo {author} {\bibfnamefont {Željko}\ \bibnamefont
  {Ivezić}} \emph {et~al.} (\bibinfo {collaboration} {LSST}),\ }\bibfield
  {title} {\enquote {\bibinfo {title} {{LSST: from Science Drivers to Reference
  Design and Anticipated Data Products}},}\ }\href {\doibase
  10.3847/1538-4357/ab042c} {\bibfield  {journal} {\bibinfo  {journal}
  {Astrophys. J.}\ }\textbf {\bibinfo {volume} {873}},\ \bibinfo {pages} {111}
  (\bibinfo {year} {2019})},\ \Eprint {http://arxiv.org/abs/0805.2366}
  {arXiv:0805.2366 [astro-ph]} \BibitemShut {NoStop}%
\bibitem [{\citenamefont {{Laureijs}}\ \emph {et~al.}(2011)\citenamefont
  {{Laureijs}} \emph {et~al.}}]{2011arXiv1110.3193L}%
  \BibitemOpen
  \bibfield  {author} {\bibinfo {author} {\bibfnamefont {R.}~\bibnamefont
  {{Laureijs}}} \emph {et~al.},\ }\bibfield  {title} {\enquote {\bibinfo
  {title} {{Euclid Definition Study Report}},}\ }\href@noop {} {\bibfield
  {journal} {\bibinfo  {journal} {arXiv e-prints}\ ,\ \bibinfo {eid}
  {arXiv:1110.3193}} (\bibinfo {year} {2011})},\ \Eprint
  {http://arxiv.org/abs/1110.3193} {arXiv:1110.3193 [astro-ph.CO]} \BibitemShut
  {NoStop}%
\bibitem [{\citenamefont {Eisenstein}\ and\ \citenamefont
  {Hu}(1998)}]{Eisenstein:1997ik}%
  \BibitemOpen
  \bibfield  {author} {\bibinfo {author} {\bibfnamefont {Daniel~J.}\
  \bibnamefont {Eisenstein}}\ and\ \bibinfo {author} {\bibfnamefont {Wayne}\
  \bibnamefont {Hu}},\ }\bibfield  {title} {\enquote {\bibinfo {title}
  {{Baryonic features in the matter transfer function}},}\ }\href {\doibase
  10.1086/305424} {\bibfield  {journal} {\bibinfo  {journal} {Astrophys. J.}\
  }\textbf {\bibinfo {volume} {496}},\ \bibinfo {pages} {605} (\bibinfo {year}
  {1998})},\ \Eprint {http://arxiv.org/abs/astro-ph/9709112}
  {arXiv:astro-ph/9709112} \BibitemShut {NoStop}%
\bibitem [{\citenamefont {Lewis}\ and\ \citenamefont
  {Bridle}(2002)}]{Lewis:2002ah}%
  \BibitemOpen
  \bibfield  {author} {\bibinfo {author} {\bibfnamefont {Antony}\ \bibnamefont
  {Lewis}}\ and\ \bibinfo {author} {\bibfnamefont {Sarah}\ \bibnamefont
  {Bridle}},\ }\bibfield  {title} {\enquote {\bibinfo {title} {{Cosmological
  parameters from CMB and other data: A Monte Carlo approach}},}\ }\href
  {\doibase 10.1103/PhysRevD.66.103511} {\bibfield  {journal} {\bibinfo
  {journal} {Phys. Rev.}\ }\textbf {\bibinfo {volume} {D66}},\ \bibinfo {pages}
  {103511} (\bibinfo {year} {2002})},\ \Eprint
  {http://arxiv.org/abs/astro-ph/0205436} {arXiv:astro-ph/0205436 [astro-ph]}
  \BibitemShut {NoStop}%
\end{thebibliography}%

\appendix
\section{Triangular plot for fixed $ f $ DNI}	
\label{sec:app1}
\begin{figure*}[h]
	\centering
	\includegraphics[width=\linewidth]{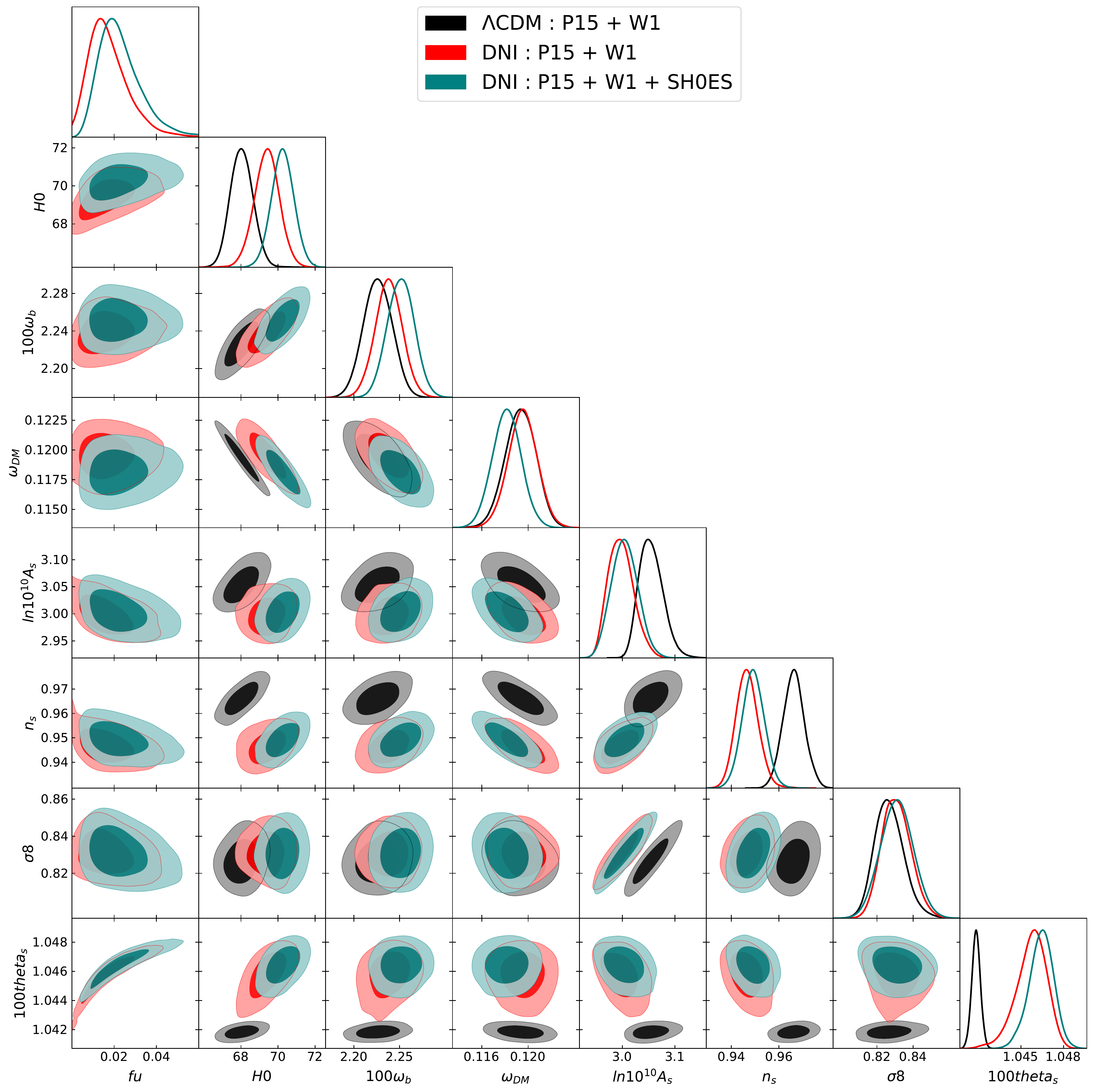}
	\caption{Triangular plot for $ \Lambda $CDM and DNI cosmologies (with $ f = 10^{-3} $) corresponding to Table.~\ref{tbl:chisq} for `P15 + W1' and `P15 + W1 + SH0ES' dataset.}
	\label{fig:u1e-1p2-vs-u1e-1p3triangle}
\end{figure*}

In Fig.~\ref{fig:u1e-1p2-vs-u1e-1p3triangle}, we show the plots of 1-D and 2-D posteriors of all the parameters for fixed-$ f ~(f=10^{-3}) $ DNI analysis both `P15 + W1' and `P15 + W1 + SH0ES' analysis corresponding to Table.~\ref{tbl:chisq} .

\end{document}